 \documentclass[preprint2]{aastex}

\slugcomment{subm. to ApJ}

\shorttitle{Bright X-ray clusters in CFHTLS}
\shortauthors{Mirkazemi et al.}

\begin{document}


\title{Brightest X-ray clusters of galaxies in the CFHTLS wide fields: \\
    Catalog and optical mass estimator}


\author{M. Mirkazemi\altaffilmark{1},
A. Finoguenov\altaffilmark{1,2,3},
M.~J. Pereira\altaffilmark{4},
M. Tanaka\altaffilmark{5},
M. Lerchster\altaffilmark{1},
F. Brimioulle\altaffilmark{6},
E. Egami\altaffilmark{4},
K. Kettula\altaffilmark{2,12},
G. Erfanianfar\altaffilmark{13,1},
H. J. McCracken\altaffilmark{7},
Y. Mellier\altaffilmark{7},
J. P. Kneib\altaffilmark{8},
E. Rykoff\altaffilmark{9},
S. Seitz\altaffilmark{1,6},
T. Erben\altaffilmark{10},
J. E. Taylor\altaffilmark{11}
}
\email{kazemi@mpe.mpg.de}

\altaffiltext{1}{Max-Planck-Institut f\"ur extraterrestrische Physik, Giessenbachstra\ss e, D-85740 Garching, Germany}
\altaffiltext{2}{Department of Physics, University of Helsinki, Gustaf H\"allstr\"omin katu 2a, FI-00014 Helsinki, Finland.}
\altaffiltext{3}{University of Maryland Baltimore County, 1000 Hilltop circle, Baltimore, MD 21250, USA}
\altaffiltext{4}{Steward Observatory, University of Arizona, 933 North Cherry Avenue, Tucson, AZ 85721, USA}
\altaffiltext{5}{National Astronomical Observatory of Japan 2-21-1 Osawa, Mitaka, Tokyo, 181-8588, Japan}
\altaffiltext{6}{Universit\"{a}tssternwarte M\"{u}nchen, Scheinerstrasse 1, 81679 M\"{u}unchen, Germany}
\altaffiltext{7}{Institut d'Astrophysique de Paris, UMR7095 CNRS, Universit\'e Pierre et Marie Curie, 98 bis Boulevard Arago, 75014 Paris, France}
\altaffiltext{8}{Laboratoire d'Astrophysique de Marseille, CNRS-Universit\'e, Pôle de l'Etoile Site de Ch\^ateau-Gombert 38, rue Fr\'ed\'eric Joliot-Curie, F-13388 Marseille Cedex 13, France}
\altaffiltext{9}{SLAC National Accelerator Laboratory, Menlo Park, CA 94025, USA}
\altaffiltext{10}{Argelander Institute for Astronomy, University of Bonn, Auf dem H\"{u}gel 71, D-53121 Bonn, Germany}
\altaffiltext{11}{Department of Physics and Astronomy, University of Waterloo, 200 University Avenue West, Waterloo, ON N2L 3G1, Canada}
\altaffiltext{12}{Helsinki Institute of Physics, P.O. Box 64, FIN-00014 University of Helsinki, Finland}
\altaffiltext{13}{Excellence Cluster Universe, Boltzmannstr. 2, 85748 Garching bei M\"{u}nchen, Germany}




\begin{abstract}

  The CFHTLS presents a unique data set for weak lensing
  studies, having high quality imaging and deep multi-band
  photometry. We have initiated an XMM-CFHTLS project to provide X-ray
  observations of the brightest X-ray selected clusters within the wide CFHTLS
  area. Performance of these observations and the high quality of
  CFHTLS data, allows us to revisit the identification of X-ray
  sources, introducing automated reproducible algorithms, based on the
  multi-color red sequence finder. We have also introduced a new
  optical mass proxy.  We provide the
  calibration of the red sequence observed in the CFHT filters and
  compare the results with the traditional single color red
  sequence and photoz. We test the identification algorithm on the subset
  of highly significant XMM clusters and identify 100\% of the
  sample. We find that the integrated z-band luminosity of the red
  sequence galaxies correlates well with the X-ray luminosity with a
  surprisingly small scatter of 0.20 dex. We further use the
  multi-color red sequence to reduce spurious detections in the full
  XMM and RASS data sets, resulting in catalogs of 196 and 32
  clusters, respectively. We made spectroscopic follow-up 
  observations of some of these systems with HECTOSPEC and 
  in combination with BOSS DR9 data.  We also describe the
  modifications needed to the source detection algorithm in order to
  keep high purity of extended sources in the shallow X-ray data.  
  We also present the scaling relation between X-ray luminosity 
  and velocity dispersion.

\end{abstract}





\section{Introduction}
\label{intro}
   
In the past two decades the accelerating expansion of the universe has
been confirmed by several experiments, such as observations of
supernovae (e.g. \citealt{Riess98}; \citealt{Perlmutter99}) and
measurements of the cosmic microwave background
(e.g. \citealt{Spergel03}).  This acceleration is thought to be a
consequence of dark energy density which, in the simplest way, can be
modelled by a non-zero Einstein's cosmological constant. Understanding
the origin of the associated phenomenon of dark energy has been set
among the most important tasks for understanding the formation and
evolution of the Universe. Galaxy clusters play an important role in
this through their sensitivity to the growth of structure. One of the
first efforts in constraining cosmology with galaxy clusters was made
by \cite{Borgani01}. They measured $\Omega_{M}$ using 103 galaxy
clusters in the ROSAT Deep Cluster Survey (RDCS; \citealt{Rosati98})
out to z$\simeq$0.85. In the subsequent study, \cite{Vikhlinin09}
obtained updated measurements of $\Omega_{M}h$, as well as the dark
energy equation-of-state, $\omega_{0}$, and the amplitude of power
spectrum, $\sigma_{8}$. For a review of cosmological constraints
obtained using galaxy clusters in the past decade, see
\cite{Weinberg12} and \cite{Allen11}.  The 2013 Planck results have
revealed a tension between a combination of CMB TT fluctuation
spectrum and baryonic acoustic oscillation (BAO) measurements versus
galaxy cluster abundance (\citealt{PlanckXX13}). The physical
interpretation of the results in view of the non-zero neutrino mass,
requires a robust understanding of the cluster scaling relations. 

From an astrophysical point of view, X-ray cluster survey data
provide an important definition of high-density environment, critical
for studies of galaxy formation e.g. \citealt{Tanaka08};
\citealp{Giodini09,Balogh11,Giodini12}) and active galactic nuclei
(AGN) (e.g. \citealt{Silverman09}, \citealt{Tanaka12a},
\citealt{Allevato12}).

The main aim of this {\it Paper} is to address the cluster identification
using CFHTLS data, and to provide the cluster sample and scaling relations
between optical and X-ray luminosity. The calibration between weak
lensing mass and X-ray observables (luminosity and temperature) will
be presented in Kettula et al. (subm.).

Optical galaxy cluster searches are often hindered by galaxy
projection effects. Several algorithms have been applied to solve this
problem. In addition to employing photometric methods such as red
sequence identification (\citealt{Gladders00}) and MaxBCG
(\citealt{Annis99}; \citealt{Koester07}), the detection of extended
X-ray sources is often a reliable indication of galaxy clusters
(\citealt{Rosati02}). With the increased number of X-ray surveys in
the past decade such as Chandra Deep Field North (CDFN;
\citealt{Bauer02}), Chandra Deep Field South (CDFS;
\citealt{Giacconi02}), Lockman Hole (\citealt{Finoguenov05}), Cosmic
Evolution Survey (COSMOS; \citealt{Finoguenov07}), XMM-Large Scale
Structure (XMM-LSS; \citealt{Pacaud07}), Canadian Network for
Observational Cosmology (CNOC2; \citealt{Finoguenov09}) and Subaru-XMM
Deep Field (SXDF; \citealt{Finoguenov10}), X-ray astronomy introduced
itself as an efficient cluster and group detection tool. In addition,
X-ray properties of clusters can be used to best characterise the
cluster mass, a requirement for precision cosmology work
(\citealt{Kravtsov06}; \citealt{Nagai07}).

In this paper, we explore the use of multi-wavelength data to identify
X-ray clusters within the RASS data. RASS data are both faint and
unresolved, so cluster confirmation is challenging. In order to
establish a reliable method, we used the highly significant extended
sources, obtained through our XMM-Newton follow-up program. We start
with a description of the XMM data reduction and detection of extended
sources in $\S$2. In $\S$3 we present the cluster identification and
validation, including spectroscopic follow-up program and velocity
dispersion measurements for a subsample of clusters. $\S 4$ provides
the X-ray cluster catalogs both for XMM and RASS and compares the
optical luminosity and X-ray luminosity of clusters. In $\S$5 we
summarise and discuss the results.

 Throughout this paper, we use the AB magnitude system and consider a
 cosmological model with $H_{0} = 72$ $\mathrm{km}$ $\mathrm{s^{-1}}$
 $\mathrm{Mpc^{-1}}$, $\Omega_{\Lambda} = 0.75$ and $\Omega_{M} =
 0.25$.

\section{Data} \label{data}
\subsection{X-ray data} \label{X-ray}

The main aim of the XMM-CFHTLS program is to efficiently find massive
galaxy clusters, through a series of short XMM-Newton follow-up
observations of faint RASS sources (\citealt{Voges99}) identified as
galaxy clusters using CFHTLS imaging data. In total, 73 observations
of cluster candidates have been performed, using 220ks of allocated
time. At the time of scheduling XMM observations, only T0005 CFHTLS
data have been publicly released, which covered 100 square degrees in
partial W1 and W4 fields and the full W2 field. In order to use the
mosaicing mode of XMM-Newton, we had to fulfil the re-pointing
constraint of 1 degree. Given the low density of RASS sources, the
number of robust clusters were rather low and for XMM snap-shot
observations, we also pointed at the RASS sources identified with a
photo-z galaxy overdensity. Performance of this program has allowed us
to both select the adequate method for cluster identification and to
perform extensive XMM studies of optically selected clusters.

The current RASS catalogs include ~122 square degrees (in W1, W2, and W4), 
while we only have observed with XMM the clusters selected from $\sim$90 square 
degrees (in W2, W4, and half of W1).
We would like to advise against using our data for studying the cluster
abundance, as our program selectively points to clusters selected from
90 square degrees, while covering 14 square degrees.
Use of our catalogs for cluster abundance studies would need to both
account for RASS sensitivity and only use our RASS source list, while
some of the bright XMM sources were filler optical clusters to ensure
repointing constraints.

In our final catalog, we also include existing serendipitous
observations, since some candidate clusters have already been
previously observed with XMM. We exclude from our survey the XMM-LSS
(and XXL) fields, where clusters are identified by the corresponding
teams (e.g. \citealt{Pacaud07}). We point out interested readers to
\cite{Gozaliasl14} where we present our catalog using the 3 square
degree overlap between XMM-LSS survey and CFHTLS.

Our survey methodology is to cover a large area of the sky with short
X-ray exposures. The detection of sources in such a shallow survey
explores the Poisson regime, so there is a need for tailored data
reduction methods. Confirming RASS sources does not require any
sophisticated modelling, given that they are typically $>20\sigma$
sources, but detection of fainter serendipitous sources requires a new
approach.

The procedure of \cite{Finoguenov07,Finoguenov09} with updates
described in \cite{Bielby10} has been further revised to store the
locally estimated background and exposure maps separately in order to
treat the Poisson noise within the source detection program (wvdetect
- \citealt{Vikhlinin98}). Furthermore, we modified the ratio of
thresholds for point and extended sources, setting the detection of
point sources to $3.3\sigma$ and that of extended sources to
$4.6\sigma$.  This choice of thresholds prevents detection of point
sources only on large spatial scales. The consideration of the
detection effect is very general, but the ratio of thresholds is
tailored for the XMM PSF and the scales of source detection we
employ. In detecting the extended source, we avoid detecting the point
sources, by detecting them on small scales and subtracting their flux
according to PSF model, so no detection occurs on any scale
anymore. The terms small and large scales are specific to XMM and
refer to scales below and above 16$^{\prime\prime}$. If the source is
not detected on small scales, but only detected on large scales, it
would be mistaken for an extended source. An example of such a
detection is a source with 3 counts in the central 16$^{\prime\prime}$
radius and 2 more counts beyond this radius.  For XMM-Newton, the PSF
model predicts 40\% of the point source flux to occur on the scales
we use for the extended source detection. The odds of not detecting
the central 60\% of the point source flux, while detecting the 100\%
of the source flux by including the outskirts are large, especially if
only a few counts suffice a detection. To beat this contamination
down, we need to increase the threshold for detecting the large
scales, so that odds of detecting the outer 40\% of the flux with a
new large threshold and not detecting the central flux of the source
with the original threshold are small, where small is set to be 1\%,
since this makes a 10\% contamination to extended sources, given that
point sources are 10 times more abundant. We also decrease the
threshold for detecting the flux on small scales. Given the PSF shape
of XMM, we find the suitable detection limits to be 3.3 $\sigma$ for
the central flux and 4.6 $\sigma$ for the outskirts. We also require
the significance of the flux determination associated with the
detection to be above 4.6 $\sigma$. The problem described above is
typical to shallow surveys, and e.g., will be important for eROSITA
\citep{Predehl10}. In deep surveys, extended source detection is
background limited, which requires more counts for large scales to be
detected at similar thresholds and so the flux on small scales is
always detected from a point source.

The 4.6 $\sigma$ threshold XMM source list is expected to have
  less than 10\% contamination of point sources to the extended source
  catalogs, which we consider acceptable, given that the highest
  identification rate for extended sources in deep fields is 90\%
  (e.g. \citealt{Finoguenov10}). The corresponding chance
  identification rate is expected to be below 2\%. These estimates are
  conservative, since all sources in this list were identified. As
in our previous work, while removing flux from point sources, we are
not going through the step of cataloguing the sources, as we model the
point-source contamination by convolving the wavelet images on small
scales with a kernel reproducing the PSF shape on large scales.

For the provisional catalog of sources found at lower X-ray $\sigma$
($<$ 4.6), the contamination from point sources increases to 50\%. The
final rate for spurious identification for such source selection is
reduced due to sparse density of matching sources (optical clusters)
and amounts to 10\%.  Given the high expected level of chance
identification, this catalog is not included in the analysis of
scaling relation between X-ray luminosity and integrated optical
luminosity.

\subsection{Optical, photometric redshift and spectroscopic data} \label{optical}

During 2003--2009, the 3.6-m Canada-France-Hawaii Telescope (CFHT)
completed a very large imaging programme known as the
Canada-France-Hawaii Telescope Legacy Survey (CFHTLS) using the 2048
$\times$ 4612 pixel wide field optical imaging camera MegaCam. With
a 0.185 arcsec pixel size, CFHT MegaCam gives a 0.96 degree $\times$
0.96 degree field of view. All the observations were done in dark and
grey telescope time ($\sim$ 2\,300 hours). Four wide fields of this
survey, with a total area amounting to 170 square degrees, were
observed in $u^\ast$, $g^\prime$, $r^\prime$, $i^\prime$ and
$z^\prime$ band down to $i^\prime$=24.5. 
In this work, we use the T0007\footnote{http://terapix.iap.fr/cplt/T0007/doc/T0007-doc.pdf} 
data release of CFHTLS and corresponding photometric redshift 
catalog\footnote{ftp://ftpix.iap.fr/pub/CFHTLS-zphot-T0007}.
The photometric redshifts were computed similar to the methods 
of \cite{Ilbert06,Coupon09} .
The photometric redshift catalog is limited to $i^\prime=24$ and
according to the report
of CFHTLS team, the achieved photometric redshift accuracy and outlier rates are
$\sigma_{\Delta{z}\setminus{1+z}}$ $\cong$ 0.07 and $\eta \cong 13 \%$
for galaxies with $22.5 \le i^\prime \le 23.5$ (almost the faintest galaxies in this survey).
We use optical data from three wide fields of CFHTLS: W1, W2 and W4.

Follow-up observations of clusters in W1, W2 and W4 fields were
performed using Hectospec on MMT. Hectospec is a 300-fiber
multi-object spectrograph with a circular field of view of 1$^{\circ}$
in diameter (\citealt{Fabricant05}). We used the 270 line grating,
which provides a wide wavelength range (3\,650 -- 9\,200 $\AA$) at 6.2
$\AA$ resolution. We reduced the spectra and measured redshifts using
the HSRED pipeline (\citealt{Cool05}).  Redshifts were determined by
comparing the reduced spectra with stellar, galaxy and quasar template
spectra and choosing the template and redshift which minimises the
$\chi^2$ between model and data. We then visually inspected the
template fits and assigned quality flags based on the certainty of the
redshift estimate.

Targets for spectroscopic follow-up were culled from the list of
candidates in the XMM-CFHTLS fields and prioritised based on a
combination of their X-ray flux and photometric redshift. High
priority clusters (with X-ray flux $> 7 \times 10^{-14}$ ergs
cm$^{-2}$ s$^{-1}$ and $0.15< z < 0.6$) dictated the
locations of the Hectospec pointings; fainter clusters or clusters
beyond these redshift limits were used as fillers, and therefore only
observed if they lay within $30^{\prime}$ of a high priority
target. AGN candidates based on the XMM-CFHTLS point source catalogs
were also used as low priority fillers. The cluster follow-up strategy
used varied according to the certainty in the red sequence redshift
estimate. For clusters with reliable redshifts, i.e. with high number of
red sequence galaxies, we use photometric redshift catalogs to
select only galaxies which lie in the photo-z slice ($dz<n \times
(1+z) \times \sigma_\mathrm{photoz}$, where $\sigma_\mathrm{photoz}$
is the photometric redshift error and $n$ is an integer number between
2 and 4).  The red sequence significance, $\alpha$, is a parameter
that shows the overdensity of galaxies in comparison to the number of
background galaxies at the cluster redshift.  This parameter will be
defined more accurately in section \ref{red_seq_method}.  This
narrower target selection means we were able to explore the infall
regions of the clusters out to larger radii. For clusters with few
number of photo-z counterparts, we performed a magnitude limited
survey at smaller radial distances, with the goal of identifying the
optical counterparts and securing a redshift for the X-ray emission.
Over the 3 fields, 32 fiber configurations were observed, mainly in W1
and W2, and secure redshifts for 6\,170 objects were measured.

In performing the analysis, we have also added spectroscopic
data in W1, W2 and W3 from SDSS-III survey (\citealt{Aihara11}). In
total, we have 13k, 3.5k and 9k spectroscopic redshifts in W1, W2 and
W4.

\begin{figure*}[]

\mbox{

\includegraphics[width=0.32\textwidth]{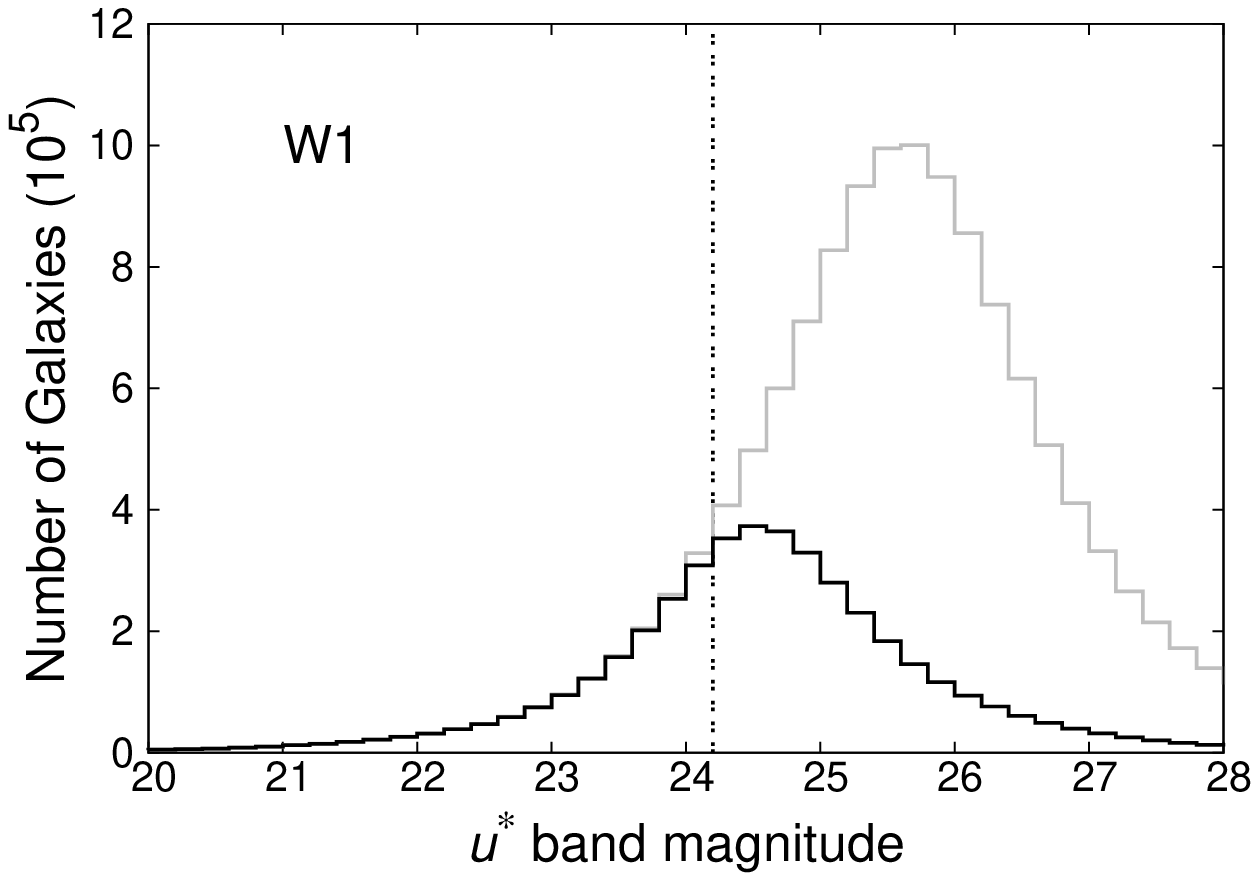}

\includegraphics[width=0.32\textwidth]{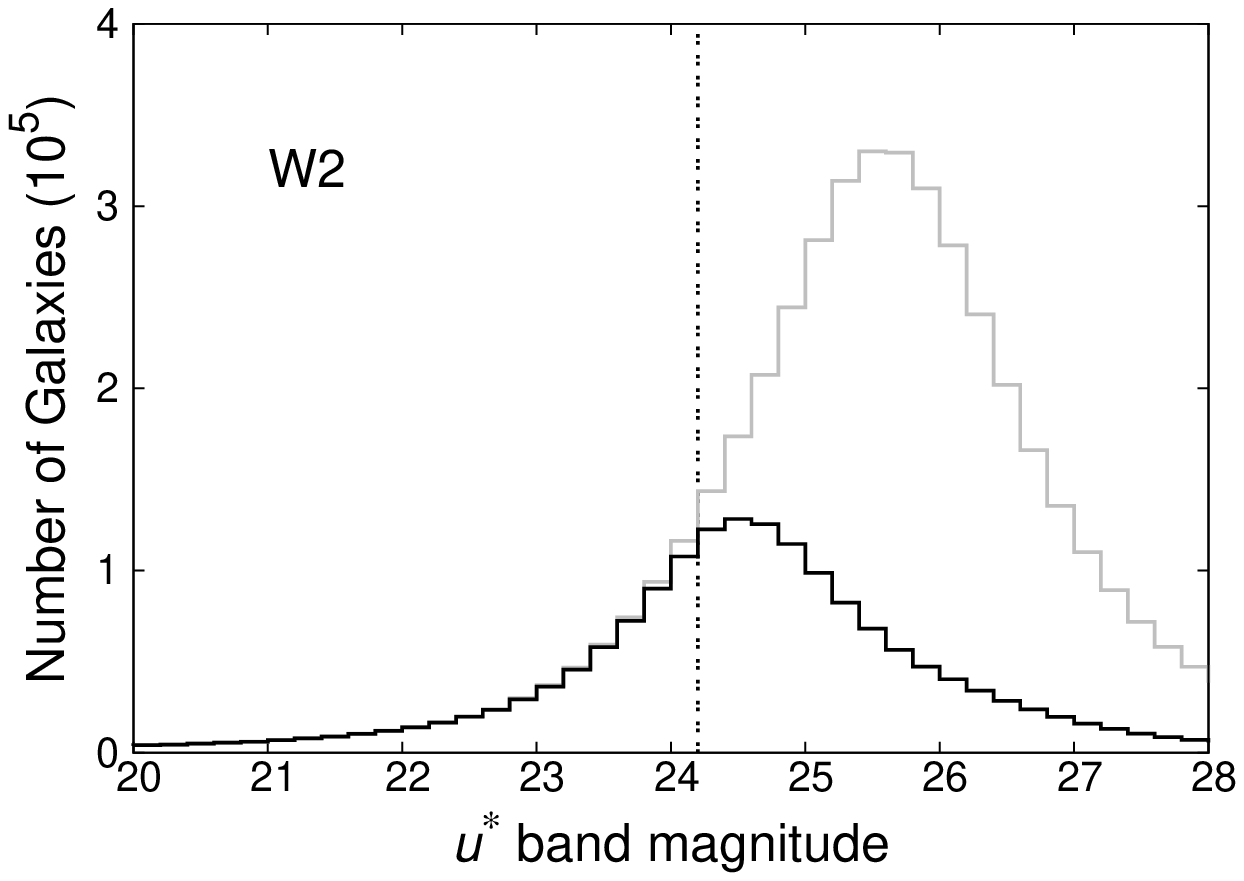}

\includegraphics[width=0.32\textwidth]{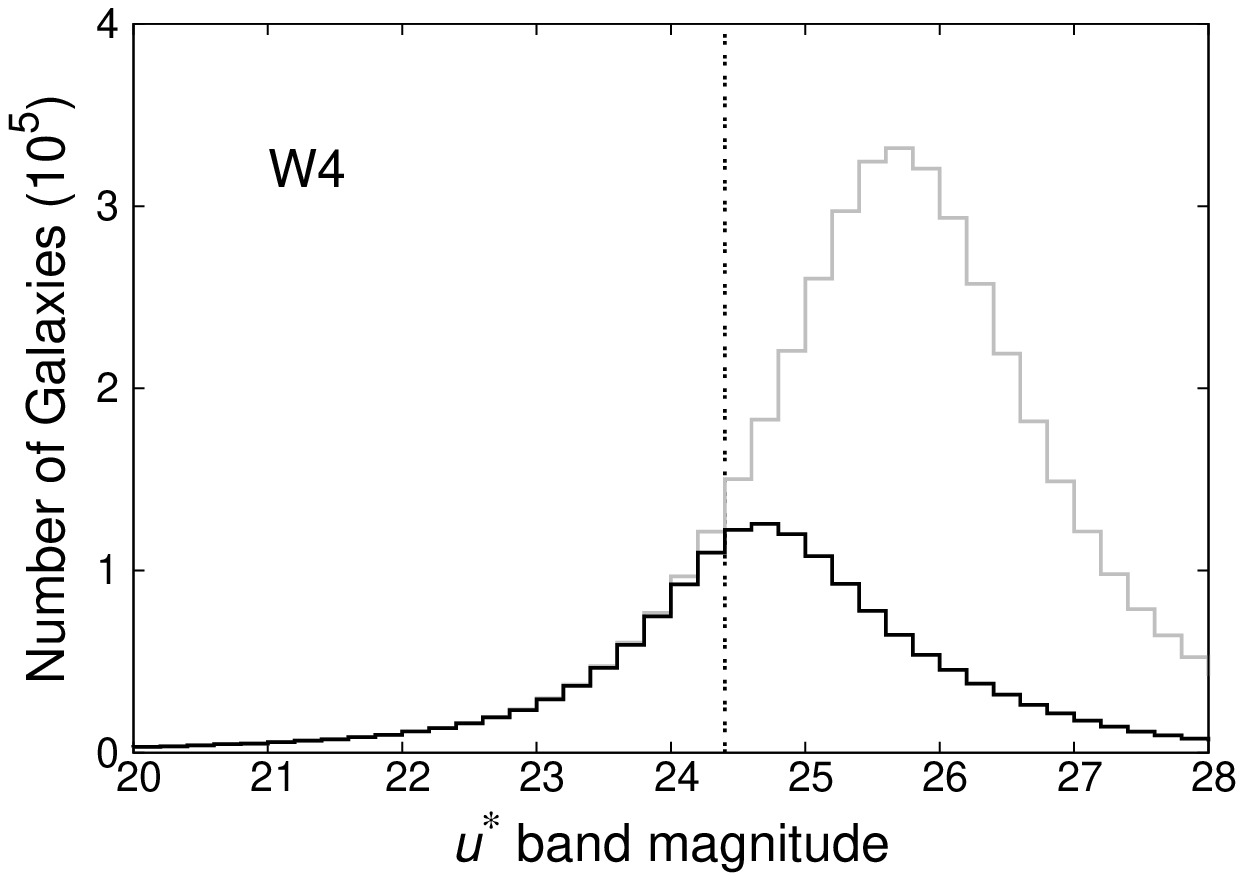}
}

\mbox{

\includegraphics[width=0.32\textwidth]{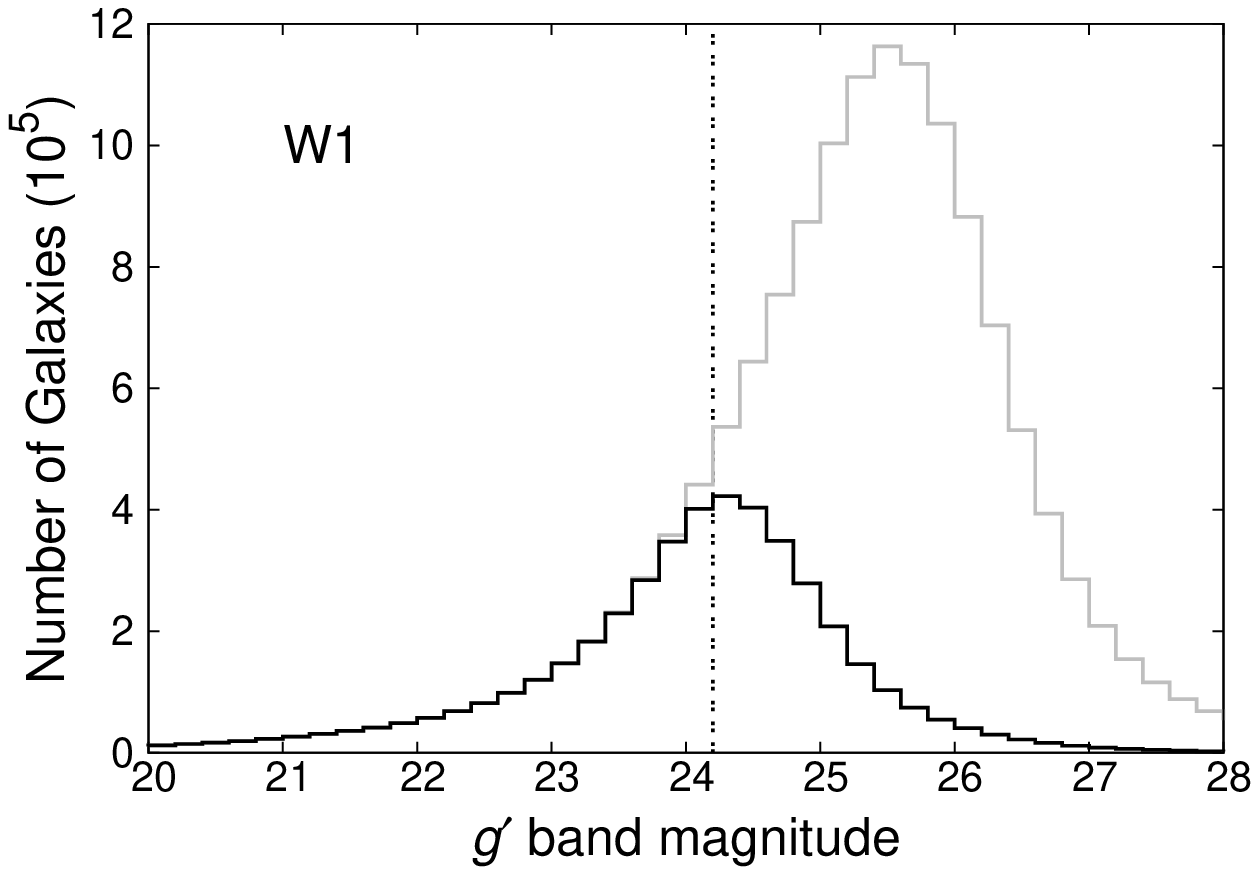}

\includegraphics[width=0.32\textwidth]{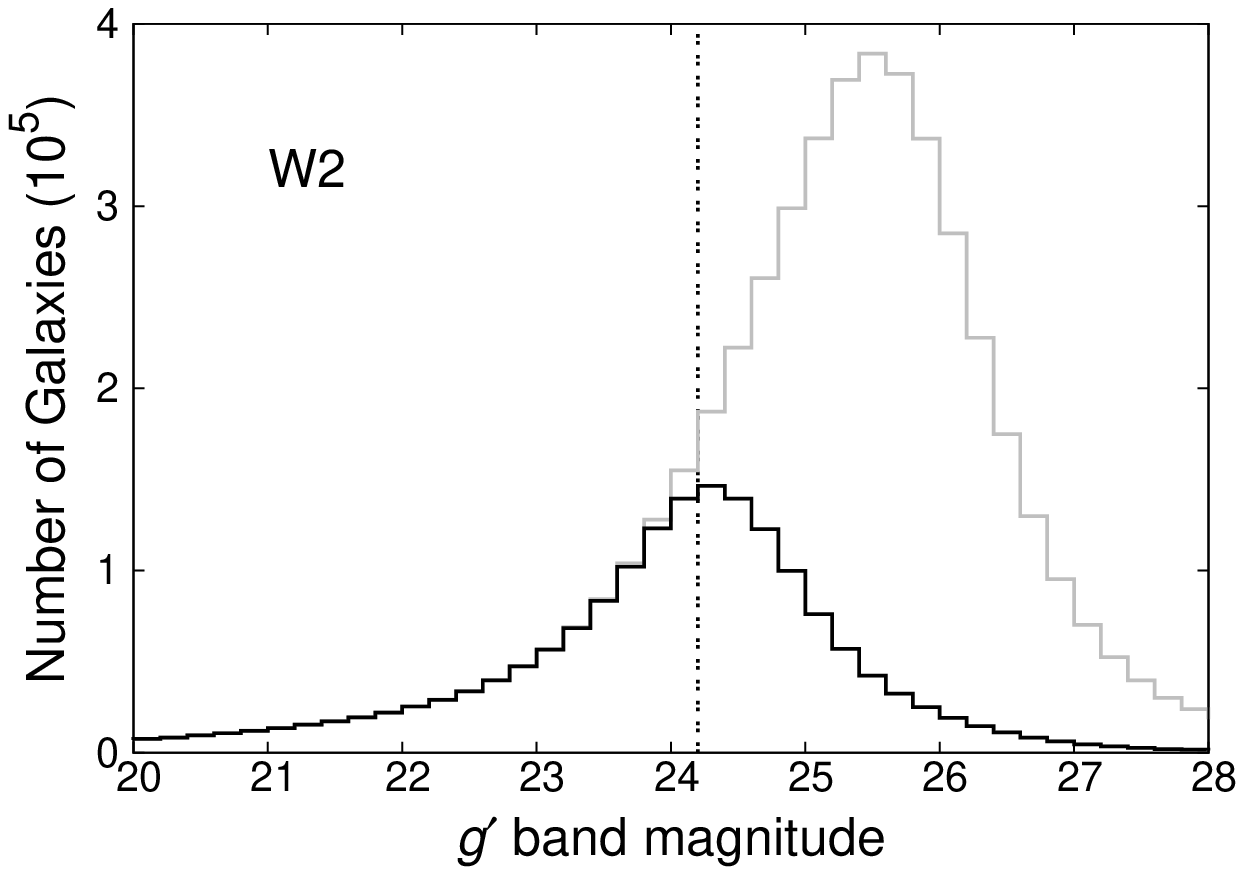}

\includegraphics[width=0.32\textwidth]{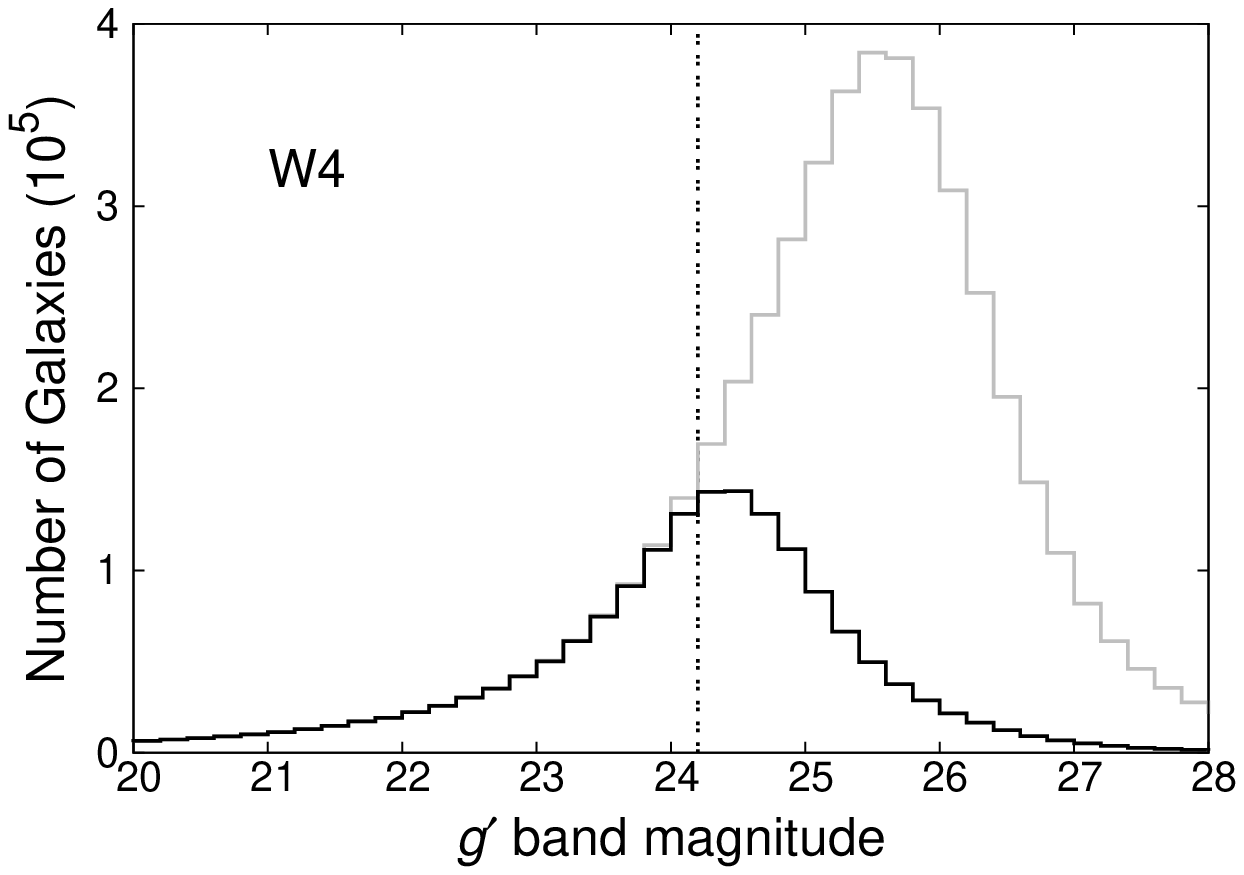}
}

\mbox{

\includegraphics[width=0.32\textwidth]{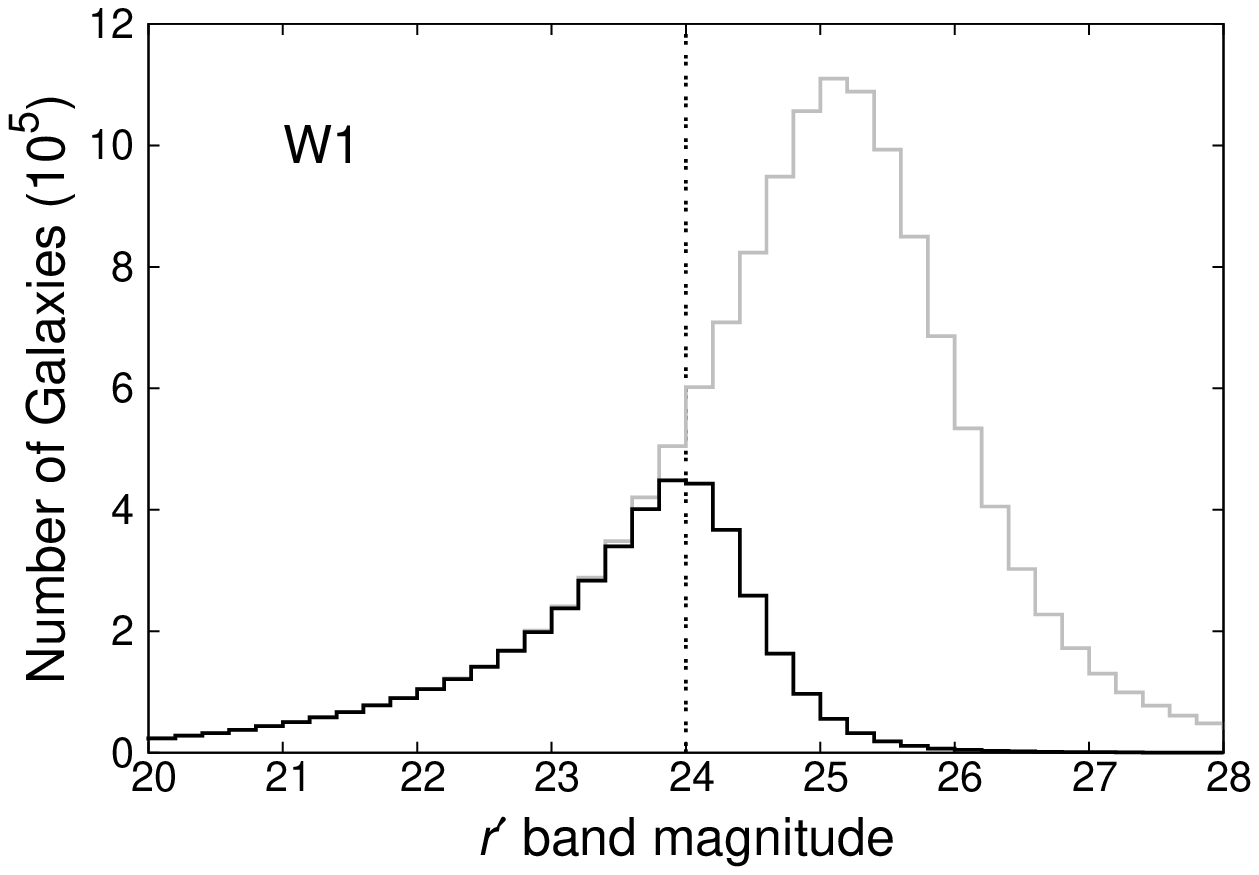}

\includegraphics[width=0.32\textwidth]{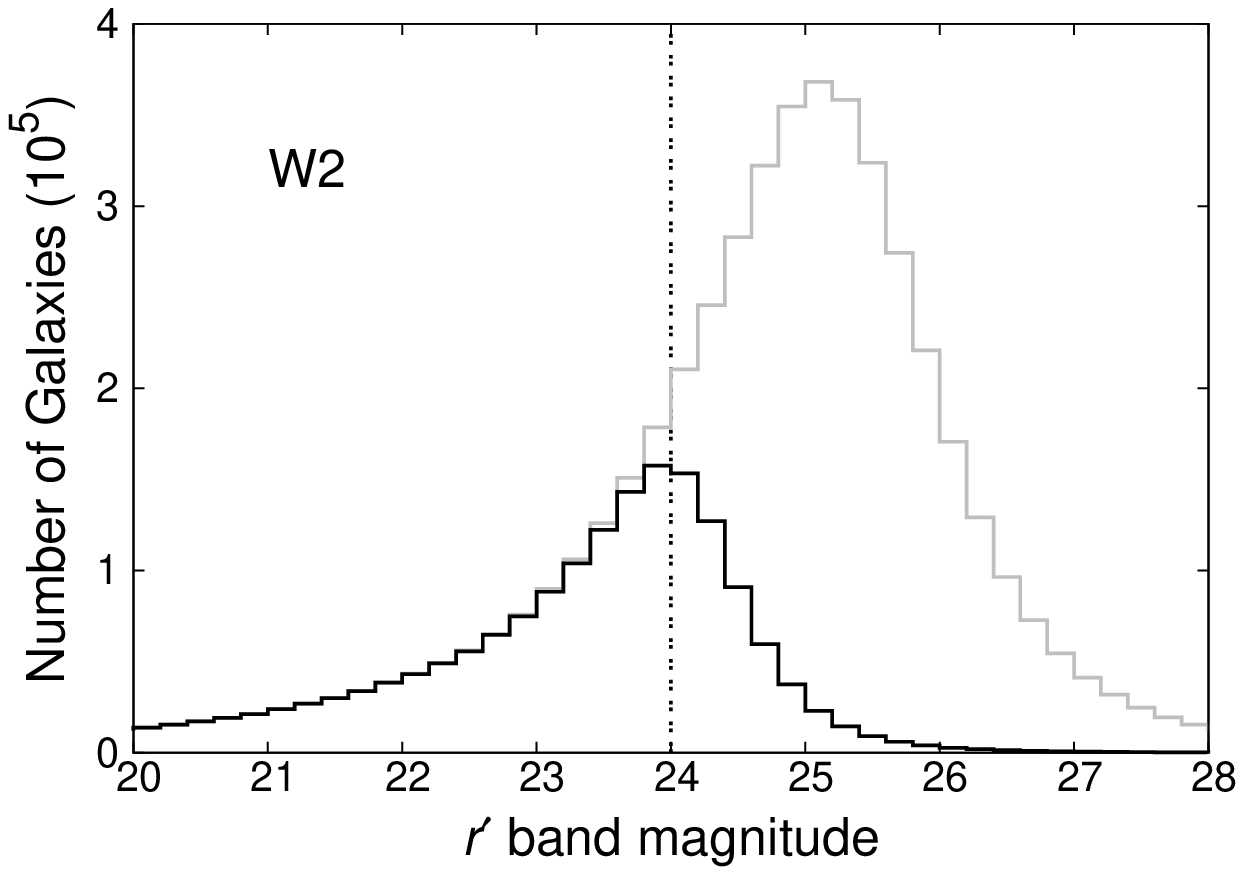}

\includegraphics[width=0.32\textwidth]{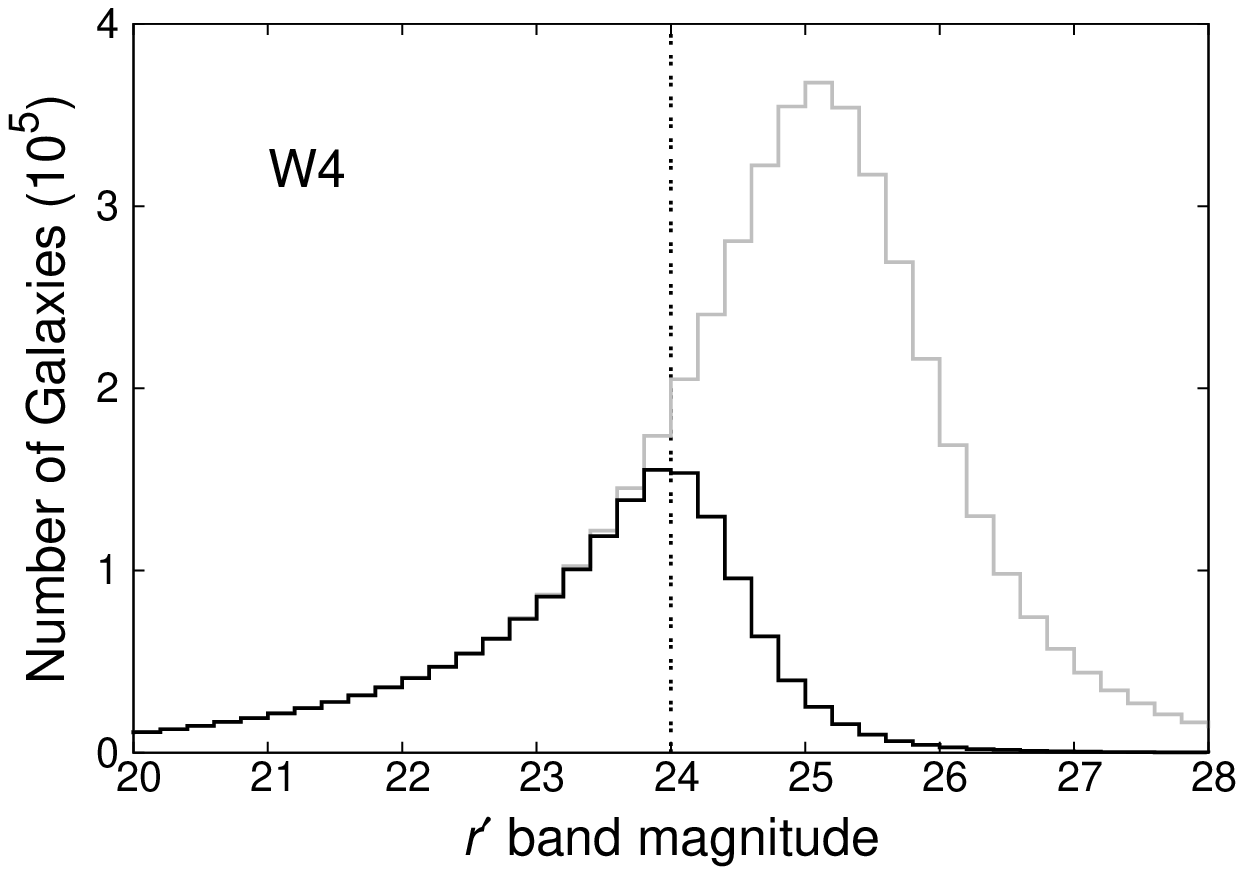}
}

\mbox{

\includegraphics[width=0.32\textwidth]{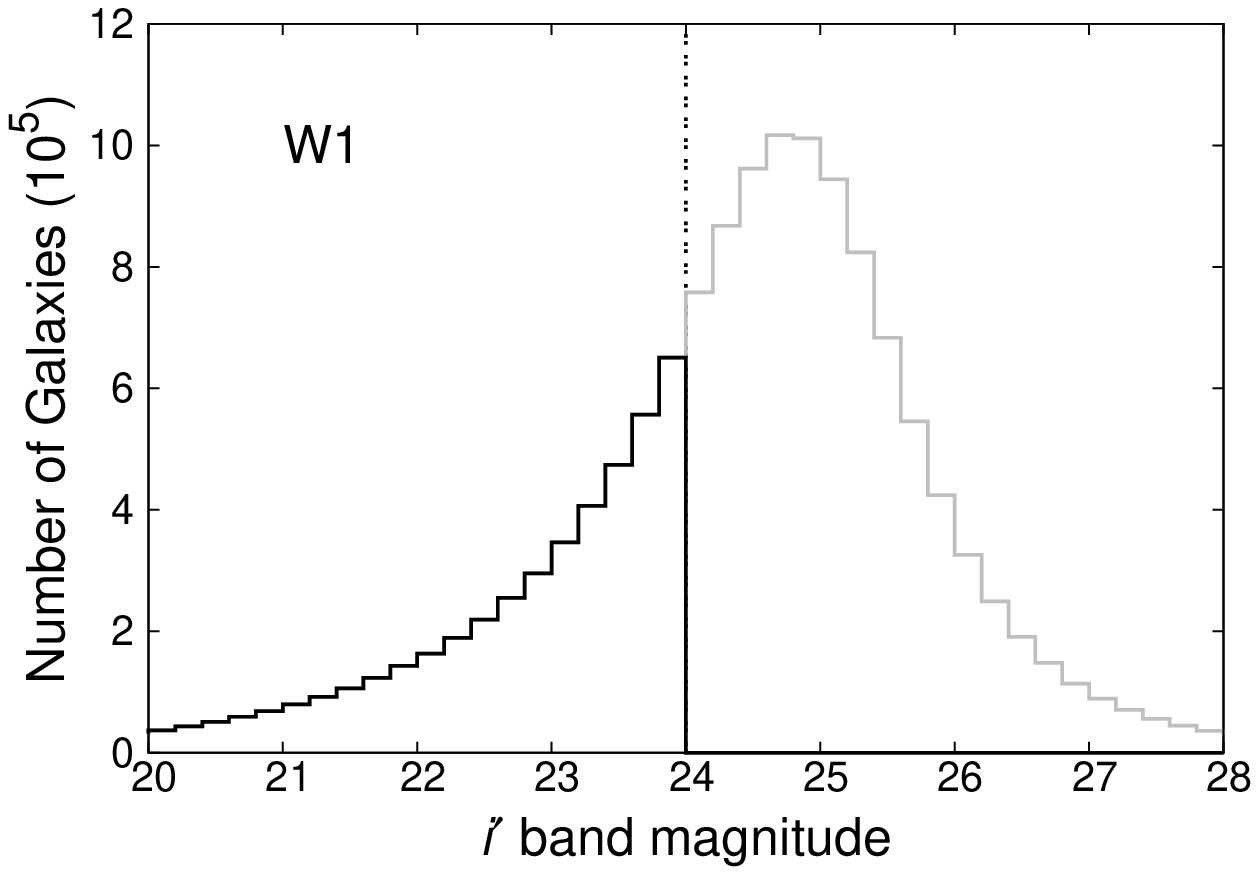}

\includegraphics[width=0.32\textwidth]{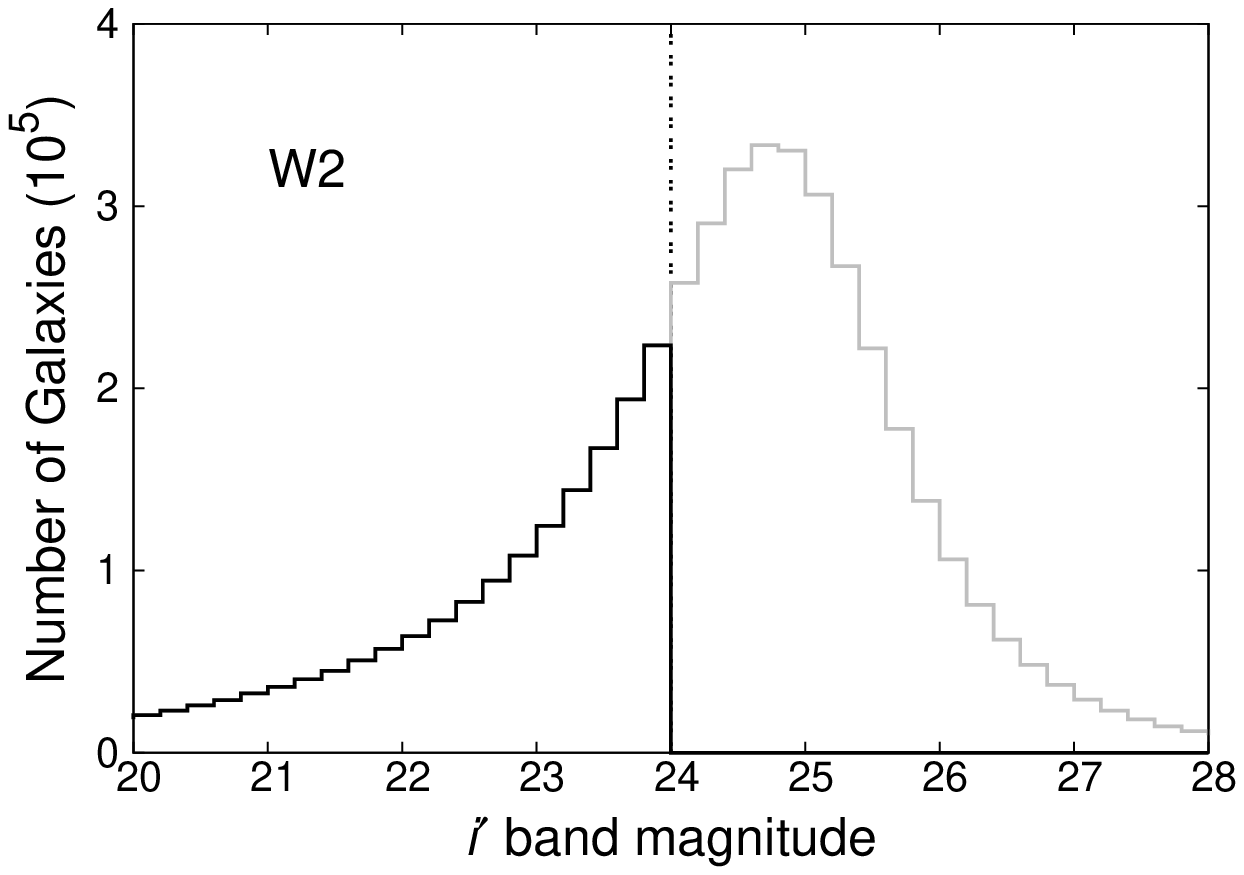}

\includegraphics[width=0.32\textwidth]{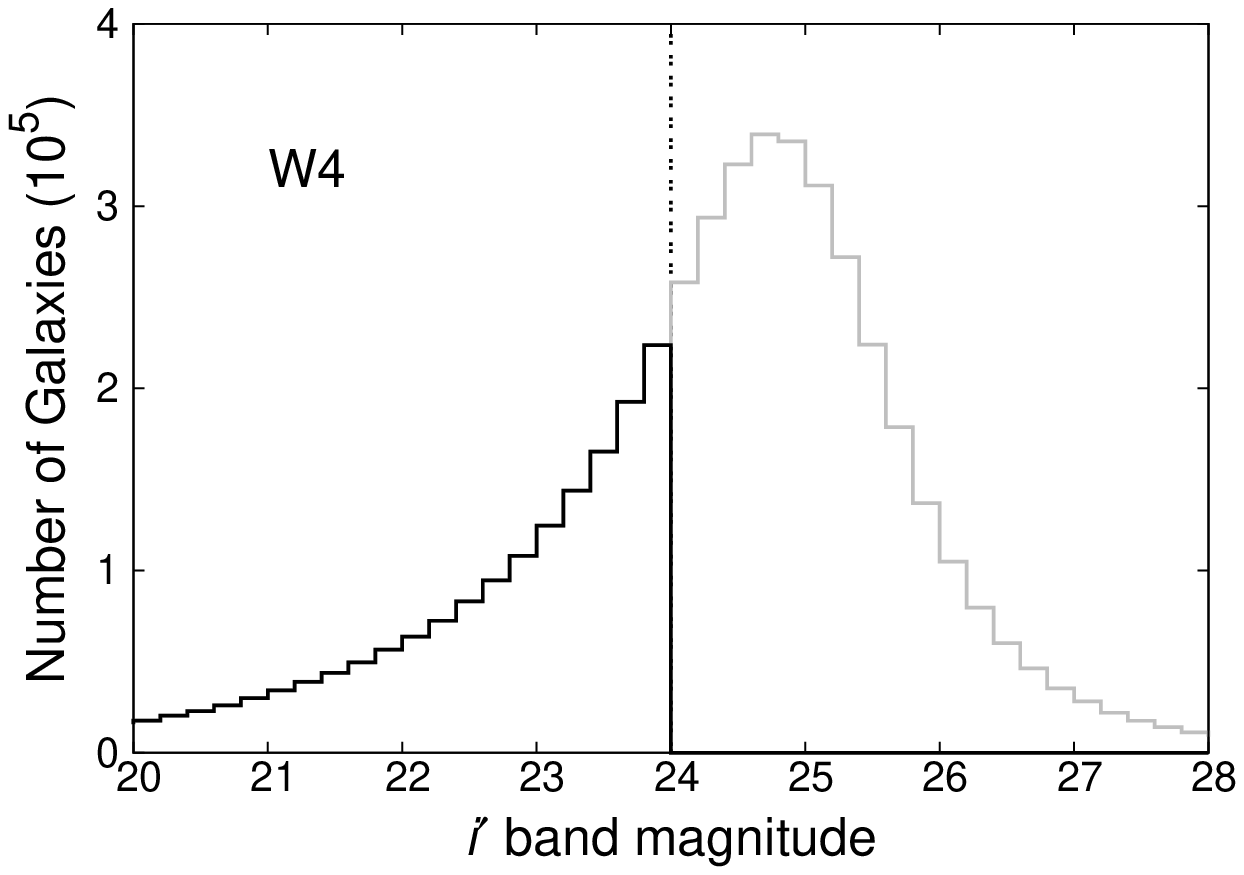}
}

\mbox{

\includegraphics[width=0.32\textwidth]{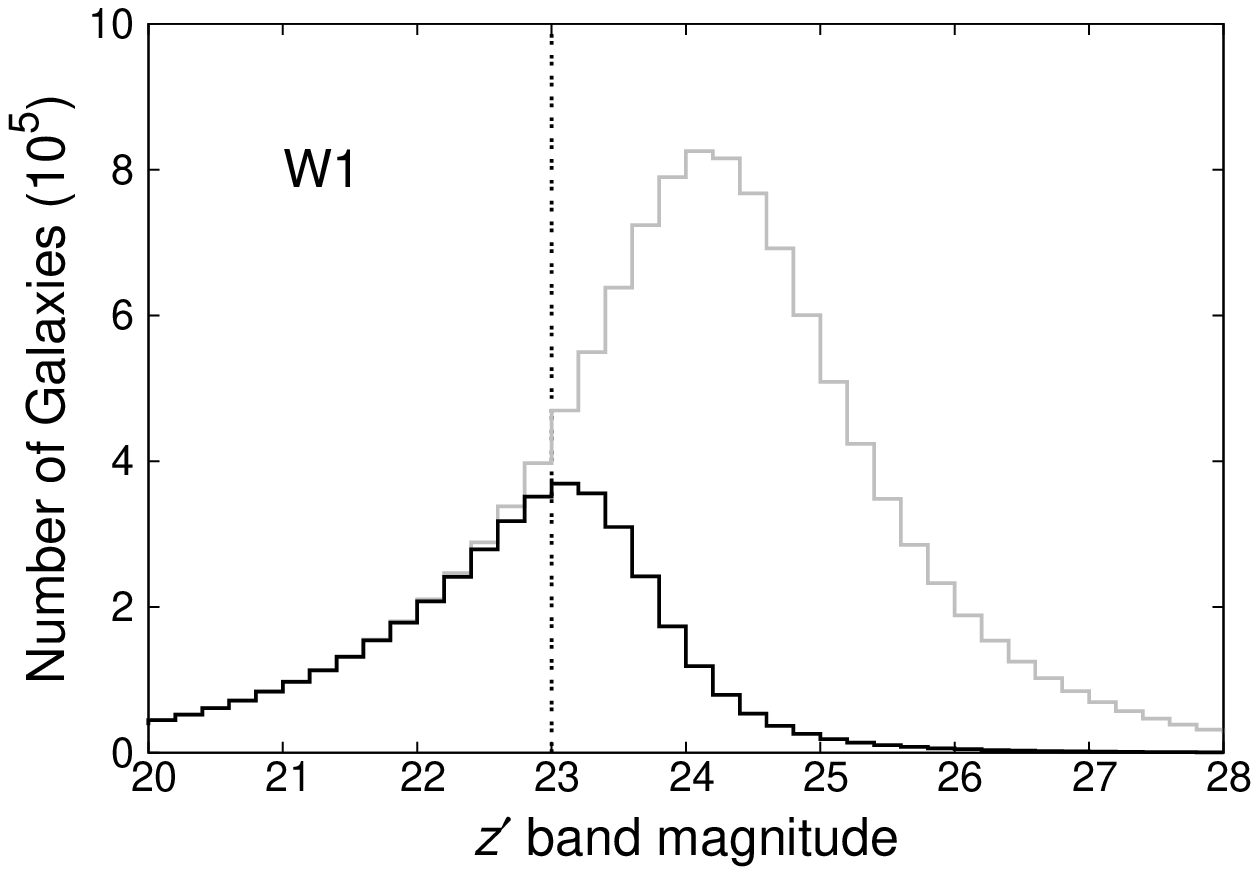}

\includegraphics[width=0.32\textwidth]{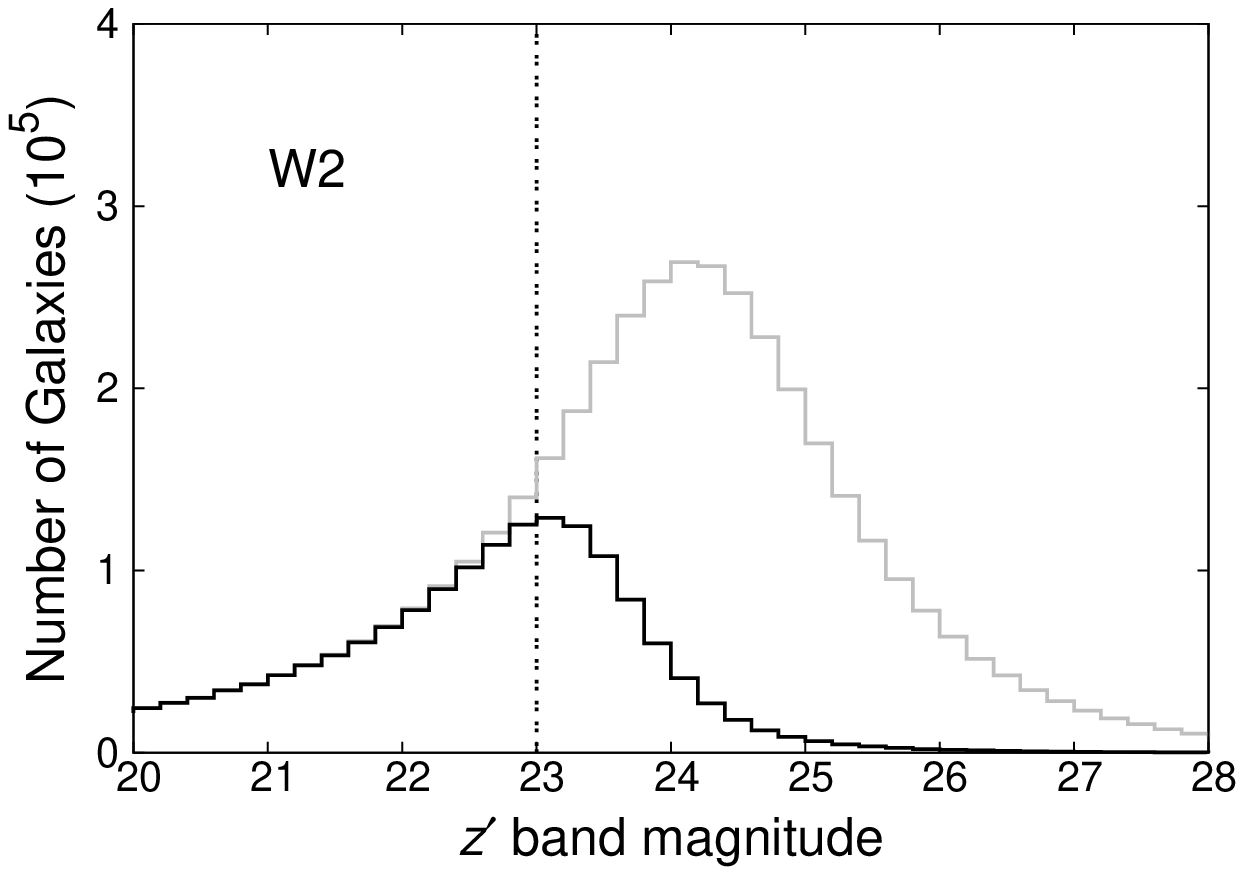}

\includegraphics[width=0.32\textwidth]{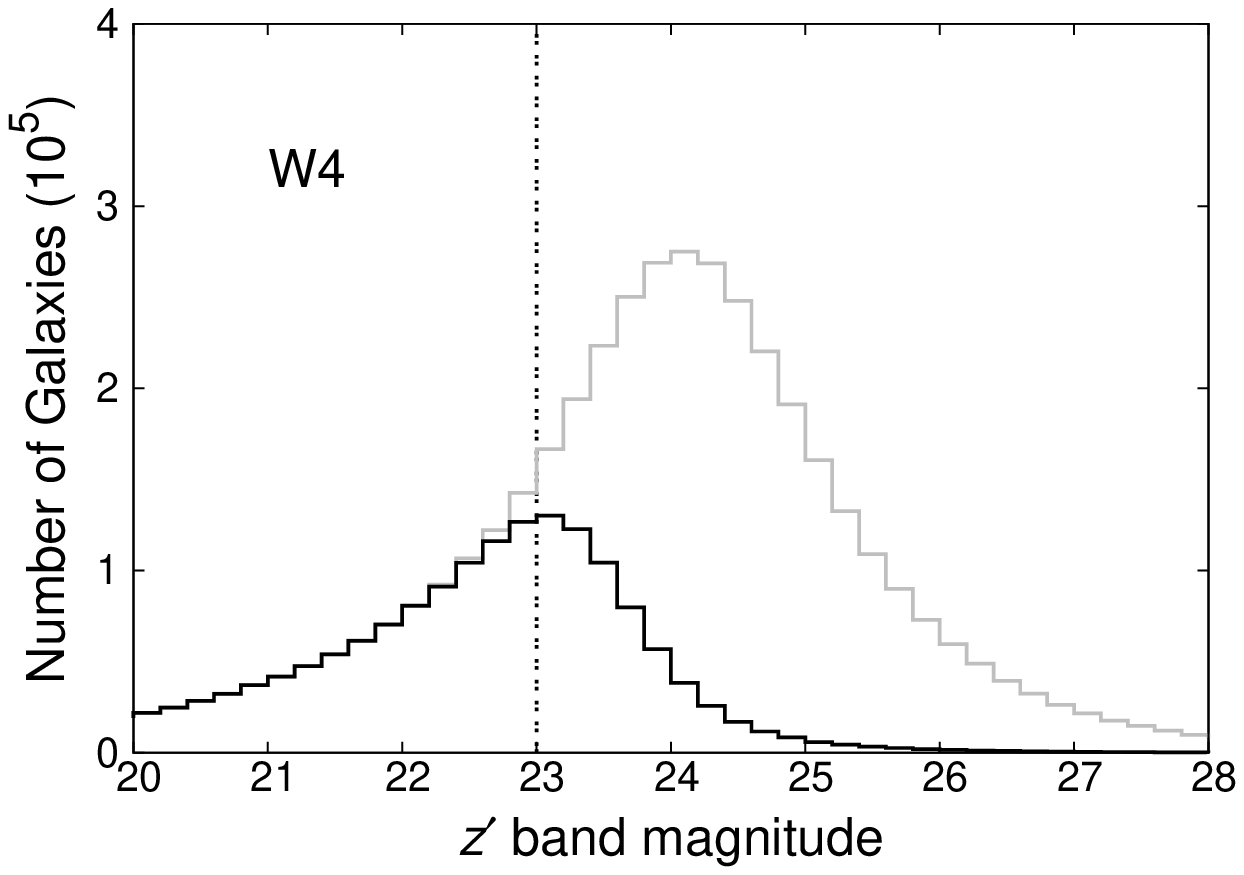}
}
\caption{Distribution of the $u^\ast$, $g^\prime$, $r^\prime$, 
$i^\prime$ and $z^\prime$ magnitudes in W1 (solid line), W2 
(dashed line) and W4 (dotted line) fields. The grey and black histograms
belong to photometry and photometric redshift catalogs. A vertical dotted 
line in each plot shows defined completeness threshold magnitude below which
the completeness in photoz catalog is less than 90\%. Since the photometric
redshift computed for galaxy $i^\prime<24$, the grey and black distributions
are identical for galaxies in this range of magnitude.
\label{all_mags_hist}}
\end{figure*}

\section{Optical counterparts for X-ray sources}\label{otp_counterpart}

\subsection{Red sequence method}\label{red_seq_method}

The red sequence (\citealt{Baum59,Bower92,Gladders00}) is a term
defining the overdensity of early-type cluster galaxies in
color-magnitude space. Usually a single color is used to find
overdensities of early-type galaxies in a limited range of
redshifts. This color is selected so that the 4\,000 $\AA$ break is
located in the bluer filter. For example, \cite{Rykoff12} used
$g^\prime$-$r^\prime$ for a redshift range between 0.1 and 0.3.
However, if we select another color, such as $r^\prime$-$i^\prime$ for
redshifts below 0.3, early-type galaxies (ETGs) in a cluster still
produce a sequence since they have similar formation redshifts and a
mostly passive evolution. While background and foreground galaxies
(e.g. a late-type galaxy at higher redshift) can have similar color to
the color of member ETGs, one can exclude them using other
filters. This approach leads to finding member ETGs with less
contamination and higher purity in selection of member galaxies, and
higher sensitivity for cluster detection. On the other hand,
multi-color selection of red sequence galaxies may miss some of the
red sequence galaxies (lower completeness). Combination of
  photometric redshift and red sequence selection can also work
  similarly. In this Paper, we will apply the multi-color selection
of red sequence galaxies to find the clusters. We will compare
  the relation between X-ray luminosity and integrated optical
  luminosity of clusters using three methods: 1) single color red
  sequence, 2) multi-color red sequence, and 3) combination of photoz
  and single color red sequence (regardless of purity and completeness
  for each method) to know which of them gives a better optical proxy
  for X-ray luminosity (or mass) of clusters.

The photometric redshifts are available in T0007 public catalog
thus we only need to calibrate the red sequence method for CFHTLS wide
survey. In the red sequence method, a model for describing the color
of galaxies and its corresponding dispersion as a function of redshift
is assumed. Then, at each redshift step, the number of red galaxies
with absolute magnitude lower than a threshold is counted (using the
model-predicted color value and its dispersion) and corrected for the
number of background red galaxies at the same redshift.  We
denote the mentioned threshold on absolute magnitude as
$M_\mathrm{cut}$.  It should be adopted according to the depth of the
survey in a way that the completeness is maintained in the whole
redshift range. This corrected number is the cluster richness, and
the redshift with the highest richness is chosen as the cluster
redshift.

As we move to higher redshifts, galaxies more luminous than
$M_\mathrm{cut}$ can still be below the completeness limit of the
sample in one or more filters. Figure \ref{all_mags_hist} shows the
magnitude distributions of CFHTLS survey of galaxies in the W1, W2 and
W4 fields in photometric catalogs in the 5 bands. We derived the
photoz completeness threshold by comparison between the photoz catalog
and photometry catalog. Figure \ref{all_mags_hist} shows magnitude
distribution for these two catalogs. We employ 0.2 magnitude bin width
in calculating the distributions. We defined the completeness in
photoz catalog as the magnitude above which the photoz catalog has a
completeness below 90\%.  We display these limits with the dotted
vertical lines in Figure \ref{all_mags_hist}.  Since the photoz is
computed for galaxies brighter than $i^\prime = 24$, the completeness
in other filters are almost the same for different fields.  Table
\ref{mag_limit} shows the magnitude completeness limits for each
field, derived this way. With the above method, we derive these
completeness thresholds for photoz catalog: $u^\ast=24.2$,
$g^\prime=24.2$, $r^\prime=24.0$, $i^\prime=24.0$ ,and
$z^\prime=23.0$.

\begin{table}
\begin{center}
\renewcommand{\arraystretch}{1.1}\renewcommand{\tabcolsep}{0.12cm}
\caption{\footnotesize
Completeness magnitude limits for each field. 
Because the photometric redshift catalog has a cut at
$i^\prime < 24$, the completeness thresholds are almost the same
for different fields.
\label{mag_limit}}

\begin{tabular}{cccc}
\hline
\hline
filter & W1 & W2 & W4  \\
\hline
$u^\ast$ & 24.2 & 24.2 & 24.4 \\
$g^\prime$ & 24.2 & 24.2 & 24.2 \\
$r^\prime$ & 24.0 & 24.0 & 24.0 \\
$i^\prime$ & 24.0 & 24.0 & 24.0 \\
$z^\prime$ & 23.0 & 23.0 & 23.0 \\
\hline
\end{tabular}
\end{center}
\end{table}

For computing any optical quantity at different redshifts, we
  need to consider an identical cut on rest frame luminosity for the
  whole redshift range. The reason is that galaxies with similar
  absolute magnitude seem fainter at higher redshifts.  This cut can
  also change the scaling relations and their scatters. For example,
\cite{Rykoff12} tested between richness and X-ray luminosity
(hereafter, $L_\mathrm{X}$) for different $L_\mathrm{cut}$ from
0.1$L_\mathrm{\ast}$ to 0.4$L_\mathrm{\ast}$, showing that the
richness-$L_\mathrm{X}$ relation of a cluster sample has the least
scatter with $L_\mathrm{cut} = 0.2 L_\mathrm{\ast}$. In addition to
minimising the scatter in the richness-$L_\mathrm{X}$ relation, we
need to check the feasibility of selecting a given value of
$L_\mathrm{cut}$, given the depths of the survey. Using
\cite{Maraston09} stellar population model and combining its spectral
energy distribution (SED) with CFHT/MegaCam filters, we derive
apparent magnitude $m_\mathrm{\ast}(z)$ for all filters and
subsequently $k$-correction model. $m_\mathrm{\ast}(z)$ is the
  apparent magnitude of a galaxy with rest frame luminosity of
  $L_\mathrm{\ast}$ at a given redshift $z$. The computations is
done by ``Le Phare`` package \cite{Ilbert06}. \cite{Maraston09} showed
that their model is in agreement with color evolution of luminous red
galaxies in SDSS. This model is based on a single-burst model with a
solar metallicity. Similar to \cite{Rykoff12}, we adopt
$L_\mathrm{\ast}$ = 2.25 $\times$ $10^{10} L_{\odot}$.

\begin{figure}[t]
\includegraphics[width=0.45\textwidth]{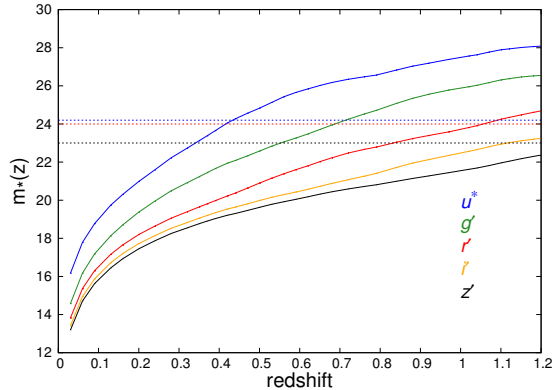}\hfill
\caption{Characteristic magnitude $m_\mathrm{\ast}(z)$ for different
filters as a function of redshift. Blue, green, red, yellow and black
solid lines correspond to the magnitudes in $u^\ast$, $g^\prime$,
$r^\prime$, $i^\prime$ and $z^\prime$ bands, respectively. The blue,
red and black dotted lines show the completeness limits of survey for
$u^\ast$/$g^\prime$, $r^\prime$/$i^\prime$, and $z^\prime$ filters
respectively.  \label{mstar_z_color}} \end{figure}

Figure \ref{mstar_z_color} shows $m_\mathrm{\ast}(z)$ for all
five filters derived from our model for redshifts below 1.2. Based on
the magnitude completeness of the survey, we estimate the maximum
redshift at which a galaxy with luminosity of 0.2, 0.4 and 1 times of
$L_\mathrm{cut}$ can be observed in each filter. Table \ref{Lstar_z}
shows the redshift limits for each $M_\mathrm{cut}$.

\begin{table}
\begin{center}
\renewcommand{\arraystretch}{1.1}\renewcommand{\tabcolsep}{0.12cm}
\caption{\footnotesize
{The maximum redshift at which galaxies with luminosity of
0.2$L_\mathrm{\ast}$, 0.4$L_\mathrm{\ast}$ and 1.0$L_\mathrm{\ast}$
have photometric redshift in T0007 catalog.}
\label{Lstar_z}}

\begin{tabular}{cccc}
\hline
\hline
filter & 0.2$L_\mathrm{\ast}$ & 0.4$L_\mathrm{\ast}$ &
1$L_\mathrm{\ast}$  \\
\hline
$u^\ast$ & 0.27 & 0.34 & 0.42 \\
$g^\prime$ & 0.48 & 0.60 & 0.71 \\
$r^\prime$ & 0.70 & 0.84 & 1.05 \\
$i^\prime$ & 0.94 & 1.1 & $>$1.2 \\
$z^\prime$ & 1.12 & $>$1.2 & $>$1.2 \\
\hline
\end{tabular}
\end{center}
\end{table}

\begin{figure}[t]
\includegraphics[width=0.48\textwidth]{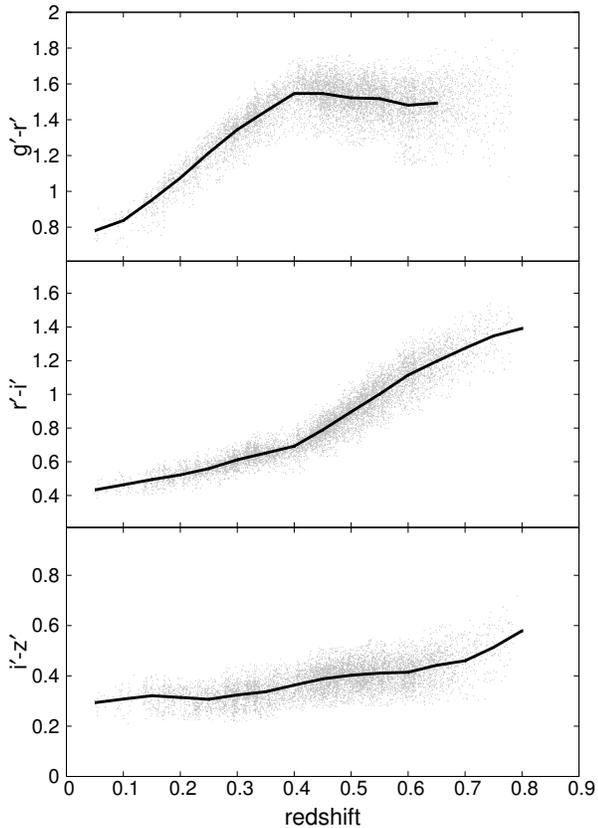}\hfill
\caption{Model colors for ETGs as a function of redshift. Grey dots
show the ETGs and solid lines are the average at each redshift.
   \label{colors_z}}
\end{figure}

\begin{figure}[t]
\includegraphics[width=0.48\textwidth]{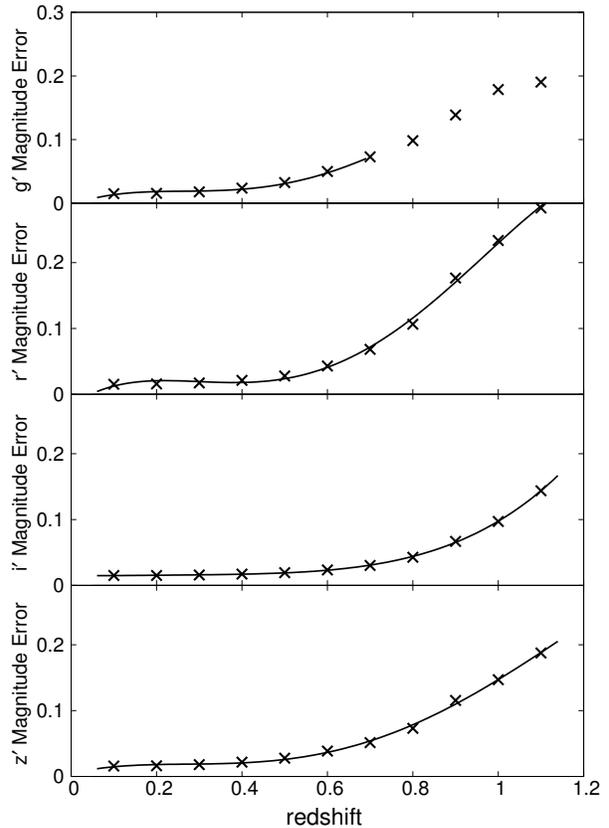}\hfill
\caption{Magnitude errors in $g^\prime$, $r^\prime$, $i^\prime$ and
   $z^\prime$ band versus redshift for galaxies brighter than
   $m_\mathrm{\ast(z)}$+1. Crosses show the mean magnitude error for each
   redshift bin and solid lines are polynomial fits to the mean
   values. \label{mag_error_z_0.4LstarZband}}
\end{figure}

\begin{figure*}[t]

\mbox{

\includegraphics[width=0.33\textwidth]{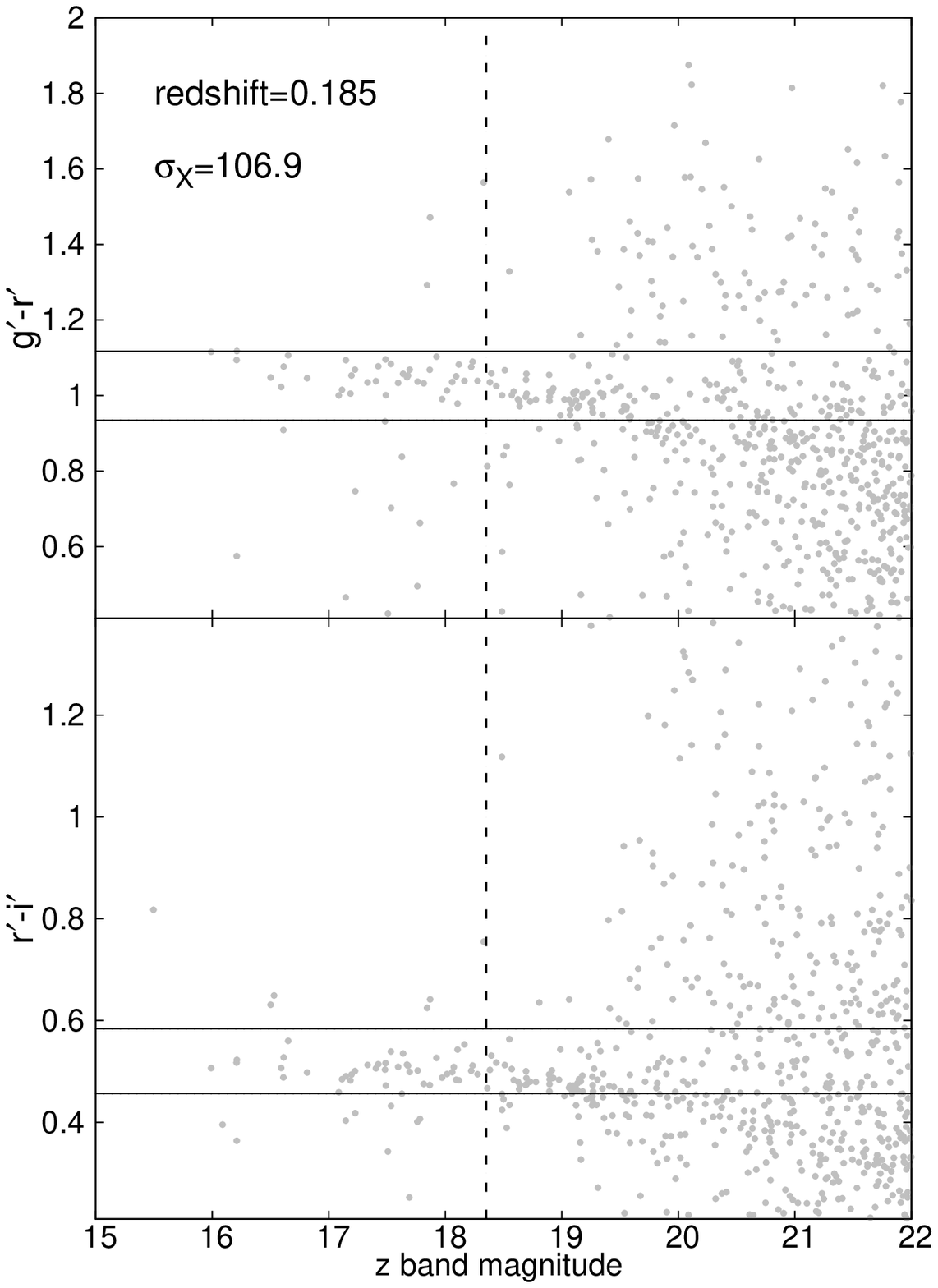}

\includegraphics[width=0.33\textwidth]{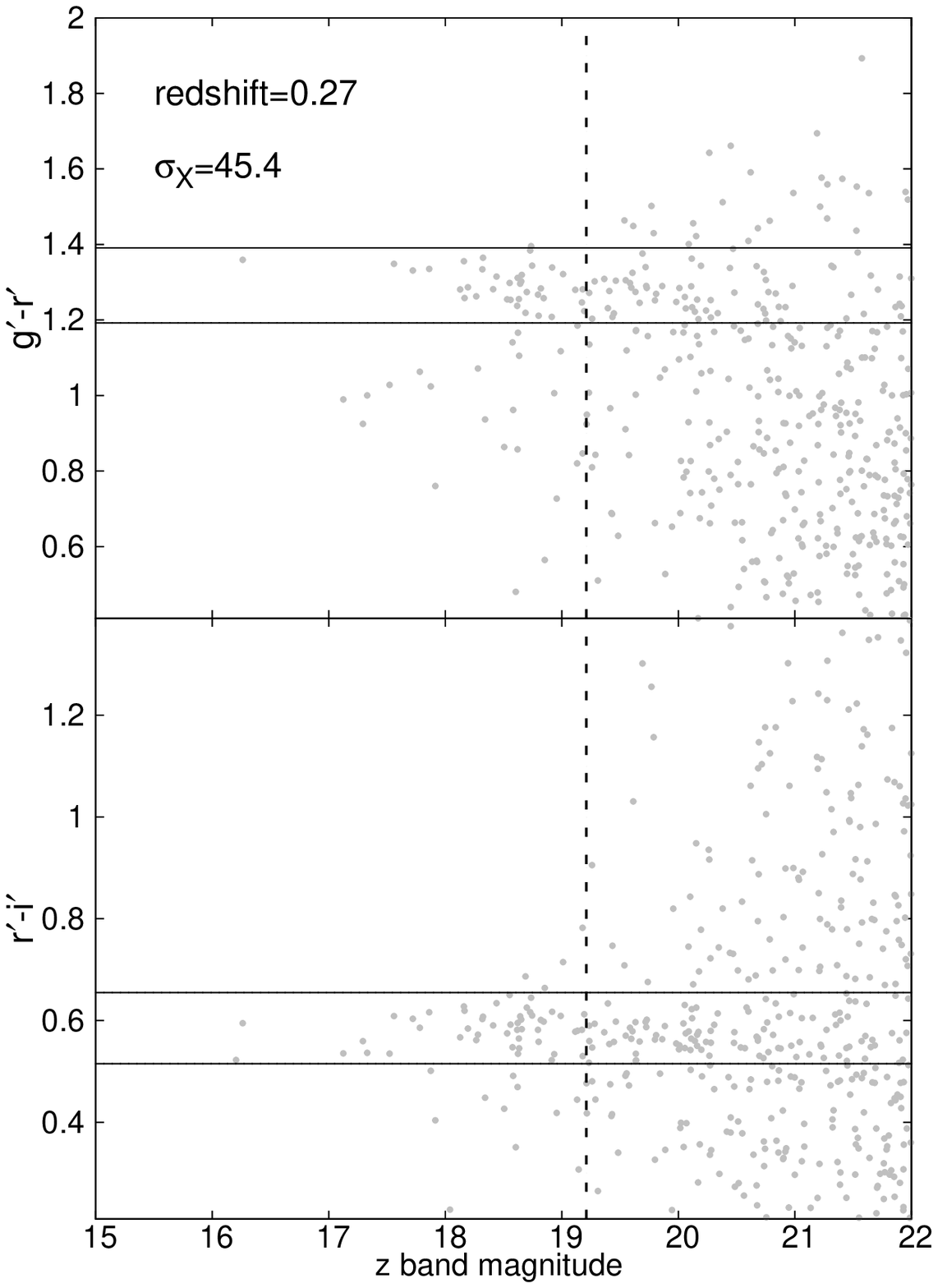}

\includegraphics[width=0.33\textwidth]{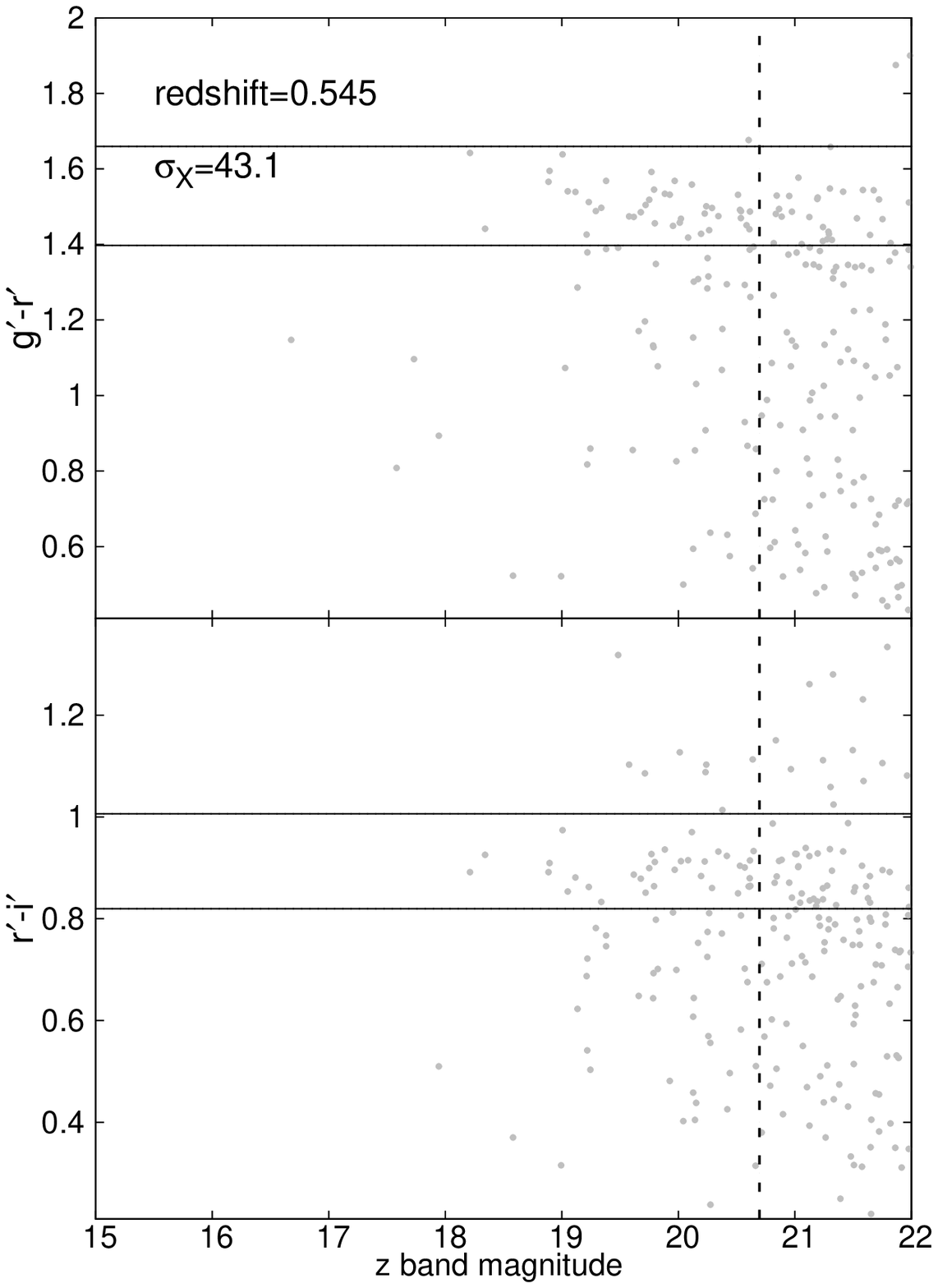}
}

\caption{\footnotesize Color -- magnitude diagrams for three clusters
with high detection level in X-ray. The solid lines are upper and lower
limits on the colors to encompass the bulk of red sequence galaxies. The
dashed line is the $m_\mathrm{\ast}$+1 at the redshift of clusters.}
\label{rs_examples}
\end{figure*}

\begin{figure*}[t]

\mbox{

\includegraphics[width=0.5\textwidth]{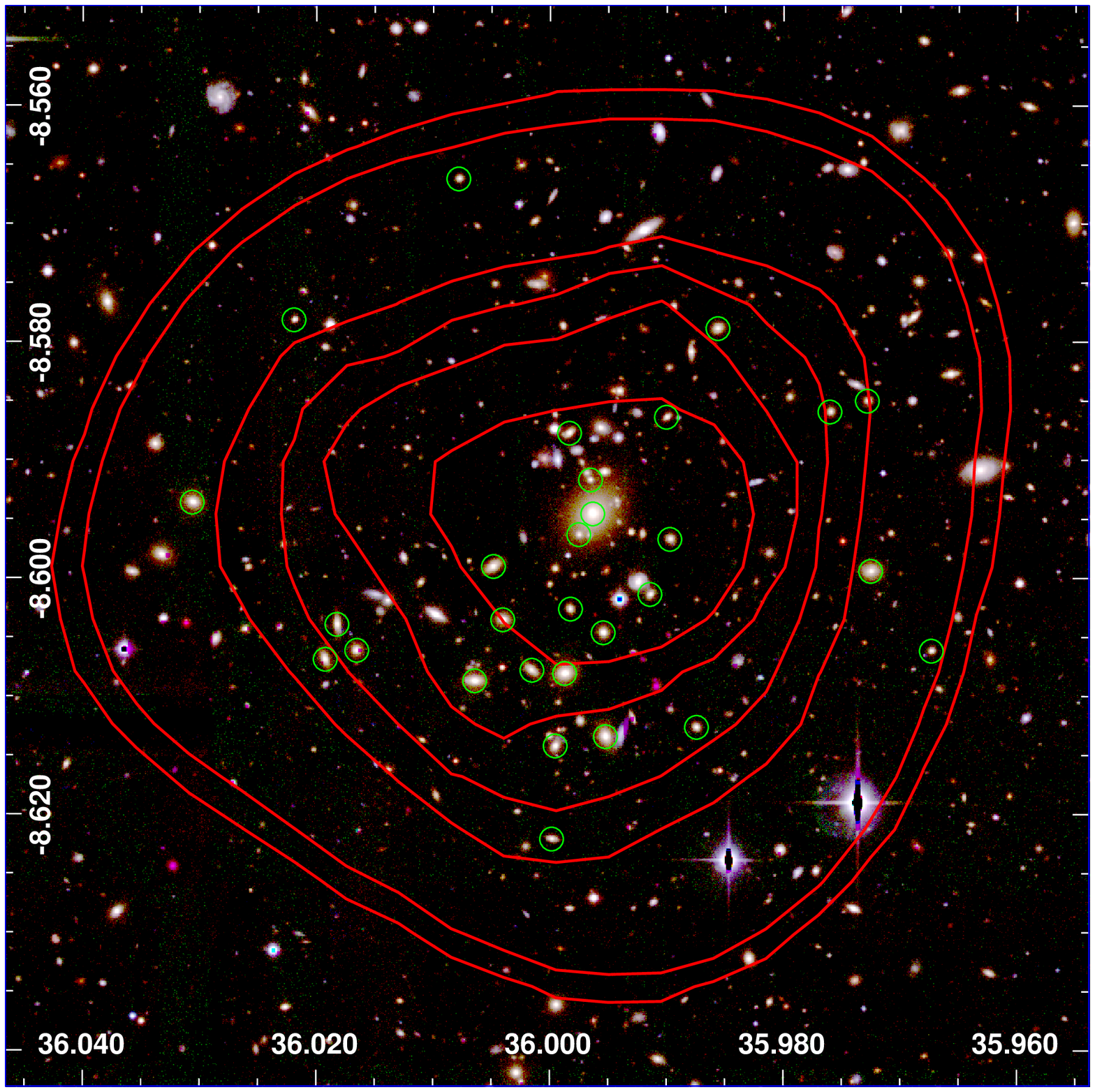}

\includegraphics[width=0.5\textwidth]{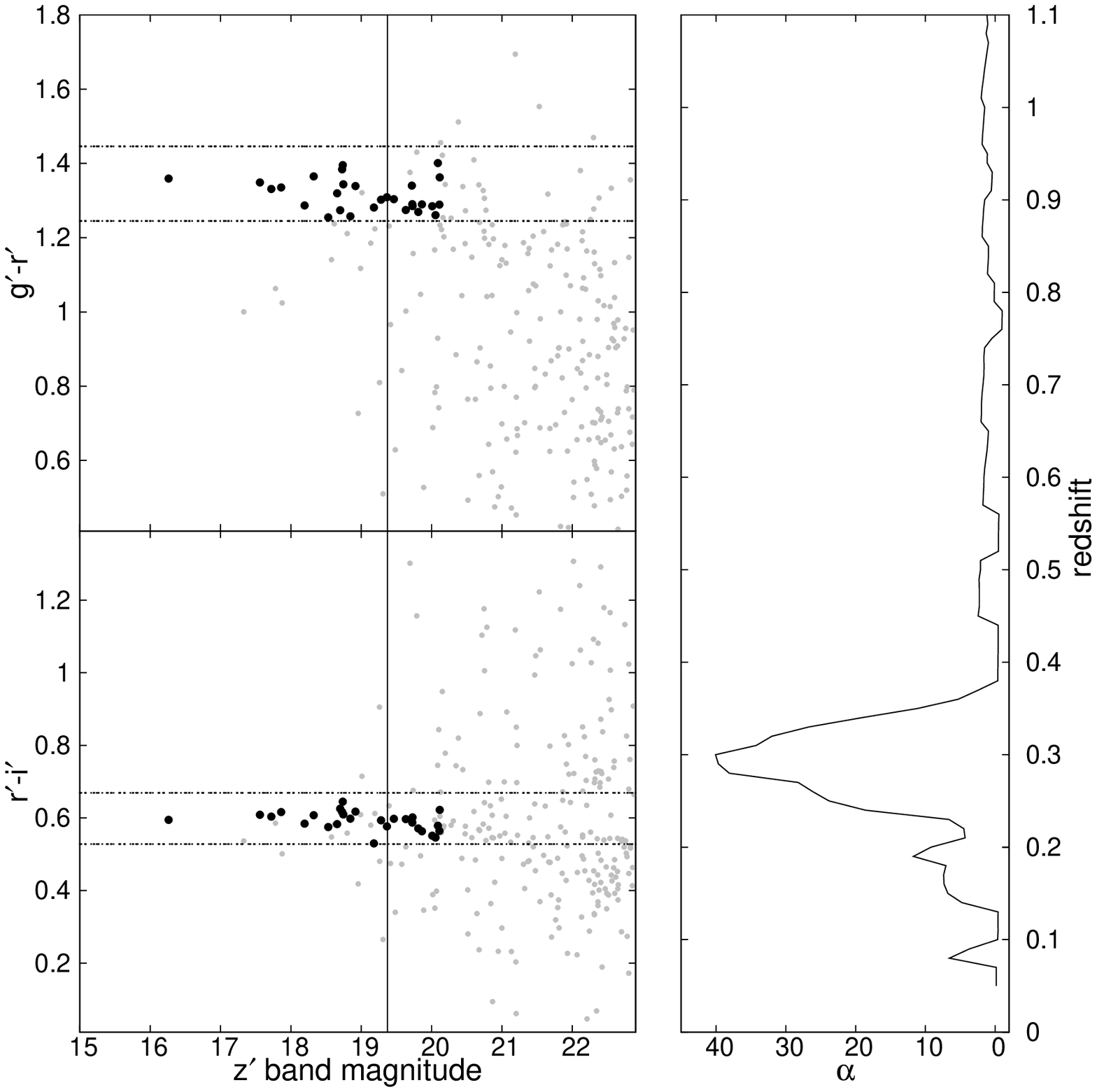}
}
\caption{\footnotesize Illustration of the red sequence finder using
  XMM cluster XCC J0224.0-0835 at a red sequence redshift of
  0.28. Left panel shows the RGB image of the cluster, where
  $i^\prime$, $r^\prime$, and $g^\prime$ images are used as red, green
  and blue components, respectively. The X-ray flux levels are
  represented by the red contours and the green circles are red sequence
  galaxies brighter than 0.2$L_\mathrm{\ast}$ within 0.5 Mpc from
  X-ray centre. The middle panels are color magnitude diagrams,
  $g^\prime$-$r^\prime$ (top) and $r^\prime$-$i^\prime$ (bottom)
  versus $z^\prime$ band magnitude. Grey points are all galaxies at
  the redshift of the cluster, located within the radius of 0.5~Mpc
  from the X-ray source centre. Black dots are red galaxies brighter
  than 0.2$L_\mathrm{\ast}$ within 0.5~Mpc. In each color magnitude diagram
  two horizontal dotted lines are upper and lower limits of color for
  selecting red galaxies according to an estimate of the color
  scatter, described in the text. The solid vertical line shows
  0.4$L_\mathrm{\ast}$ at the redshift of 0.3. The middle panels show
  the corresponding color--magnitude diagrams. The horizontal dashed
  lines are the lower and upper limits on the color of red sequence
  galaxies at a redshift of 0.3 and the solid vertical line is
  $L_\mathrm{cut}$=0.4$L_\mathrm{\ast}$ at the same redshift. The grey
  dots are all the galaxies with projected distances of 0.5 Mpc from
  the X-ray source centre. The black dots are the galaxies with green
  circles in left panel. The right panel is the variation of $\alpha$
  as a function of redshift with a maximum at redshift of 0.3.  The
  red sequence significance, $\alpha$ as a function of redshift is
  shown on the right panel and exhibit a maximum at a redshift of
  0.3.}

\label{xmm110460}

\end{figure*}

Given that $u^\ast$ band is not deep enough to cover at
least half of the redshift range of 0.05 to 1.1, this
filter is not used in this work. We chose the following
set of redshift ranges, filters and $L_{cut}$ for red
sequence algorithm:
\\
\\
0.05$\le$ z $\le$0.6 : $L_\mathrm{cut}$=0.4$L_\mathrm{\ast}$ and
$g^\prime$,$r^\prime$,$i^\prime$
\\
\\
0.6$<$ z $\le$1.1 : $L_\mathrm{cut}$=0.4$L_\mathrm{\ast}$ and
$r^\prime$,$i^\prime$,$z^\prime$
\\
\\

The $r^\prime$ band detections become incomplete at redshifts beyond
0.84, so the identification there has to rely on a single color. As
shown in Figure \ref{mstar_z_color} and Table \ref{Lstar_z},
$z^\prime$ band has the deepest imaging. We have therefore adopted
$z^\prime$ band for the magnitude parameter in color-magnitude
space. Hereinafter we use $m$ to denote the $z^\prime$ magnitude.

A galaxy is assumed to be on the red sequence at a redshift $z$ if:

\begin{equation}
  | GC_\mathrm{a-b}-MC_\mathrm{a-b}(z) | <2\times\sigma_\mathrm{a-b}(z),
\end{equation}
where $a-b$ represents a color ($g^\prime$-$r^\prime$,
etc). $GC_\mathrm{a-b}$ and $MC_\mathrm{a-b}(z)$ are galaxy color and
model color for red sequence galaxies at redshift $z$,
respectively. $\sigma_\mathrm{a-b}(z)$ is the dispersion of the
observed galaxy $a-b$ color around the model color.
$\sigma_\mathrm{a-b}(z)$ is a total dispersion, given by the sum in
quadrature of two other parameters, the magnitude errors and the
intrinsic width of the color. In the following, we consider
these two parameters in detail.

\begin{figure*}[t]

\mbox{

\includegraphics[width=0.33\textwidth]{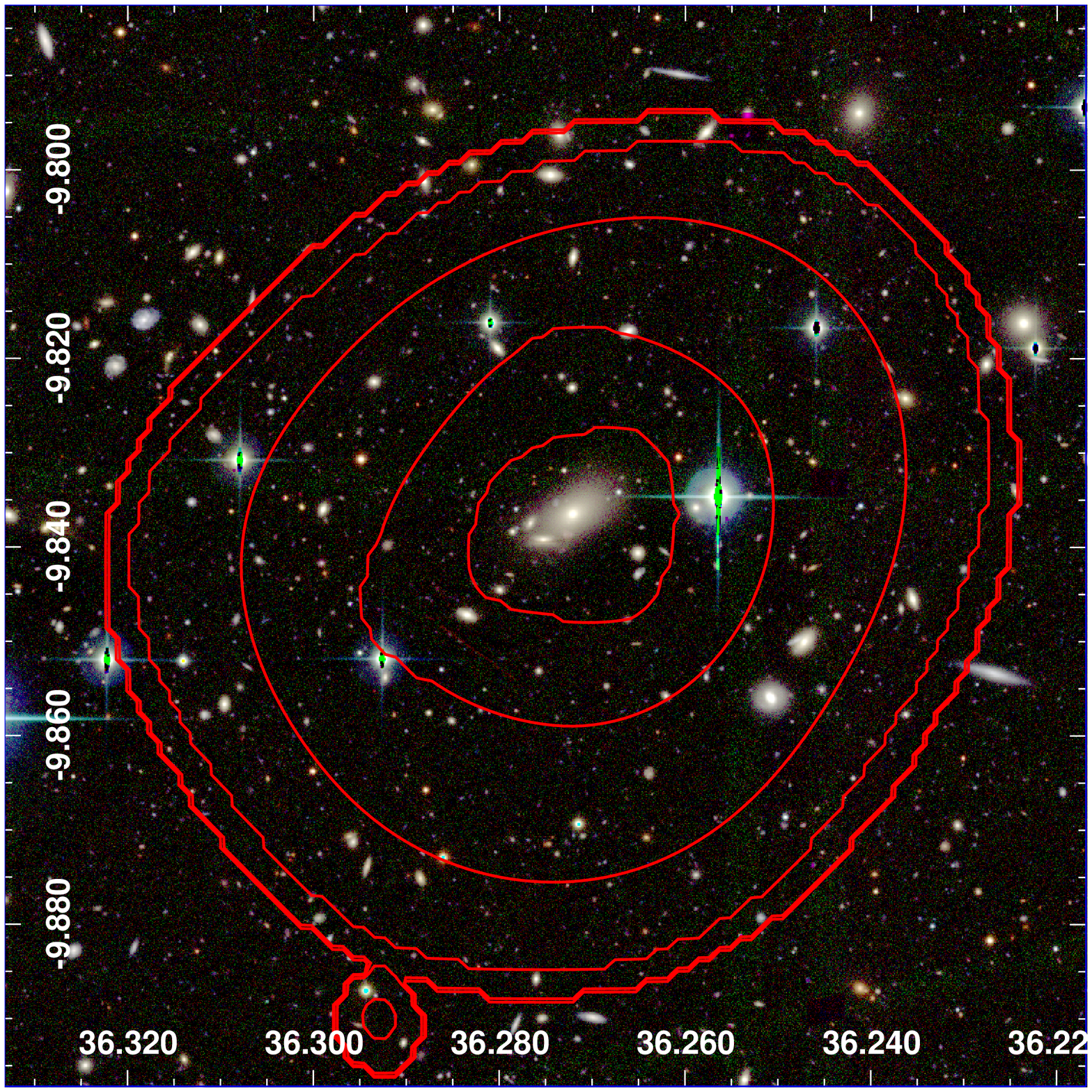}

\includegraphics[width=0.33\textwidth]{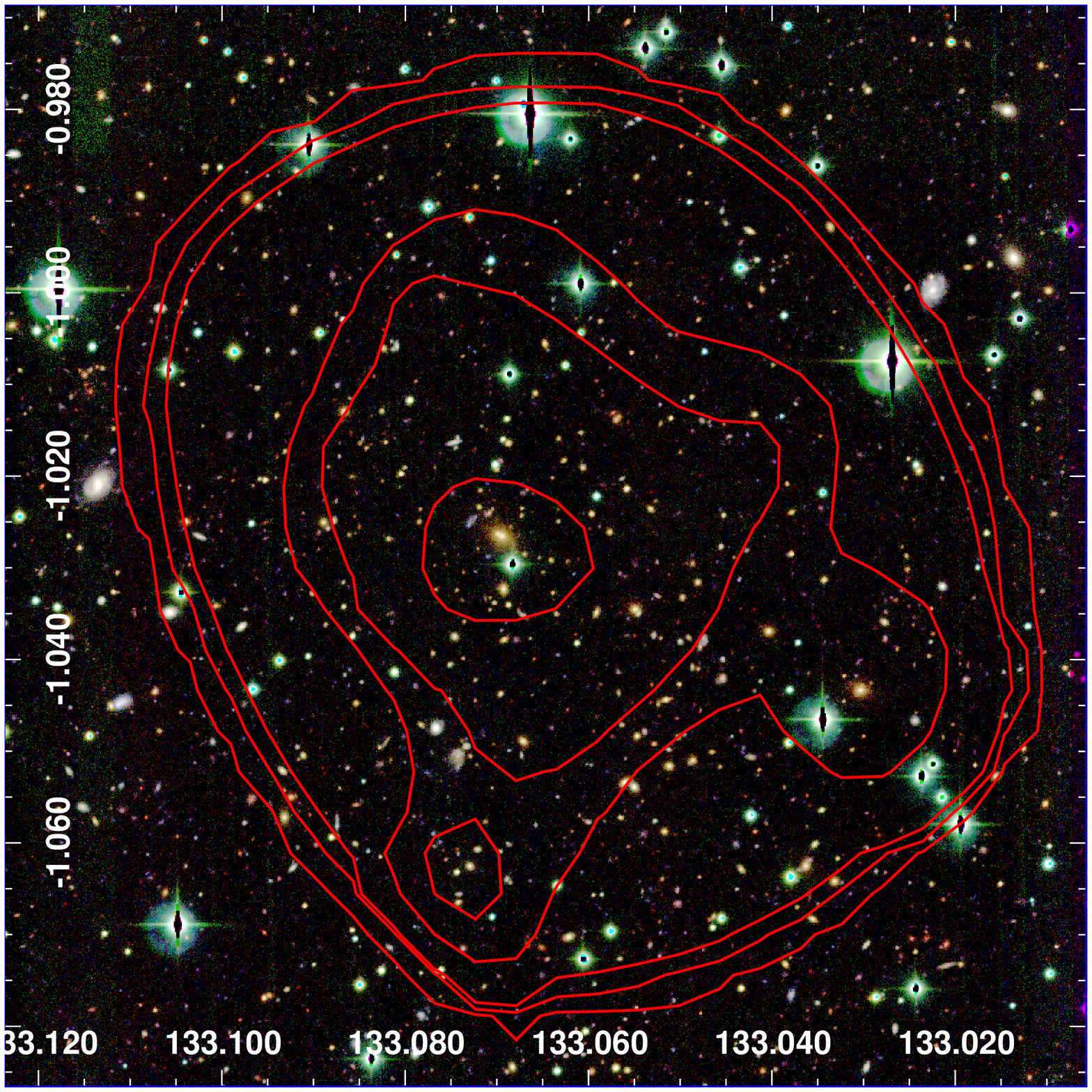}

\includegraphics[width=0.33\textwidth]{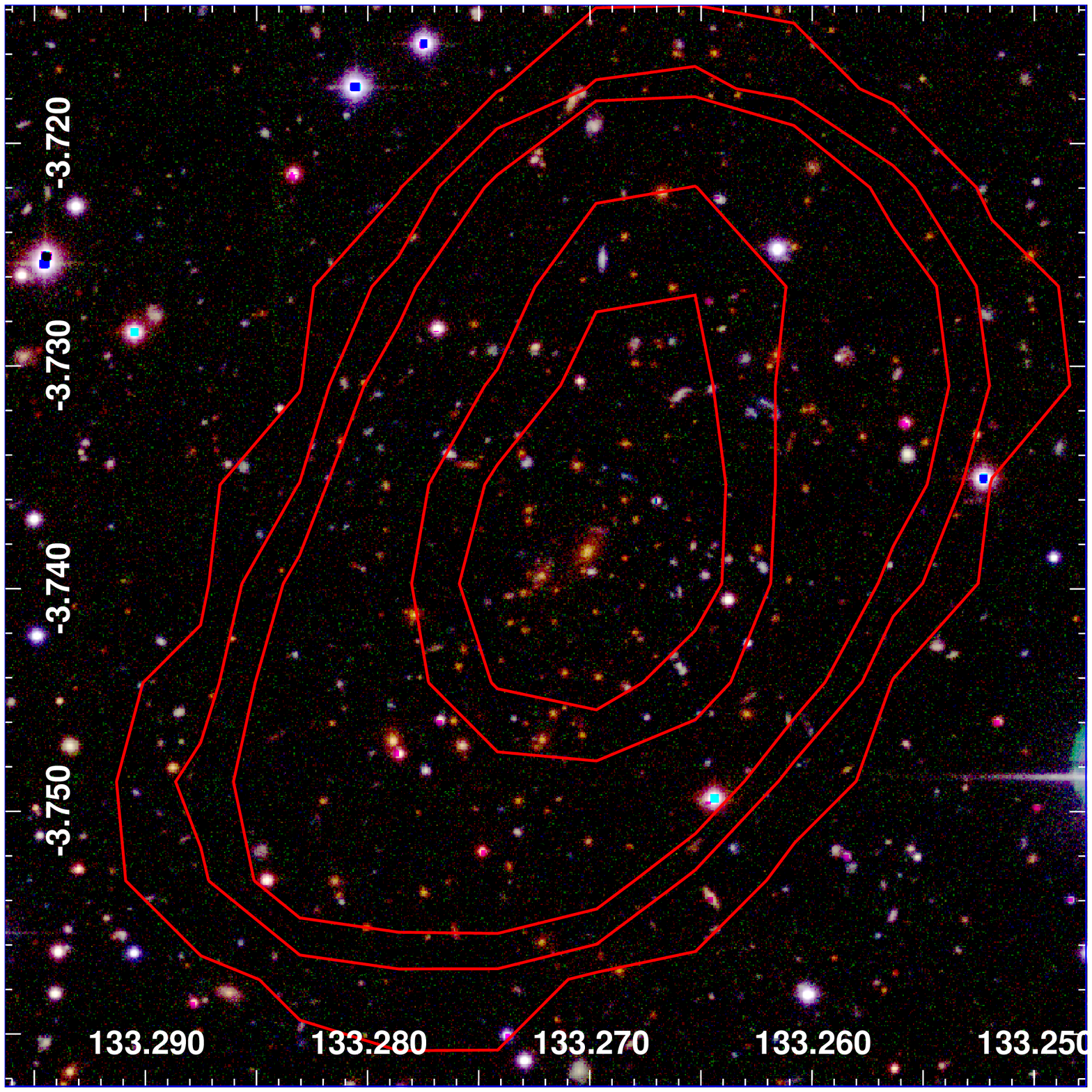}

}

\vspace*{0cm}
\mbox{

\includegraphics[width=0.33\textwidth]{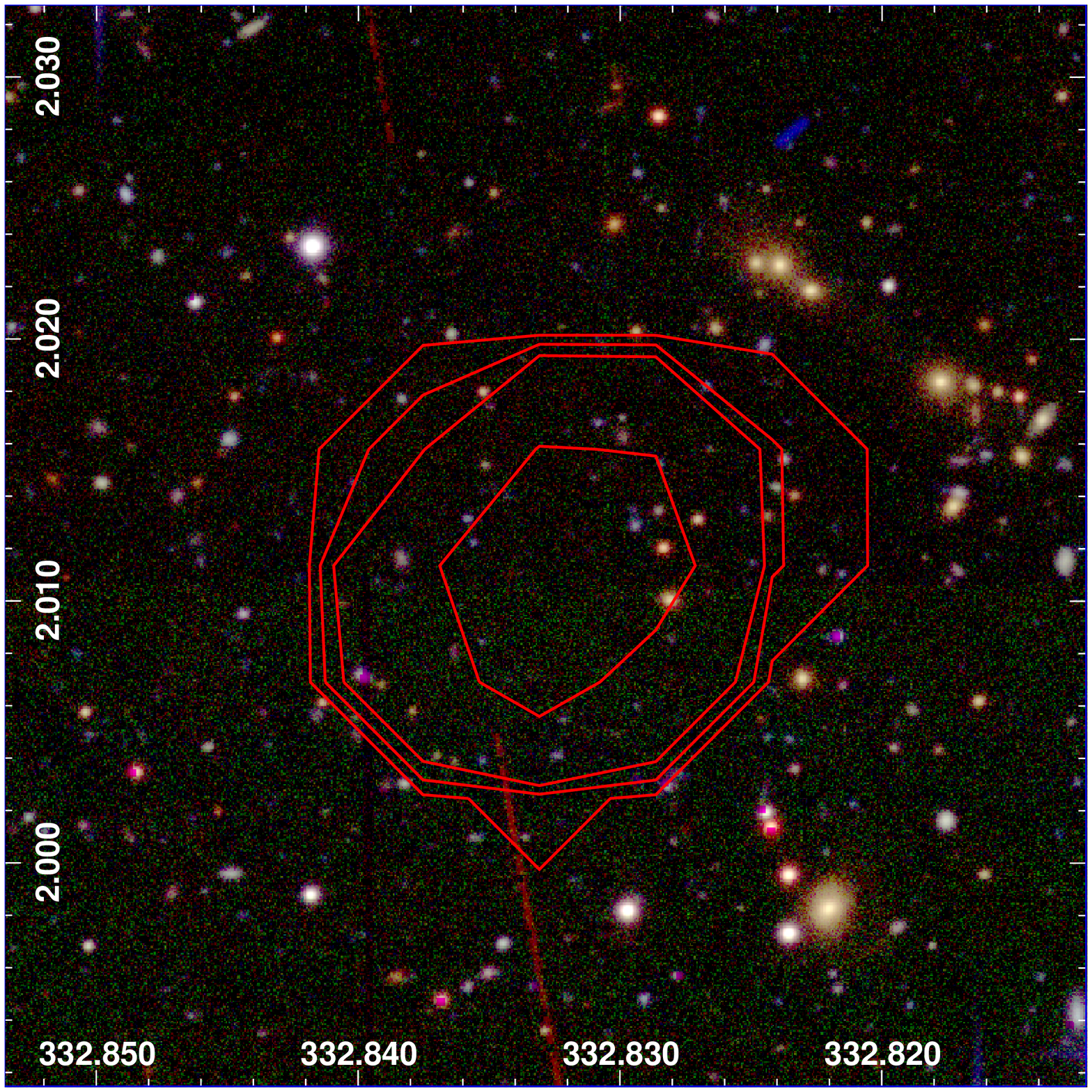}

\includegraphics[width=0.33\textwidth]{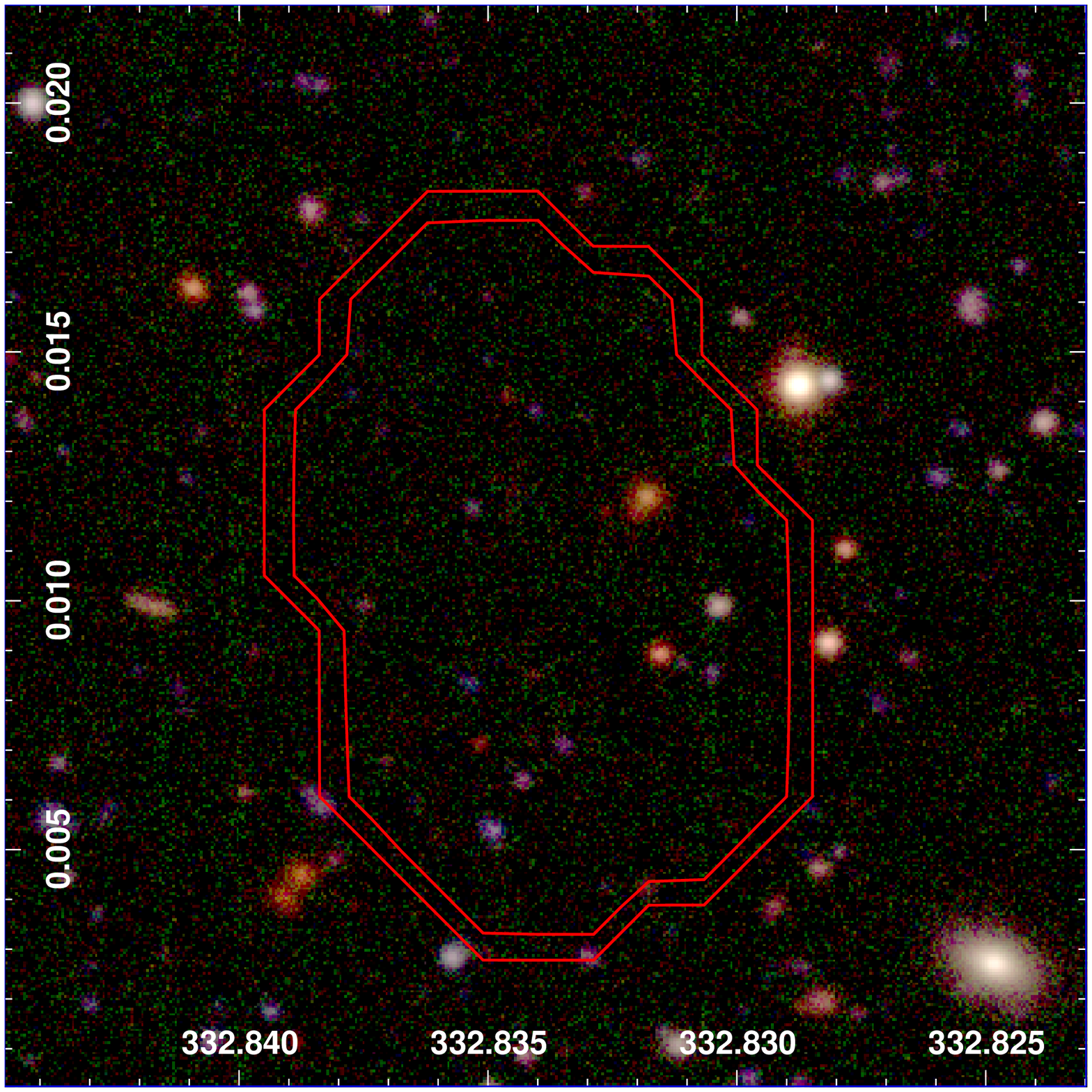}

\includegraphics[width=0.33\textwidth]{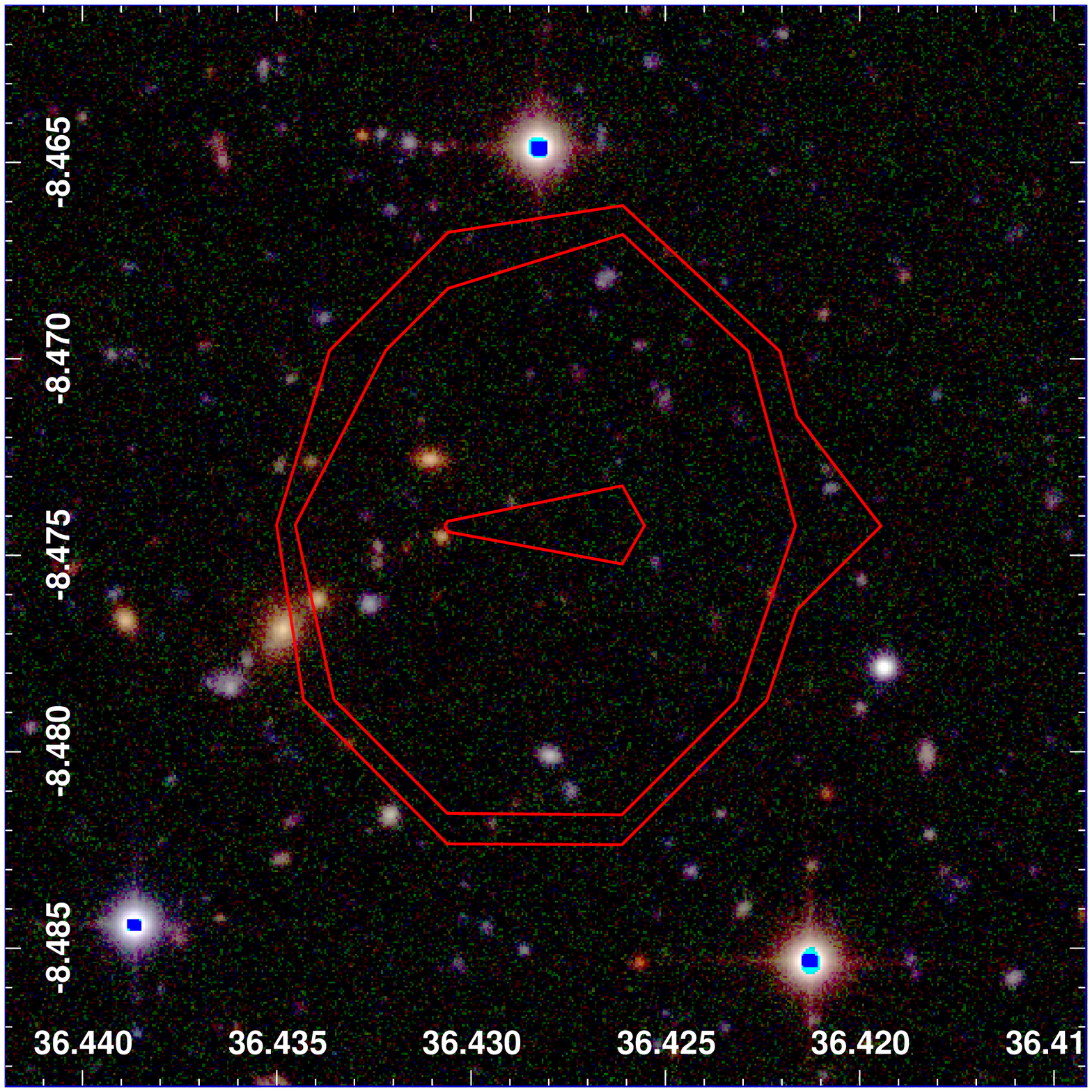}
}

\caption{\footnotesize Examples of clusters with different visual flags. 
  Top panels are examples of CFHTLS clusters with
  visual flag=1 at z=0.16, 0.46 and 0.92 (from left to right) and
  bottom panels are clusters with visual flag=2 at z=0.46, 0.83 and
  0.55 (from left to right). We use $g^\prime$ band image as blue,
  $r^\prime$ -- as green and $i^\prime$ -- as red component of RGB
  image. The red contours show the X-ray emission.  The upper clusters,
 from left to right, have X-ray signal significance of 31.03, 43.06, 8.77, 
 and lower ones have 4.51, 2.15, 3.58  }
\label{vf12_image}
\end{figure*}

In order to derive the observed color evolution of red sequence
galaxies, we use our spectroscopic sample of galaxies at low redshifts
and a stellar population model at high redshifts.  For low redshift,
we select galaxies brighter than $m_\mathrm{\ast}(z)$+1 (or
$\le$0.4$L_\mathrm{\ast}$) and exclude those with AGN or star-forming
classification in spectroscopic data or non-ET spectral energy
distribution (SED), yielding a sample of 7\,160 early-type
galaxies out of the full spectroscopic redshift catalog in W1, W2, and
W4. Second, we calculate the average color values and their standard
deviation for these galaxies in 16 spectroscopic redshift bins from
0.05 to 0.80 with the bin size of 0.05. For each bin, we discard the
galaxies with color offset from the average value exceeding two
standard deviations and repeat the calculation of the mean. Figure
\ref{colors_z} shows the $g^\prime$-$r^\prime$, $r^\prime$-$i^\prime$
and $i^\prime$-$z^\prime$ colors of ETGs and derived color model as a
function of redshift (solid lines). Given that the sample of galaxies
brighter than 0.4$L_\mathrm{\ast}$ is incomplete in g band for
redshifts above 0.6, the modelling of $g^\prime$-$r^\prime$ color is
limited to $z$ of 0.6.

At higher redshifts, above the redshift of 0.75, the
spectroscopic sample of ETGs becomes poor, so we derive
$MC_\mathrm{a-b}(z)$ from \cite{Maraston09} model for early-type
galaxies, the same model for $m_\mathrm{\ast}(z)$ model in Figure
\ref{mstar_z_color}.

In order to determine the dispersion of the red-sequence color,
$\sigma_\mathrm{a-b}(z)$, We assume that it has two components, an
intrinsic dispersion, $\sigma_\mathrm{a-b,int}(z)$, and a color error, 
$\sigma_\mathrm{a-b,color}(z)$. In estimating
$\sigma_\mathrm{a-b,color}(z)$, we selected the galaxies with
photometric redshift below 1.2 and brighter than
$m_\mathrm{\ast}(z)$+1 (similar to the original work of
\cite{Gladders00}). Using the redshift bin width of 0.1, we determine
the mean magnitude error for each band, and approximate it with the
fourth order polynomials.  Figure \ref{mag_error_z_0.4LstarZband}
illustrates the magnitude errors and the polynomial curves as
functions of redshift. The total color dispersion is calculated as a
sum of the color errors (derived from the magnitudes errors) and the
intrinsic color dispersion in quadrature.

The red sequence is known to exhibit a tilt in the color-magnitude
space due to the age-metallicity relation (\citealt{Nelan05}). Since
we work with both low-mass and high-z clusters, where the
age-metallicity relation can be different, we prefer to consider the
tilt as part of color scatter. We note that a similar approach is
adopted in RedMapper \citep{Rykoff13}. In estimating the intrinsic
color dispersion, we assume that the variation of color in cluster
ETGs can be modelled by a variation in metallicity. We use PEGASE.2
stellar population/galaxy formation models to estimate the intrinsic
color dispersion. For the reference model, a unit solar metallicity is
considered (similar to \citealp{Eisenstein01,Rykoff12}) and we model
the evolution of the dispersion, by selecting the metallicity that
reproduces the observed color scatter for a subsample of well observed
clusters) and high number ($>10$) of spectroscopic redshifts. We model
$r^\prime$-$i^\prime$ and $i^\prime$-$z^\prime$ colors between
redshifts 0.05 and 1.2 and $g^\prime$-$r^\prime$ between 0.05 and
0.66. In Appendix \ref{app1}, it is shown that a linear evolution for
intrinsic color dispersion of ETGs is a reasonable assumption
especially for $g^\prime$-$r^\prime$ and $i^\prime$-$z^\prime$
colors. Thus the intrinsic color dispersions at redshifts between the
two models were derived by interpolating the model points. We check
the color-magnitude diagram for the training sample with different
$\sigma_\mathrm{a-b}$ associated with different
$\sigma_\mathrm{a-b,int}$ and realise that the metallicity of 0.75
solar is appropriate for the second model to enclose the bulk of the
red sequence galaxies within two times $\sigma_{a-b}$.  Figure
\ref{rs_examples} illustrates color -- magnitude diagrams for three
clusters at different redshifts with metallicity of 0.75 and 1 for
modelling the intrinsic color dispersion. We do not optimise the width
of red sequence for minimising the contamination or maximising the
number of member galaxies.

The derived intrinsic dispersion of colors as functions of redshift are:
\\
\\
\begin{equation}
\sigma_\mathrm{g^{\prime}-r^{\prime},int} (z)=0.029+0.044\times z
\end{equation}
\begin{equation}
\sigma_\mathrm{r^{\prime}-i^{\prime},int} (z)=0.011+0.046\times z
\end{equation}
\begin{equation}
\sigma_\mathrm{i^{\prime}-z^{\prime},int} (z)=0.021+0.035\times z
\end{equation}.
\\

When running the red-sequence finder, we consider a fixed physical
radius for galaxy selection and vary the redshift of red sequence from
0.05 and 1.1 with a step of 0.01. At each redshift, we calculate the
number of red sequence galaxies brighter than 0.4$L_{\ast}$,
$N_{0.4,R}(z)$. Using 294 random areas in three optical fields we
estimate the background, $NB_{0.4,R}(z)$, and its standard deviation,
$\sigma_{NB}(z)$.  At each redshift we compute the red sequence
significance, $\alpha$, as

\begin{equation}
\alpha=\frac{N_\mathrm{0.4,R}(z)-NB_\mathrm{0.4,R}(z)}{\sigma_\mathrm{NB}(z)}
.
\end{equation}

The overdensity with the highest red sequence significance is adopted
as the X-ray counterpart. The uncertainty in $\alpha$ is estimated by
randomly changing the magnitudes of catalog galaxies according to the
corresponding photometric errors.

\begin{figure}[t]
\centering
\includegraphics[width=0.45\textwidth]{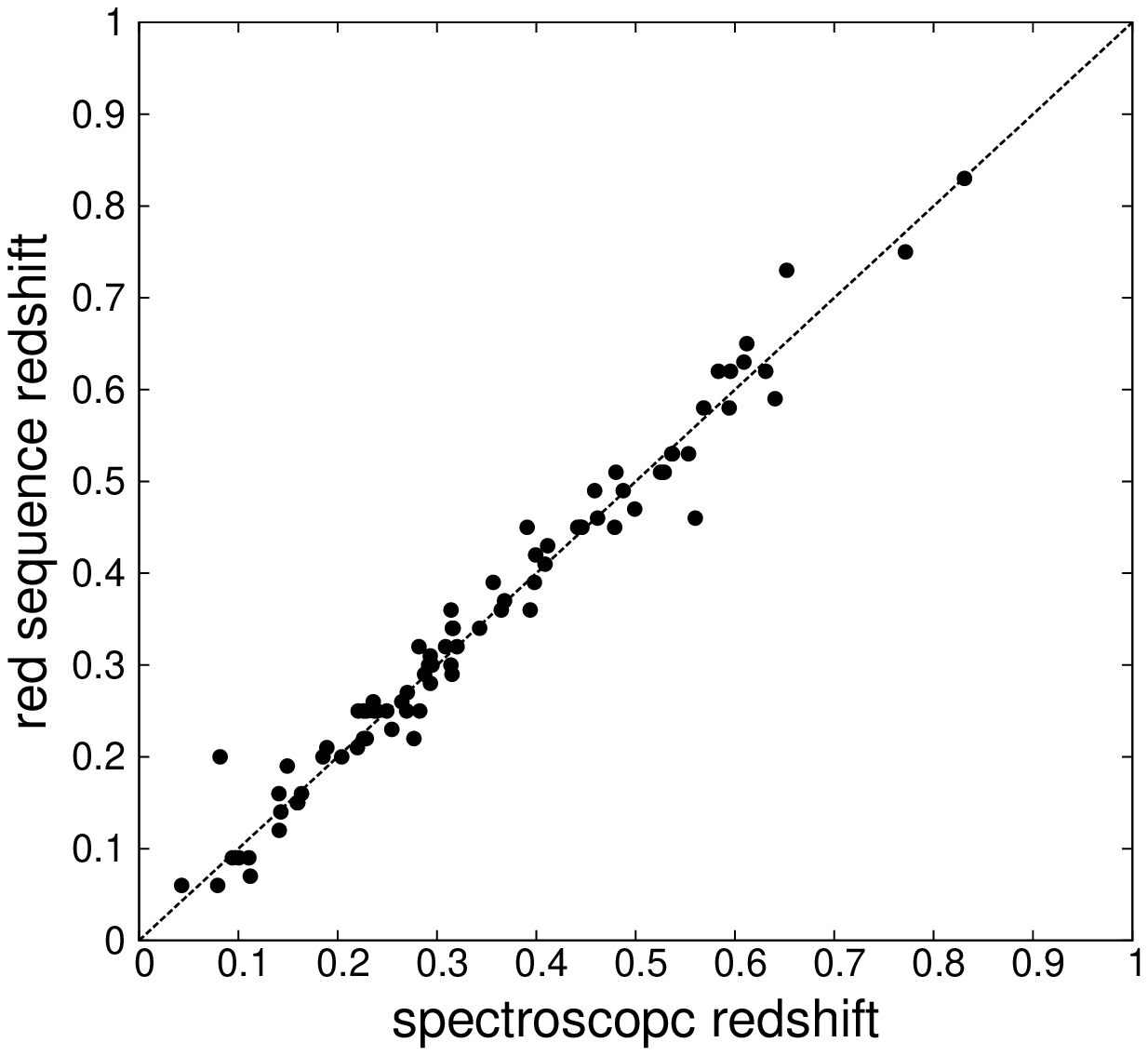}\hfill
\caption{Red sequence redshifts versus spectroscopic redshifts for 82
 clusters with spectroscopic counterparts. The dashed line shows a 1:1 correspondence.
\label{z_compare}}
\end{figure}

\begin{figure}[t]
\includegraphics[width=0.45\textwidth]{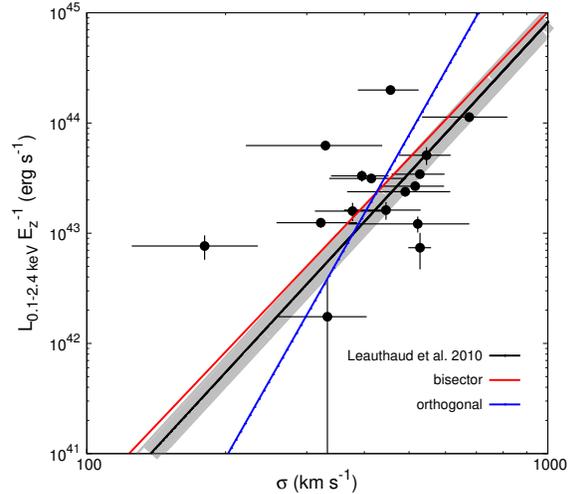}\hfill
\caption{X-ray luminosity versus the velocity dispersion for XMM
clusters with more than ten spectroscopic members from gapper
estimator method. The black line shows expected $L_\mathrm{X}-\sigma$
from scaling relation. The grey area marks the 20\% uncertainty in the
mass estimate using the $L_\mathrm{X}-M_{200c}$ relation.  The red and
blue lines are fitted lines with bisector and orthogonal
methods. Their equations are
$log(L_\mathrm{X})=(31.77\pm4.41)+(5.49\pm2.07) log(\sigma)$ and
$log(L_\mathrm{X})=(24.01\pm7.37)+(10.17\pm3.84) log(\sigma)$
respectively.  
\label{sigma_lx}}
\end{figure}

\subsection{Applying the red sequence finder to identify XMM-Newton extended sources} \label{red_seq_xmm}

We utilize our red sequence finder to identify the counterparts for
133 XMM-Newton extended sources in our $primary$ sample with a 4.6
$\sigma$ detection limit. We use a galaxy selection radius of 0.5 Mpc,
as the centers of XMM extended sources correspond well to the cluster
center (deviations are less than 15 arcseconds,
\citealt{George12}). Figure \ref{xmm110460} illustrates the results of
applying the red sequence on a cluster at a redshift of 0.28. After
applying the red sequence finder on all the X-ray sources, we visually
inspect them to compare the correspondence of a two-dimensional
distribution of X-ray photons and location of galaxies, presence of
secondary peaks in X-rays and optical quality of the images. The
photometric and spectroscopic galaxy catalogs are fully utilised
during visual inspection for optical counterparts of the X-ray
sources. Obvious cluster candidates are marked with a visual flag$=1$
in the catalog. Visual flag$=2$ is assigned to X-ray sources which
have low significance of the optical counterpart or concentration of
galaxies almost on the edge or out of X-ray source, indicative of a
confused X-ray source. Figure \ref{vf12_image} illustrates clusters
with different visual flags. It is worth mentioning that visual
flag (or quality flag) has no utility in this paper and we provide it
for others who will use this sample of cluster.  We provide an
identification to all XMM sources with flux significance above 4.6
sigma. During this inspection, we also visually checked faint sources
with detection levels below 4.6, discarding the sources revealing no
visual concentration of galaxies. We added 63 clusters from the lower
X-ray detection threshold sample, we arrive at a sample of 196
clusters with assigned RS redshift.

81 clusters among 196 clusters have spectroscopic redshift. In
defining the spectroscopic redshift, we first visually select the
redshift of the brightest galaxy with spectroscopic redshift close to
the red sequence redshift of a cluster and assume it as an initial
redshift of a cluster. Then we select all galaxies within 0.5Mpc from
X-ray centre and the sigma clipping is done within $\pm0.005(1+z)$
around the initial redshift. Finally, the mean of spectroscopic
redshifts is computed. The number of spectroscopic counterparts per
cluster varies from 1 to 10 member galaxies. In Figure \ref{z_compare}
we compare the red sequence redshift with mean of spectroscopic
redshift of member galaxies.  The average difference between the
red sequence and spectroscopic redshift is 0.002 with a standard
deviation of $0.02(1+z)$.

\begin{figure}[t]
 \includegraphics[width=0.45\textwidth]{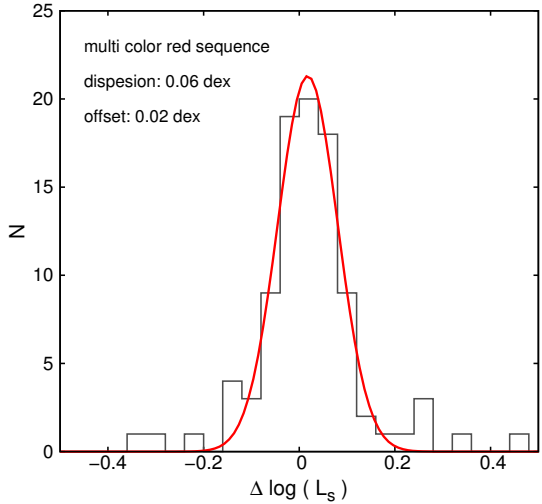} 
\caption{Convergence test for calculating $L_\mathrm{S}$. The plot 
shows the distribution of $L_\mathrm{S}$ calculated within three 
times of $\sigma_{a-b}(z)$ subtracted by the $L_\mathrm{S}$ calculated 
within two times of $\sigma_{a-b}(z)$. The distribution is fitted by a 
Gaussian with a standard deviation of 0.06
dex. The peak offset is 0.02 dex.
\label{Ls_3sig_Ls_2sig}}
 \end{figure}

\begin{figure}[t]
\centering
 \includegraphics[width=0.45\textwidth]{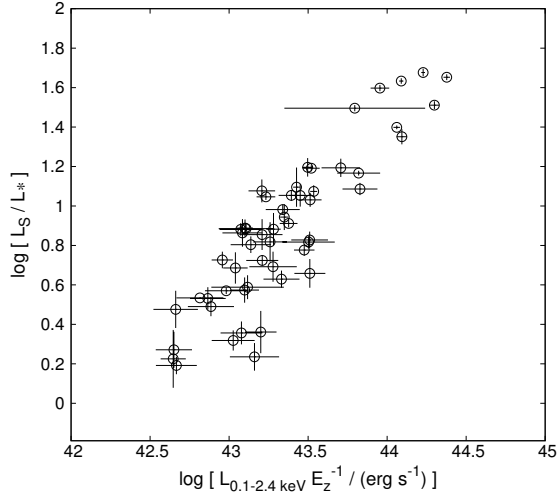}\hfill
\caption{Integrated stellar luminosity in $z^\prime$ band versus 
$L_\mathrm{X}$ for clusters with X-ray detection level above 4.6
and $0.1<z<0.6$. 
\label{Lx-Ls_1}}
 \end{figure}

\subsection{Velocity dispersion} \label{Velocity_dispersion}

We can also use velocity dispersion measurements as an independent
confirmation for the existence of a galaxy cluster and a
characteristic for the system. Such a calculation is only reliable for a
high number of member galaxies (typically more than 10), though we
provisionally calculate dispersions down to systems with 5 member
galaxies and present them in the catalog. We limit the sample for
relation between X-ray luminosity and velocity dispersion to the
clusters with more than 10 member galaxies (N$_{\sigma}$ $\le$ 10)
because of lower error in velocity dispersion measurement.

We follow the analysis of \cite{Connely12}. In detail, we select
galaxies iteratively, starting with an initial guess for the observed
velocity dispersion of $\sigma (z)_\mathrm{obs}$=$500(1+z)$ km
s$^{-1}$ as

\begin{equation}
  \delta (z)_\mathrm{max}=2 \frac{ \sigma (z)_\mathrm{obs}}{c}
\end{equation}

We then calculate the spatial distribution associated with $\delta
(z)_{max}$:
\begin{equation}
  \delta (r)_\mathrm{max}= \frac{c \delta (z)_\mathrm{max}}{b h_\mathrm{71}(z)}
\end{equation}
where b=9.5 is the aspect ratio. We use the peak of the X-ray emission
as the cluster center. The observed velocity dispersion, $\sigma
(z)_\mathrm{obs}$ is then calculated for galaxies within $\delta
(r)_\mathrm{max}$ using the $gapper$ estimator method
(\cite{Wilman05I,Beers90}), and the new value is then used to
re-estimate $\delta (z)_\mathrm{max}$ and $\delta (r)_\mathrm{max}$.
The procedure is repeated until convergence is achieved. The
rest-frame velocity dispersion $\sigma (z)_\mathrm{rest}$ and
intrinsic velocity dispersion $\sigma (z)_\mathrm{int}$ are finally
given by

\begin{equation}
\label{eqn:optfilt}
\sigma(v)_\mathrm{rest}=\frac{\sigma(v)_\mathrm{obs}}{1+z}
\end{equation}
\begin{equation}
\label{eqn:optfilt}
\langle\Delta(v)\rangle^2=\frac{1}{N}\sum_\mathrm{i=1}^N\Delta({v})_i^2
\end{equation}
\begin{equation}
\label{eqn:optfilt}
\sigma(v)_\mathrm{intr}^2=\sigma(v)_\mathrm{rest}^2-\langle\Delta(v)\rangle^2
\end{equation}

where $\Delta(v)$ is the uncertainty in the spectroscopic velocity measurement.
For computing velocity dispersion, we use galaxies with spectroscopic
redshift error less than $3 \times 10^{-4}$.

\begin{figure*}
\mbox{

\includegraphics[width=0.5\textwidth]{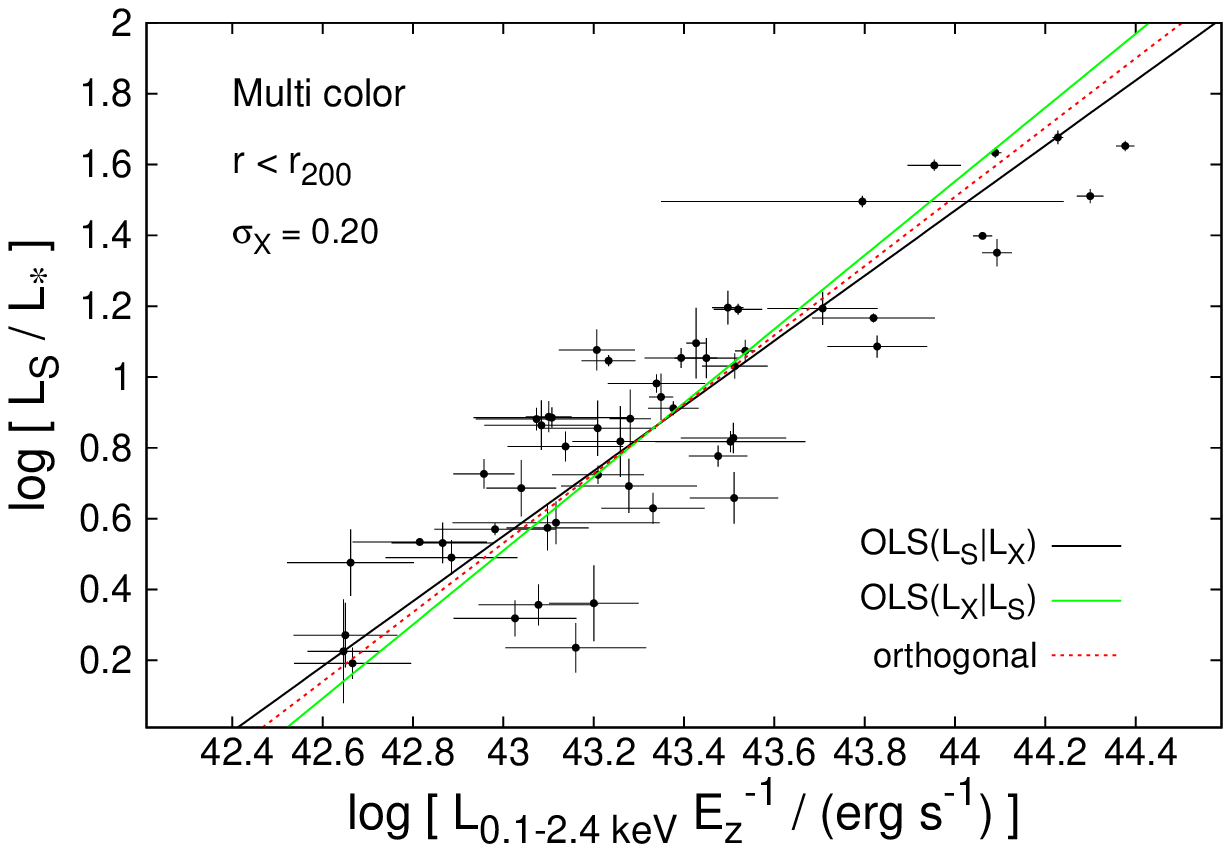}

\includegraphics[width=0.5\textwidth]{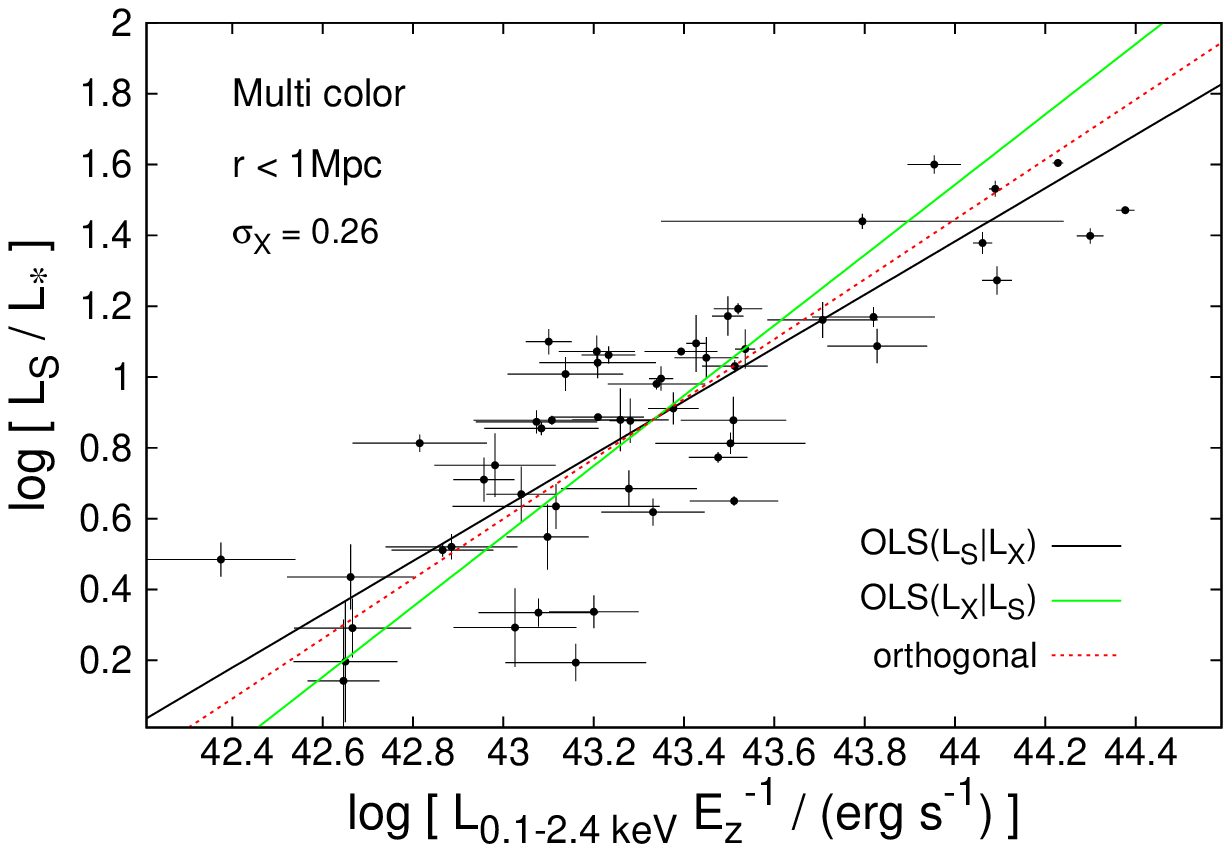}
}

\mbox{

\includegraphics[width=0.5\textwidth]{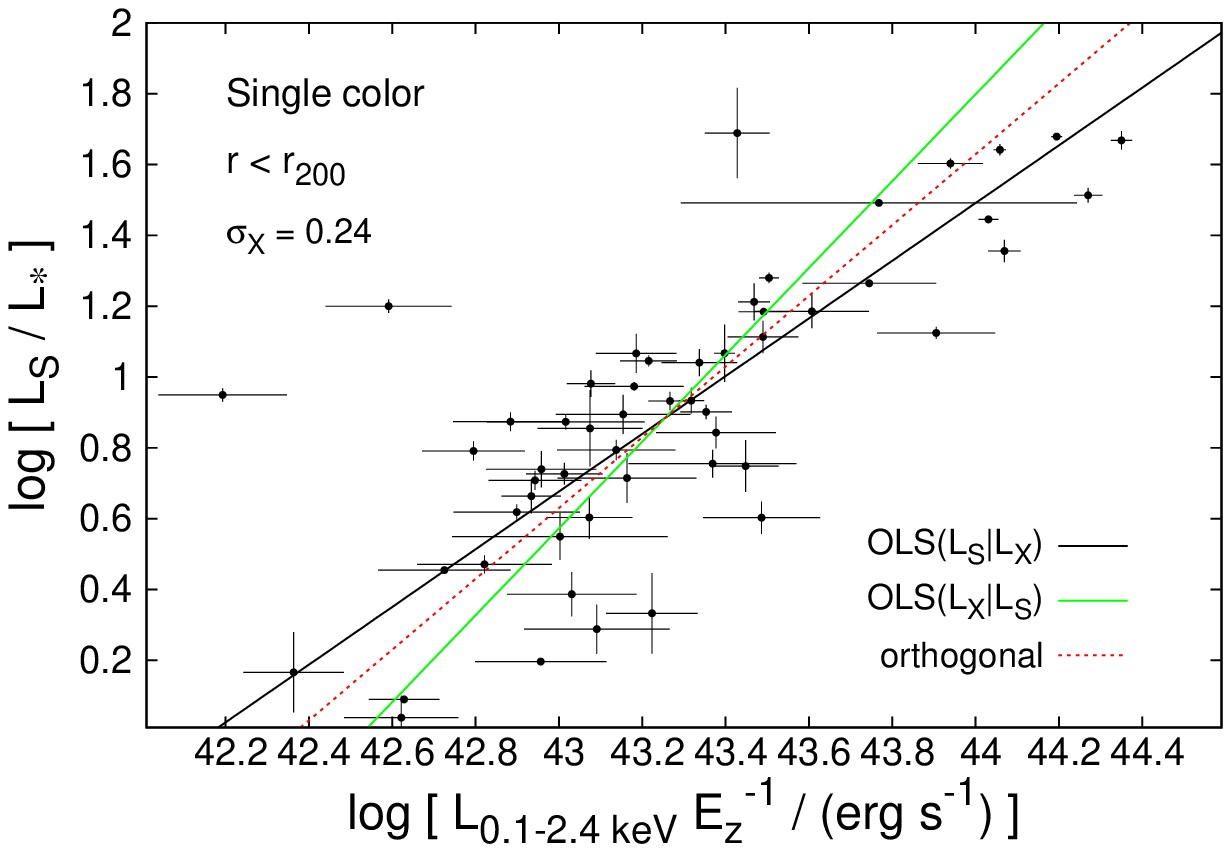}

\includegraphics[width=0.5\textwidth]{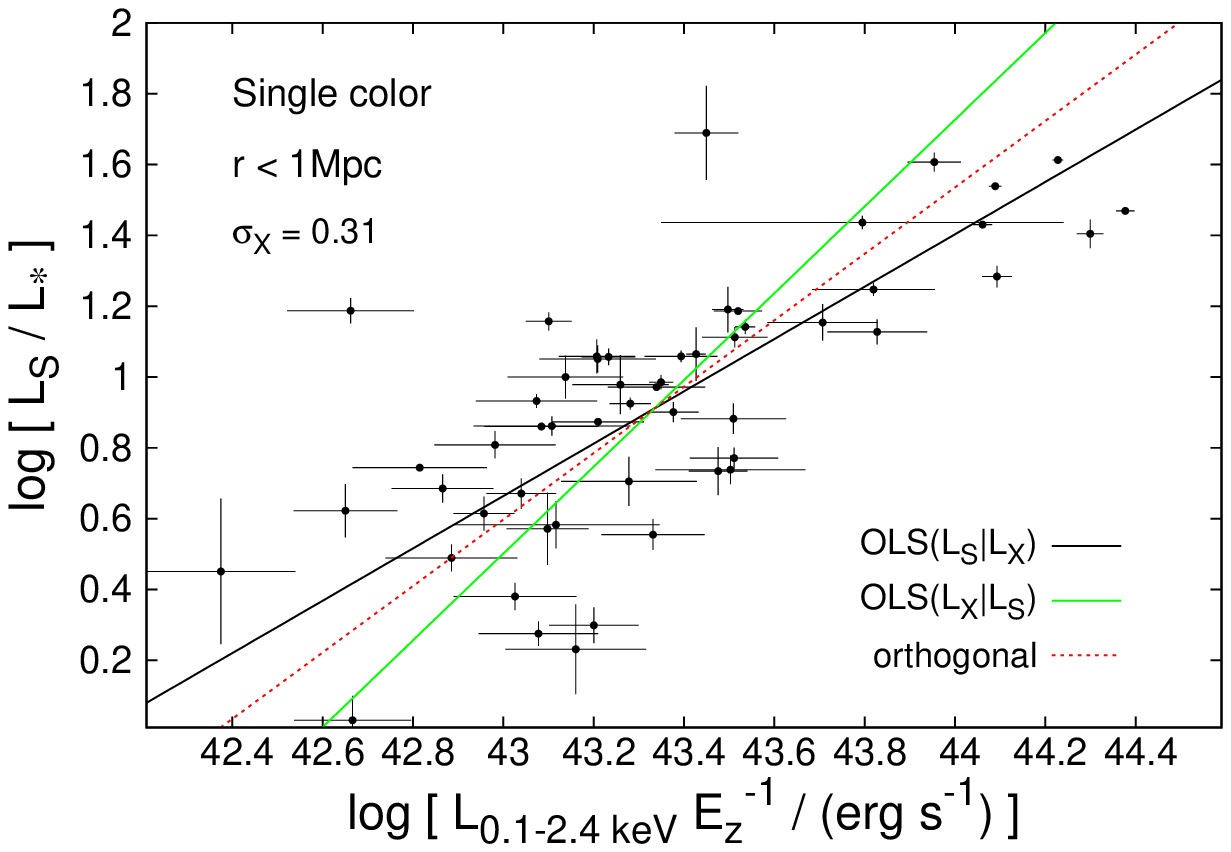}
}
\mbox{

\includegraphics[width=0.5\textwidth]{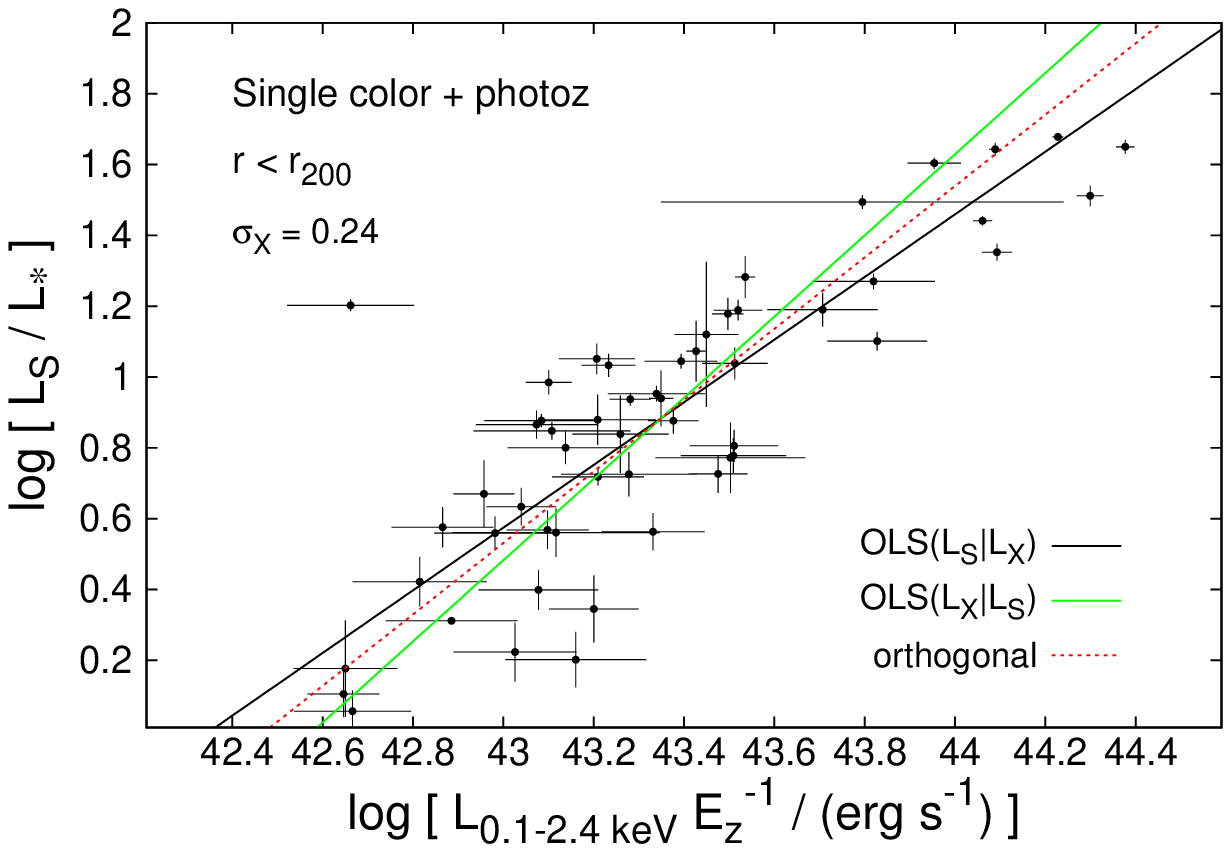}

\includegraphics[width=0.5\textwidth]{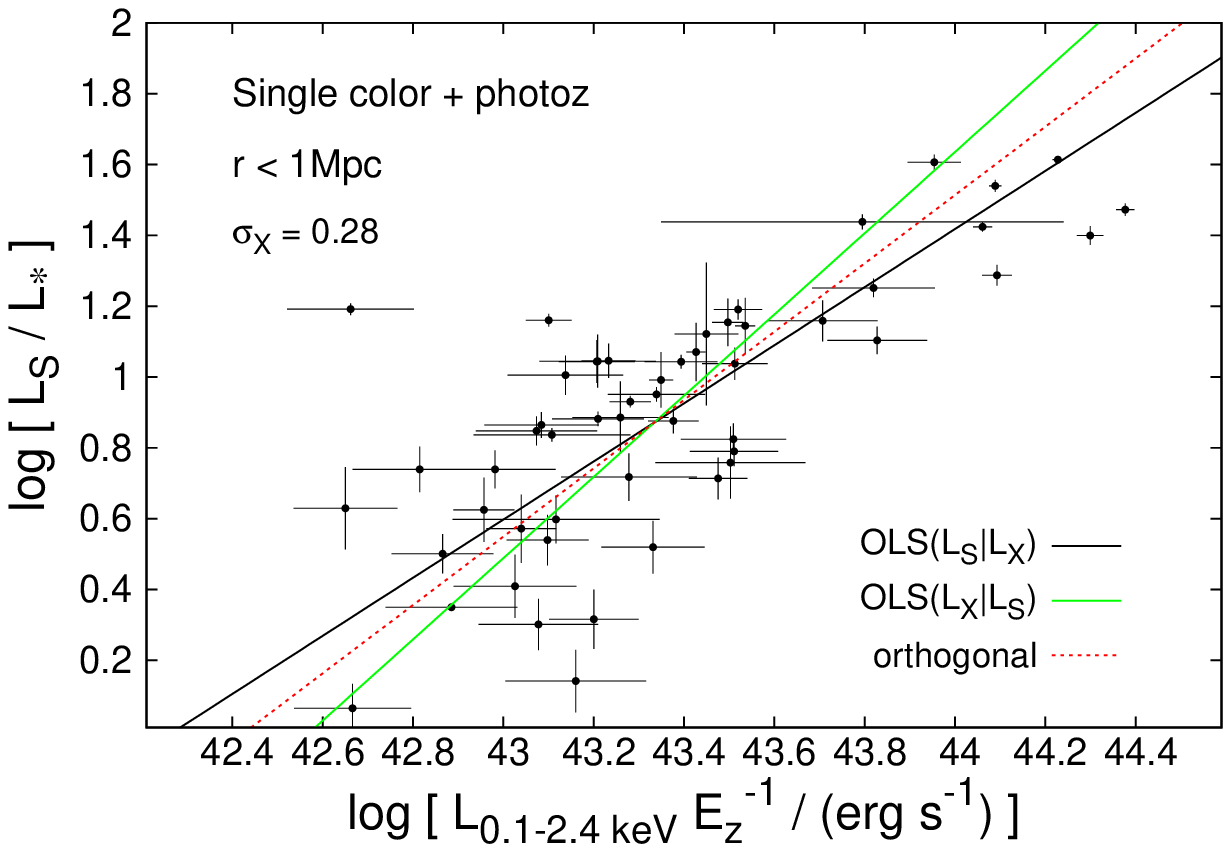}

}
\caption{Integrated stellar luminosity in $z^\prime$ band versus 
$L_\mathrm{X}$ for clusters at $0.1<z<0.6$. The
right and left panels show the results within $r_\mathrm{200c}$ and
1~Mpc, respectively. Upper panels use galaxy selection from multi-color
red sequence, middle panels are the single-color red sequence and
bottom panels belong to selection by combination of photometric redshift and single color red sequence. 
The solid black, solid green and dashed red lines show
OLS($L_\mathrm{S}$|$L_\mathrm{X}$), OLS($L_\mathrm{X}$|$L_\mathrm{S}$)
and orthogonal fits, respectively.  In each panel, $\sigma_{X}$ is the
scatter in $L_\mathrm{X}$ direction for the orthogonal fits.
The fitting parameters are summarised in Table \ref{fitting_result_Lx_Ls}.
\label{Lx-Ls}}
 \end{figure*}

The intrinsic velocity dispersion is calculated by subtracting the
contribution of redshift errors from the rest frame velocity
dispersion. To assess the velocity dispersion error associated with
galaxy sampling, a Jackknife method is applied \citep{Efron82} and
the associated error is computed as
$[\frac{N}{N-1}\sum(\delta_i^2)]^\frac{1}{2}$, where
$\delta_i=\sigma(v)_\mathrm{obs}-\sigma(v)_\mathrm{obs,excluding\,
i_\mathrm{th}\, member}$, for a cluster with $N$ member galaxies.
\cite{Connely12} showed that for calculation of velocity dispersion, 
applying luminosity weighted recentering can change the center up 
to 0.18 arcminutes but it does not change the velocity dispersion 
value. For more detailed description of velocity dispersion calculation, 
see \cite{Connely12} and \cite{Erfanianfar13}. 

\begin{figure}[t]
\centering
\includegraphics[width=0.45\textwidth]{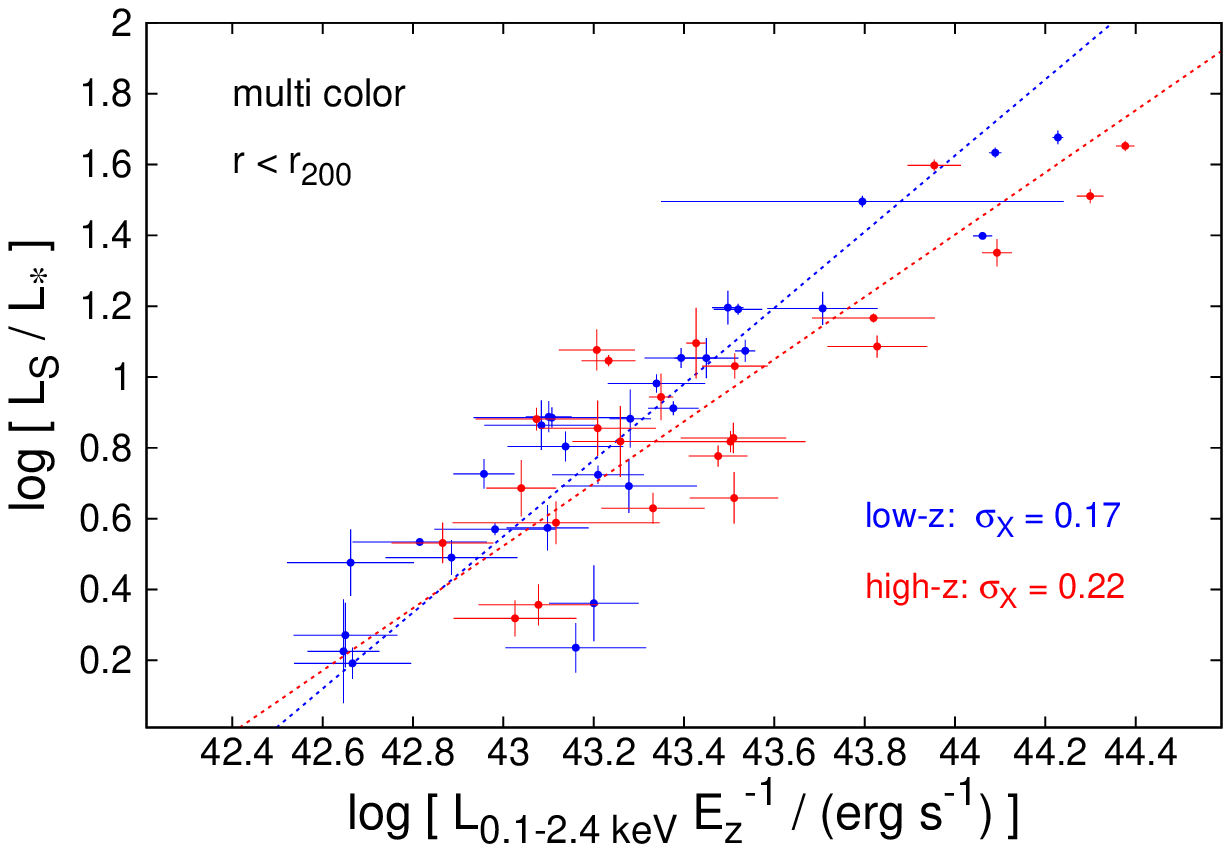}\hfill
\caption{Stellar luminosity versus X-ray luminosity of clusters for low redshift 
(blue dots) and high redshift (red dots) subsample. The blue and
red lines show the orthogonal fitting results for each subsample, with the parameters presented in Tab.\ref{fitting_result_Lx_Ls}. 
The fitting parameters are summarised in Table \ref{fitting_result_Lx_Ls_z}
\label{Lx-Ls_z}}
\end{figure}  

 \begin{figure}[t]
\includegraphics[width=0.45\textwidth]{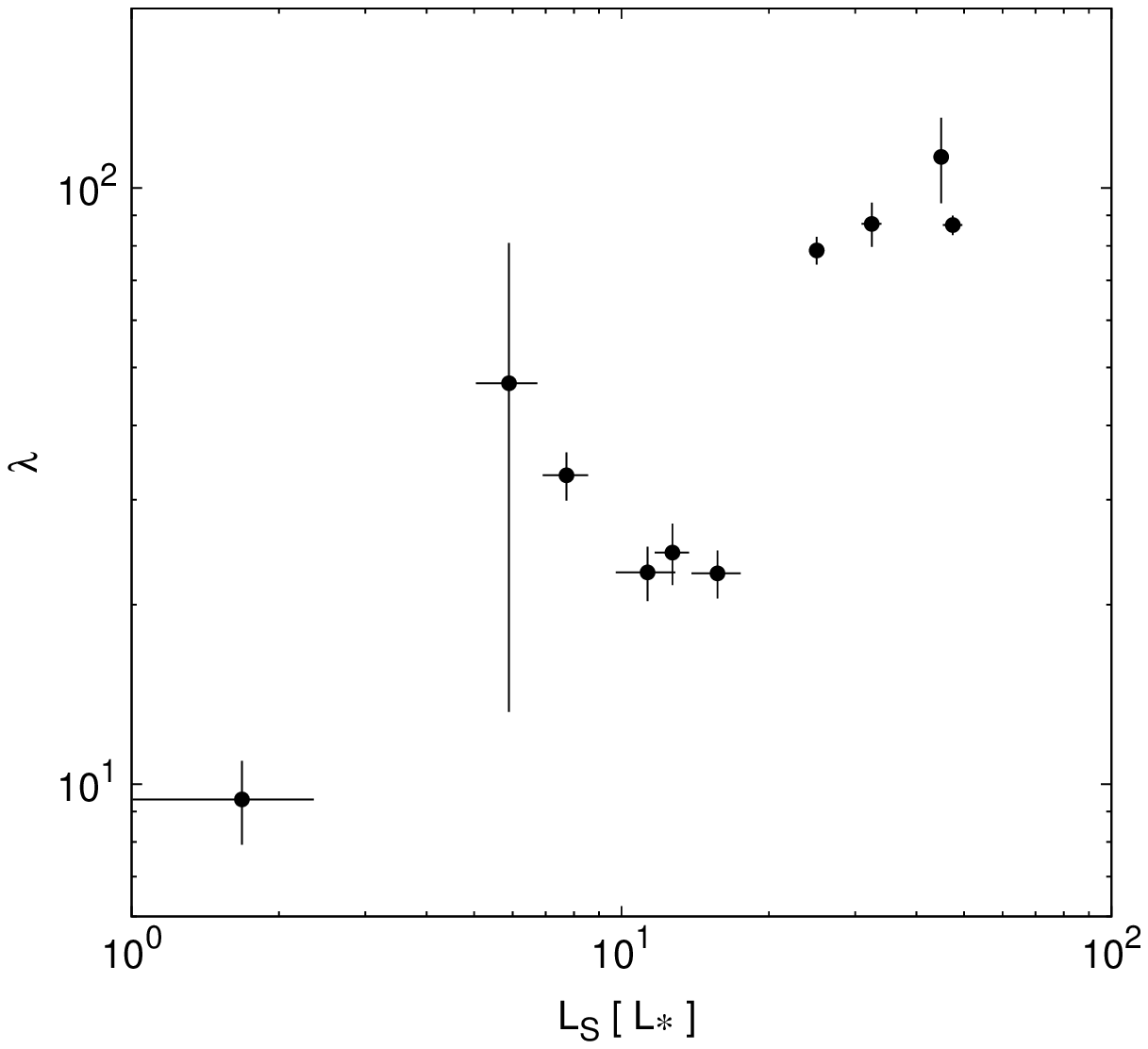}\hfill
 \caption{redMaPPer richness parameter $\lambda$ (calculated using SDSS data) 
versus stellar luminosity within $r_{200c}$, $L_\mathrm{S}$. 
A good correspondence between two measurements is observed. 
   \label{ls_lambda}}
\end{figure}

To investigate the results of our red sequence finder and velocity
dispersion calculation, let us compare ${\sigma_\mathrm{v}}$ to
$L_\mathrm{X}$.  Figure \ref{sigma_lx} shows the X-ray luminosity as a
function of velocity dispersion for 16 XMM clusters with more than ten
spectroscopic counterparts.  The black line shows the expected
relation between velocity dispersion and X-ray luminosity from
\cite{Leauthaud10}. The grey area also shows a 20$\%$ error on mass
estimate from using the $L_\mathrm{X}$-$M_\mathrm{200c}$ relation
\citep{Allevato12}. We do not account for the intrinsic scatter
between velocity dispersion and $M_{200c}$. The blue and red lines are
fitted lines using bisector and orthogonal fitting methods
(\cite{Akritas96}). The bisector method minimises the square distance
independently in X and Y directions. The orthogonal method minimises
the squared orthogonal distances. The result of bisector fitting
method is very close to the relation, expected from the weak lensing
calibration.  While most of the clusters are close to the predicted
relation, three of them have significantly larger $L_\mathrm{X}$ than
the values of $L_\mathrm{X}$ predicted by the scaling relation. Since
this offset is about one order of magnitude in $L_\mathrm{X}$, a
significant contribution of unresolved X-ray point sources can be
ruled out. Among these three clusters, two less luminous ones have
N$_{\sigma}$ of 12 and 13 and the more luminous one has 20. As
discussed in \cite{Ruel13}, the low number of spectroscopic members can
be a reason for these deviations. The compatibility between our
${\sigma_\mathrm{v}}$--$L_\mathrm{X}$ relation and the scaling
relation of \cite{Leauthaud10} indicates that although
\cite{Leauthaud10} relation was derived using a sample of clusters
mostly with $L_\mathrm{X} < 10^{43} erg s^{-1}$, it is still reliable
for mass estimation of more luminous clusters.


\begin{table*}[t]
\begin{center}
\renewcommand{\arraystretch}{1.1}\renewcommand{\tabcolsep}{0.12cm}
\caption{Fitting parameters of $log(L_\mathrm{X})$--$log(L_\mathrm{S})$ relation. Col.(1) indicates the type of selection of red galaxies; col.(2) is the radius within which $L_\mathrm{X}$ is
    calculated. The fitting procedure is listed in col.(3).  The
    cols. (4--5) present the intercept and slope of fittings
    respectively. The scatter in log$L_\mathrm{X}$ and
    log$L_\mathrm{S}$ direction are in cols. (6--7).
\label{fitting_result_Lx_Ls}}
\centering
\begin{tabular}{cc|ccccc}
\hline
\hline
red sequence & radius           & Fitting                               & intercept       & slope & $L_\mathrm{X}$ scatter & $L_\mathrm{S}$ scatter \\
	     &                   &                                      &                 &               & dex       & dex  \\
\hline
                          &                         	 	& OLS($L_\mathrm{S}|L_\mathrm{X}$)   & -38.97$\pm$2.97 & 0.92$\pm$0.07 & 0.21 & 0.19 \\
multi-color  & $r_\mathrm{200c}$ 	& OLS($L_\mathrm{X}|L_\mathrm{S}$)   & -44.33$\pm$3.21 & 1.04$\pm$0.07 & 0.20 & 0.21  \\    
                          &            	              	& orthogonal      		                  & -41.51$\pm$3.11 & 0.98$\pm$0.07 & 0.20 & 0.20  \\
\hline             
                          &            	              	& OLS($L_\mathrm{S}|L_\mathrm{X}$)   & -40.46$\pm$5.13 & 0.95$\pm$0.12 & 0.28 & 0.27 \\
single color & $r_\mathrm{200c}$ 	& OLS($L_\mathrm{X}|L_\mathrm{S}$)   & -54.12$\pm$5.39 & 1.27$\pm$0.12 & 0.25 & 0.32 \\ 
                          &            	              	& orthogonal      		                  & -47.50$\pm$5.03 & 1.12$\pm$0.12 & 0.26 & 0.29  \\
\hline
                                       &            	 	& OLS($L_\mathrm{S}|L_\mathrm{X}$)   & -37.47$\pm$4.19 & 0.88$\pm$0.10 & 0.26 & 0.23  \\
single color + photoz & $r_\mathrm{200c}$ & OLS($L_\mathrm{X}|L_\mathrm{S}$) & -48.82$\pm$4.46 & 1.15$\pm$0.10 & 0.23 & 0.27  \\ 
                                       &            	 	& orthogonal      		                               & -42.82$\pm$3.57 & 1.01$\pm$0.08 & 0.24 & 0.24  \\
\hline
             &            		&              OLS($L_\mathrm{S}|L_\mathrm{X}$)    & -31.71$\pm$2.95 & 0.75$\pm$0.07 & 0.28 & 0.21   \\
multi-color  & 1 Mpc      		& OLS($L_\mathrm{X}|L_\mathrm{S}$)    & -42.15$\pm$4.33 & 0.99$\pm$0.10 & 0.25 & 0.24  \\    
             &            		& orthogonal     		                                             & -35.77$\pm$3.58 & 0.85$\pm$0.08 & 0.26 & 0.22  \\
\hline
             &            	                          	& OLS($L_\mathrm{S}|L_\mathrm{X}$)    & -31.12$\pm$4.35 & 0.74$\pm$0.10 & 0.37 & 0.27   \\
single color & 1 Mpc     		& OLS($L_\mathrm{X}|L_\mathrm{S}$)    & -52.18$\pm$7.32 & 1.23$\pm$0.17 & 0.29 & 0.36   \\    
             &           	                          	& orthogonal     		                        & -39.75$\pm$5.99 & 0.94$\pm$0.14 & 0.31 & 0.29  \\
\hline
             &                         		& OLS($L_\mathrm{S}|L_\mathrm{X}$)    & -37.47$\pm$4.19 & 0.88$\pm$0.10 & 0.32 & 0.26   \\
single color + photoz & 1 Mpc     	& OLS($L_\mathrm{X}|L_\mathrm{S}$)    & -48.82$\pm$4.46 & 1.15$\pm$0.10 & 0.27 & 0.31  \\    
             &           	                          	& orthogonal     		                                & -42.82$\pm$3.57 & 1.01$\pm$0.08 & 0.28 & 0.27   \\
\hline
\end{tabular}
\end{center}
\end{table*}

\begin{table*}
\begin{center}
\renewcommand{\arraystretch}{1.1}\renewcommand{\tabcolsep}{0.12cm}
\caption{Fitting parameters of $log(L_\mathrm{X})$--$log(L_\mathrm{S})$ relation for low and high redshift subsamples.
Stellar luminosity computed using multi-color selection.
 Col.(1) indicates redshift ranges of redshift ranges; col.(2) is the radius within which $L_\mathrm{X}$ is
    calculated. The fitting procedure is listed in col.(3).  The
    cols. (4--5) present the intercept and slope of fittings
    respectively. The scatter in log$L_\mathrm{X}$ and
    log$L_\mathrm{S}$ direction are in cols. (6--7).
\label{fitting_result_Lx_Ls_z}}
\centering
\begin{tabular}{cc|ccccc}
\hline
\hline
redshift        & radius           & Fitting                                             & intercept       & slope & $L_\mathrm{X}$ scatter & $L_\mathrm{S}$ scatter  \\
	            &                        &                                                             &                 &               & dex  & dex \\
\hline
                     &            	 	& OLS($L_\mathrm{S}|L_\mathrm{X}$)                              &  -44.56$\pm$4.10 & 1.05$\pm$0.09 & 0.17 & 0.18 \\
0.1 $<$ z $<$ 0.3  & $r_\mathrm{200c}$ 	& OLS($L_\mathrm{X}|L_\mathrm{S}$)  &  -46.69$\pm$3.10 & 1.10$\pm$0.09 & 0.17 & 0.19 \\    
                     &            	 	& orthogonal      		                                                     &  -45.69$\pm$3.89 & 1.08$\pm$0.09 & 0.17 & 0.18 \\
\hline
                     &            	 	& OLS($L_\mathrm{S}|L_\mathrm{X}$)                              & -33.61$\pm$3.59 & 0.79$\pm$0.08 & 0.23 & 0.19 \\
0.3 $<$ z $<$ 0.6  & $r_\mathrm{200c}$ 	& OLS($L_\mathrm{X}|L_\mathrm{S}$)  & -42.49$\pm$5.13 & 0.10$\pm$0.12 & 0.21 & 0.21 \\    
                     &            	 	& orthogonal      		                                                     & -37.27$\pm$4.22 & 0.88$\pm$0.10 & 0.22 & 0.19 \\
\hline
                     &            	 	& OLS($L_\mathrm{S}|L_\mathrm{X}$)                     & -35.54$\pm$3.81 & 0.84$\pm$0.09 & 0.24 & 0.21 \\
0.1 $<$ z $<$ 0.3  & 1 Mpc   	& OLS($L_\mathrm{X}|L_\mathrm{S}$)            & -43.27$\pm$5.65 & 1.02$\pm$0.13 & 0.23 & 0.23 \\    
                     &            	 	& orthogonal      		                                            & -38.91$\pm$4.38 & 0.92$\pm$0.10 & 0.23 & 0.21 \\
\hline
                     &            	 	& OLS($L_\mathrm{S}|L_\mathrm{X}$)                     & -29.42$\pm$4.14 & 0.69$\pm$0.10 & 0.29 & 0.20 \\
0.3 $<$ z $<$ 0.6  & 1 Mpc   	& OLS($L_\mathrm{X}|L_\mathrm{S}$)            & -42.79$\pm$6.85 & 1.01$\pm$0.16 & 0.24 & 0.25 \\    
                     &            	 	& orthogonal      		                                            & -34.25$\pm$5.50 & 0.81$\pm$0.13 & 0.26 & 0.21 \\

\hline
\end{tabular}
\end{center}
\end{table*} 

\subsection{Stellar luminosity as a $L_\mathrm{X}$ estimator} \label{Lstellar}

We calculate the integrated $z^\prime$-band luminosity,
$L_\mathrm{S}$, of red sequence galaxies (brighter than
$0.4L_\mathrm\ast$) within $r_\mathrm{200c}$, for clusters in the
redshift range of 0.1$<$z$<$0.6 and the X-ray detection threshold
above 4.6. The $r_\mathrm{200c}$ is also calculated from
$M_\mathrm{200c}$ (see section \ref{catalogs}). The luminosity of red
sequence galaxies are added to each other and subtracted by background
luminosity at the same redshift. The background is the mean of
integrated luminosity of the red sequence galaxies at random points in sky
and within the similar radius. In \S \ref{red_seq_method}, we
mentioned that we define the width of the red sequence to enclose the bulk
of bright red sequence galaxies. Here we show that the adopted width
does not affect the measured stellar luminosity of the clusters.  For
this purpose, we increase the widths of all colors in the red sequence
selection to three times the $\sigma_{a-b}(z)$ (1.5 times the previous
width) and re-computed the stellar luminosity. The background
computation was also repeated for changing the width of red
sequence. Figure \ref{Ls_3sig_Ls_2sig} illustrates the variation of
$L_\mathrm{S}$ after increasing the width of red sequence by
50$\%$. The change in $L_\mathrm{S}$ is 0.02 dex with a standard
deviation of 0.06 dex. We conclude that the obtained $L_\mathrm{S}$
values have converged.

\begin{figure*}[t] 

\mbox{

\includegraphics[width=0.5\textwidth]{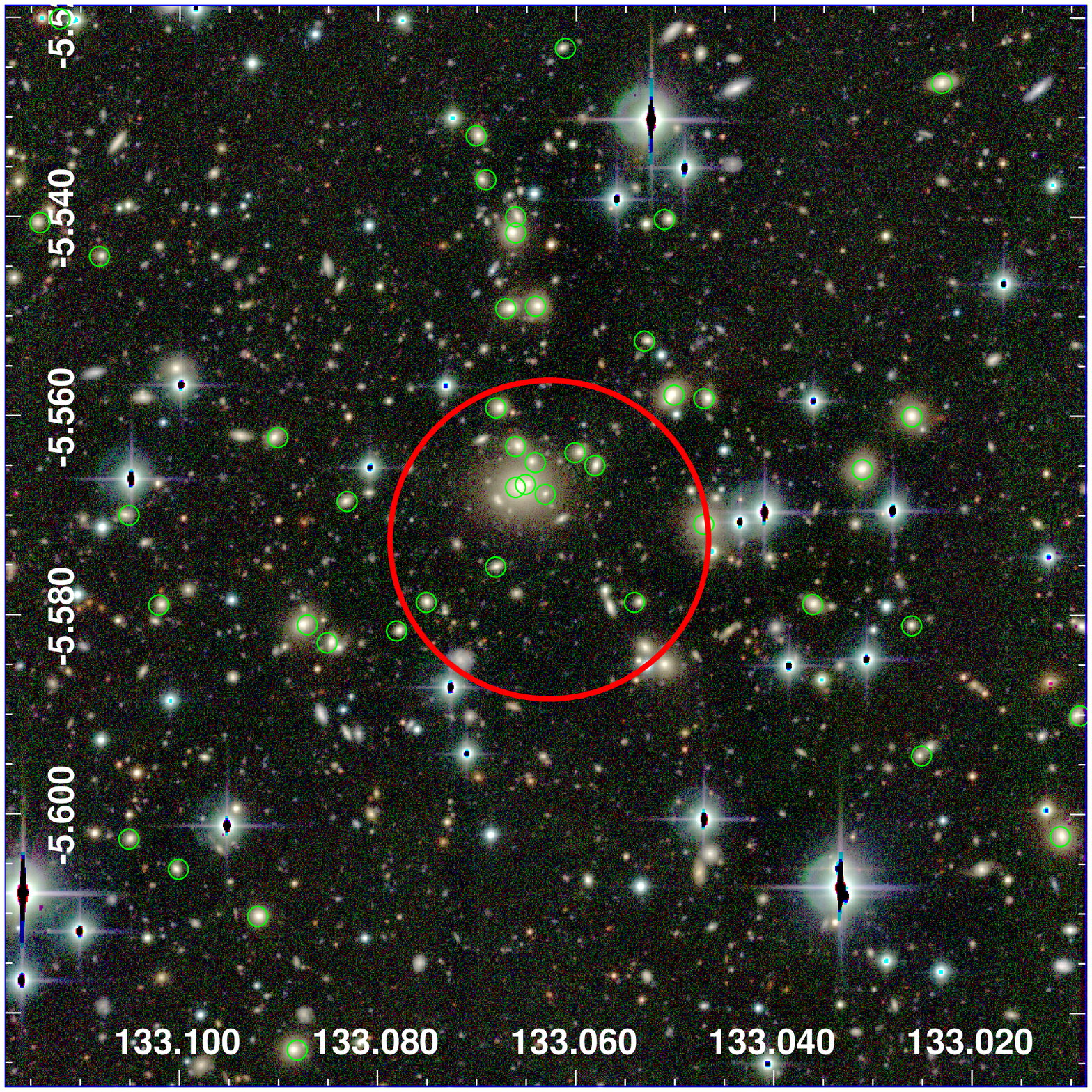}

\includegraphics[width=0.5\textwidth]{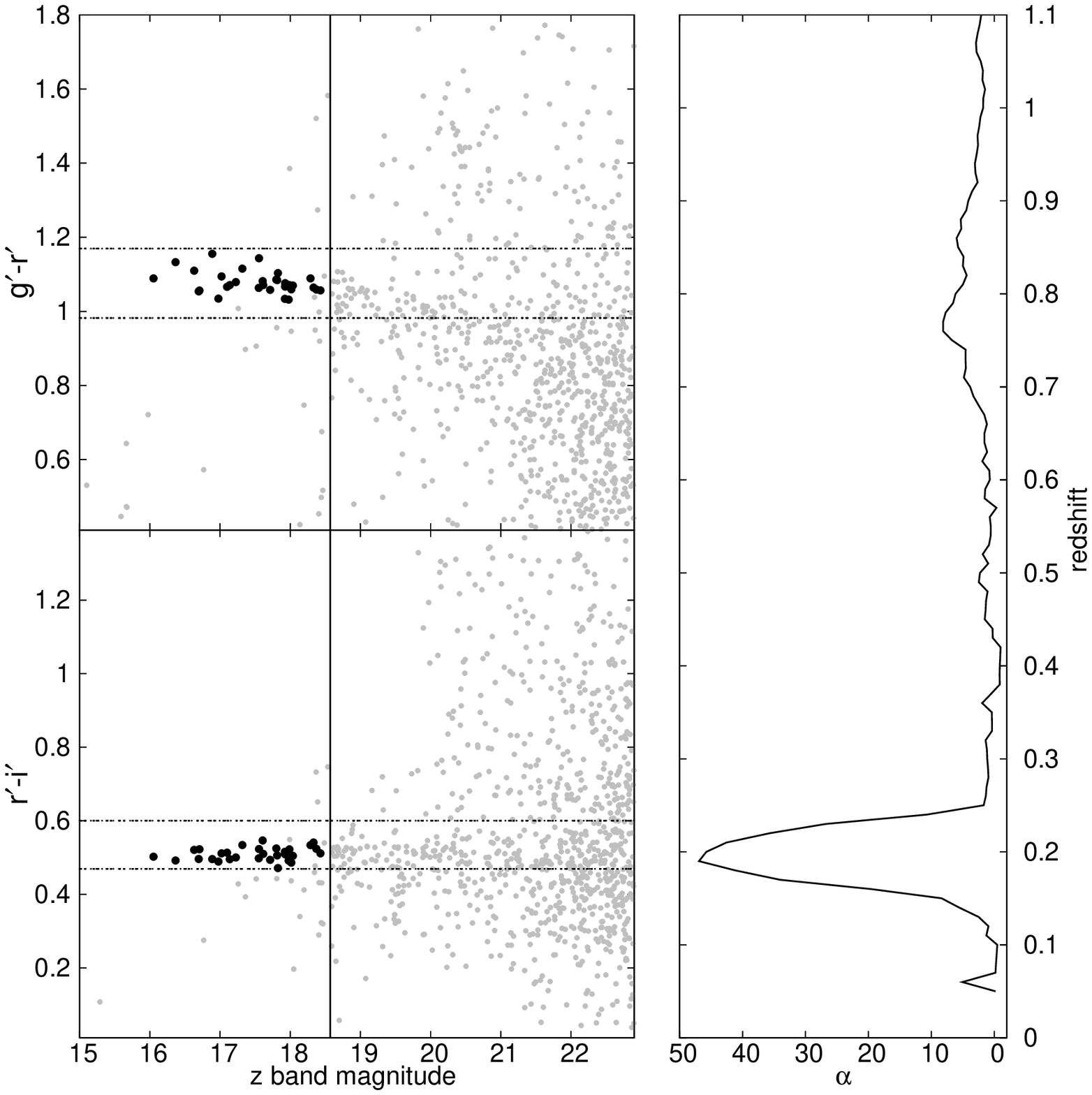}
}
\caption{\footnotesize As in Figure \ref{xmm110460}
but for a RASS cluster at the redshift of 0.19. The position of the
RASS source is shown by a large red circle with a radius of one
arcminute. }

\label{361_4paper.eps}

\end{figure*}

\begin{figure*}[t]
\centering
 \includegraphics[width=0.9\textwidth]{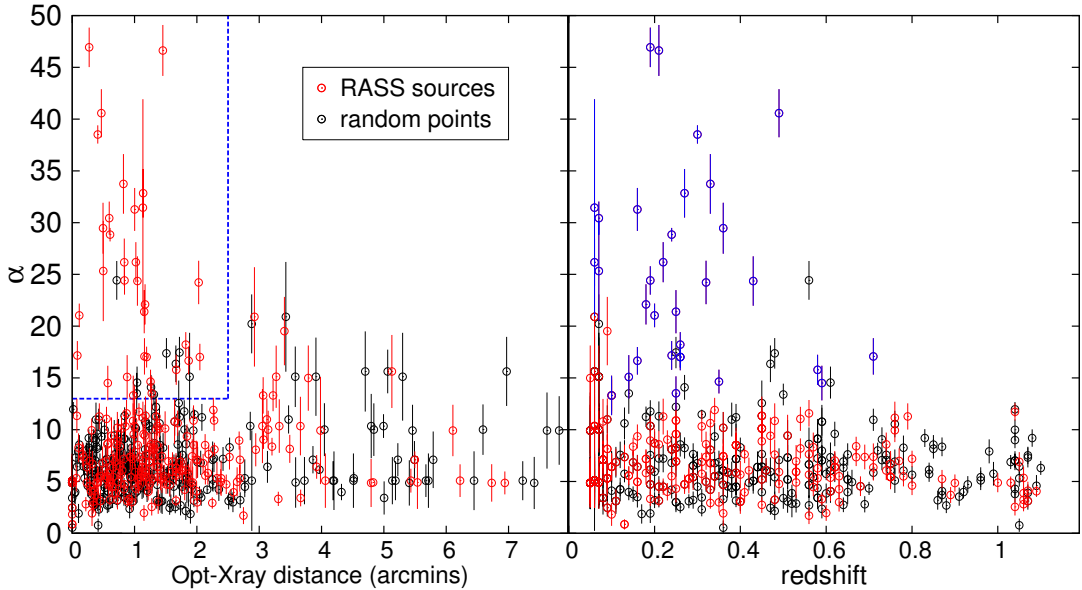}\hfill
 \caption{{\it Left:}  Red sequence significance, $\alpha$, versus
     redshift for RASS X-ray sources. {\it Right:} Red sequence
     significance, $\alpha$, versus offset between optical and X-ray
     cluster centers. Red circles mark the RASS X-ray sources and
     black circles are random sources in CFHTLS. A blue dashed line
     shows the criteria for selection of clusters among the RASS
     sources. The right-hand panel shows $\alpha$ versus the redshift
     for the RASS and random sources. In this panel, blue circles are
     the RASS clusters.}

\label{rass_sigma_z_dist.eps} \end{figure*}

 \begin{figure}
\includegraphics[width=0.45\textwidth]{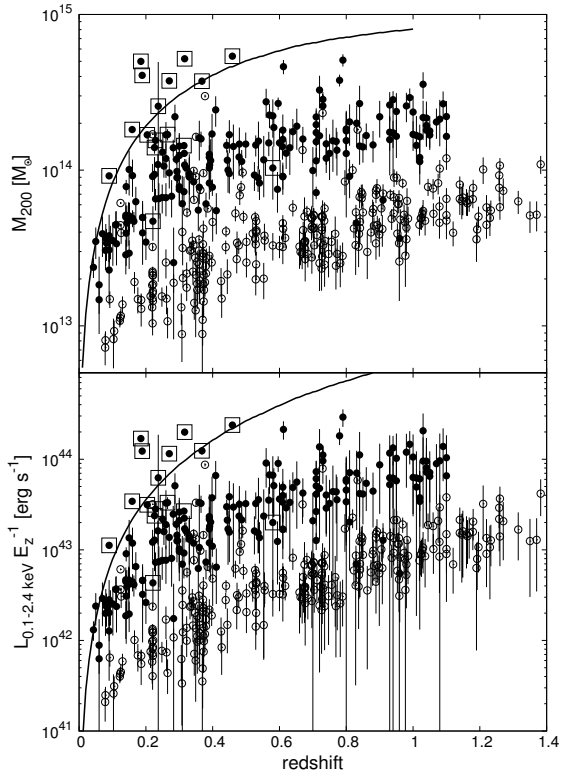}\hfill
\caption{X-ray mass (top) and X-ray luminosity (bottom) as functions of
  redshift. Black dots and open circles show XMM-CFHTLS and COSMOS
  X-ray selected galaxy clusters. The errors are calculated with
  statistical errors in the X-ray flux measurements. 16 XMM clusters
  in common to RASS clusters are marked as squares. In both panels,
  the solid curves show the detection limits in luminosity- and
  mass-redshift spaces corresponding to X-ray flux limit of
  $2\times10^{-13} erg$ $cm^{-2} s^{-1}$.  \label{xmass_xlum_z}}
\end{figure}   

 \begin{figure}[t]
\includegraphics[width=0.45\textwidth]{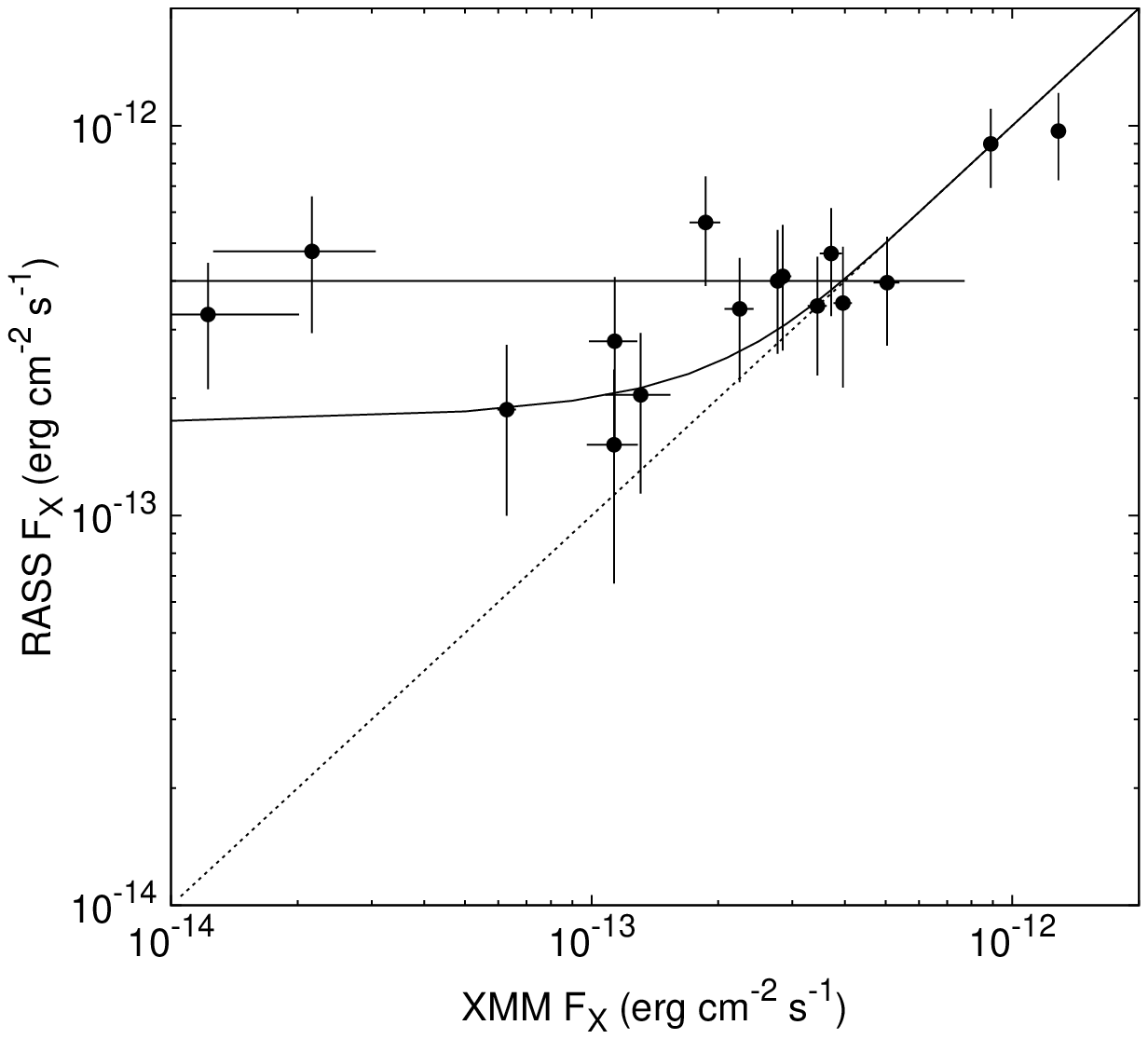}\hfill
\caption{RASS X-ray flux versus XMM-Newton X-ray flux for 16 clusters in
 overlap between the RASS and XMM cluster samples. A solid curve shows the
the prediction for Malmquist bias in RASS flux measurement.
 \label{xmmf_rassf}}
\end{figure}  

In some
cases, bright stars affect the photometry. We discard the affected
clusters from determination of $L_\mathrm{X}$.  Figure
\ref{Lx-Ls_1} illustrates the relation between $L_\mathrm{S}$ and
$L_\mathrm{X}$ for the sample of clusters in the redshift range of
0.1$<$z$<$0.6 and the X-ray detection threshold above 4.6.  There is a
strong correlation between $log(L_\mathrm{S})$ and $log(L_\mathrm{X})$ for
the bulk of the sample. The Spearman test coefficient for this relation is
0.640 with the zero value for the probability of null hypothesis of
null correlation between two quantities.

The good relation between $L_\mathrm{S}$ and $L_\mathrm{X}$ is a
motivation for using $L_\mathrm{S}$ as an estimator for $L_\mathrm{X}$
and, consequently, the cluster mass. For this purpose, besides of
$L_\mathrm{S}$ within $r_\mathrm{200c}$, we also measure the
$L_\mathrm{S}$ within 1 Mpc from the X-ray centre. Figure \ref{Lx-Ls}
illustrates the relation between $L_\mathrm{S}$ and $L_\mathrm{X}$ for
the sample of clusters. The upper left and upper right panels show the
$L_\mathrm{S}$ computed within $r_\mathrm{200c}$ and 1 Mpc,
respectively. The later is useful in the situations when the
measurement of the virial radius is not possible or noisy.  In Figure
\ref{Lx-Ls}, the lines show the power-law fits to the relation. The
procedures of \cite{Akritas96} ordinary least square (OLS) and
bi-variate correlated errors and intrinsic scatter (BCES) estimators
are used to produce the fits. The ordinary least square estimators in
$L_\mathrm{X}$ direction (OLS($L_\mathrm{X}$|$L_\mathrm{S}$)) and
$L_\mathrm{S}$ direction (OLS($L_\mathrm{S}$|$L_\mathrm{X}$)) are
shown as black and green solid curves, respectively. The red dashed
lines are the results of BCES orthogonal fitting method, which
minimises the squared orthogonal distances. The parameters of the
plotted relations are listed in Table \ref{fitting_result_Lx_Ls}.

For comparison to the multi-color red sequence, we compute the
$L_\mathrm{S}$ with a single-color selection of red sequence galaxies
($g^\prime-r^\prime$ for 0.05$\le$ z $\le$0.4 and $r^\prime-i^\prime$
for 0.4$<$ z $\le$0.6). We also compute the $L_\mathrm{S}$ with
a combination of photometric redshift and single-color selection of
red sequence galaxies. In this method, $L_\mathrm{S}$ is computed for
galaxies that satisfy both conditions of photoz range and
single-color. We need to adopt a suitable redshift range for photoz
selection. A suitable redshift range is the sum in quadrature of two
redshift errors, uncertainty in measurement of cluster redshift and
errors in photometric redshift of galaxies. In $\S$ \ref{red_seq_xmm},
we show that our red sequence technique has an uncertainty of
$0.02(1+z)$ in cluster redshift measurement.  The accuracy in
photometric redshift varies with galaxy magnitude.  We assume the
worst photoz accuracy which belongs to the galaxies with brightness of
$m_\mathrm{\ast}$+1 at redshift 0.6. According to the Figure
\ref{z_compare}, the z-band magnitude of such galaxy is $21.1$. We
compute the photometric redshift error for galaxies with z-band
magnitude of between 20.6 and 21.1. For 886 galaxies with such
magnitude and with spectroscopic redshift in the three fields of
CFHTLS, the photometric redshift error is $0.031(1+z)$. For selection
of member galaxies, we adopt the redshift interval of $\pm 0.07(1+z)$
around the mean redshift of the cluster.

The best method (among multi-color, single-color and
single-color-photoz) is sought to provide the lowest scatter versus
$L_\mathrm{X}$. The results are compared in Figure \ref{Lx-Ls}, using
$r_\mathrm{200c}$ and 1 Mpc as an extraction radius. The middle and
bottom panels of Figure \ref{Lx-Ls} show the relation between the
cluster X-ray luminosity versus $L_\mathrm{S}$ computed using the
single color and single-color-photoz methods, respectively. These
relations are fitted with power-law models, and the results of fitting
are shown in Table \ref{fitting_result_Lx_Ls}.

For all $L_\mathrm{X}-L_\mathrm{S}$ scaling relations the scatter for
the multi-color red sequence finder is smaller than or equal to the
single-color and single-color-photoz values, independent of the selection radius and the
fitting method. For example, for $L_\mathrm{S}$ computed with
$r_{200c}$, the orthogonal relation has a
scatter of 0.20, 0.29, and 0.24 dex in $L_\mathrm{X}$ for multi-color,
single color, and a combination of single color and photoz respectively. 
The reduction of scatter is even more
significant in the case of a fixed 1~Mpc radius. For instance, the
orthogonal scatter is 0.26 dex in $L_\mathrm{X}$ for the
multi-color, 0.31 dex for the single-color and 0.28 dex for the combination of single-color and photoz. 
Our results on the tight relation between the $L_\mathrm{S}$ and other
mass proxies, such as $L_\mathrm{X}$ are in line with the low redshift
studies of \cite{Rykoff12} at 0.1$<$z$<$0.3 and \cite{Andreon12} for
z$<$0.14.

In Figure \ref{Lx-Ls_z} we consider the redshift evolution of the
$L_\mathrm{S}$--$L_\mathrm{X}$ relation. Using two subsamples with
0.1$<$z$<$0.3 and 0.3$<$z$<$0.6, we find a difference in the relation
to X-ray luminosity ($L_\mathrm{X}$ $>$ 42.5 ergs s$^{-1}$). The low
redshift relation is within the errors of high redshift relation.  
The parameters of the relation are presented in Table \ref{fitting_result_Lx_Ls_z}. 
The scatter in $L_\mathrm{X}$ reduces down to 0.17dex for low redshift sample.

To compare our red sequence finder to other work, Figure \ref{ls_lambda}
shows $L_\mathrm{X}$ versus richness parameter $\lambda$ used in
redMaPPer (next generation of MaxBCG method, \citealt{Rykoff13}),
designed to find clusters in SDSS data. Briefly, redMaPPer applies a
red sequence model and assumes radial and luminosity filters to
calculate the probability that a given galaxy belongs to a
cluster. The parameter $\lambda$ is the sum of mentioned
probabilities. There are 10 RASS clusters in overlap between SDSS and
the CFHTLS fields. The large errors in $\lambda$ for a few clusters
are caused by the shallow depths of SDSS data.
\cite{Rozo14} reported a scatter of 0.23 dex in X-ray temperature at 
fixed $\lambda$. Figure \ref{ls_lambda} shows that $L_\mathrm{S}$ and
$\lambda$ correlate.

\subsection{Applying the red sequence finder to identify RASS sources} \label{red_seq_rass}

Performance of our XMM program was based on the identification of RASS
sources as galaxy clusters. This lead to the development and
verification of the source identification methods reported above. It
allows us to present a consistent identification of RASS sources using
the same method, which allows us to both characterise the target
selection, and to report the clusters which we have not observed,
since we include the full CFHTLS dataset in this analysis, covering
180 square degrees.
 
We apply the red sequence finder to identify clusters associated with
245 RASS sources within the three CFHTLS fields in our study. 
According to the $log(N)-log(S)$
relation, clusters make up only 10\% of X-ray sources
(\citealt{Finoguenov07}; \citealt{Cappelluti07}), making cluster
identification difficult using unresolved X-ray sources in RASS
data. The radius for galaxy selection has been set to 0.5 Mpc at each
redshift plus two arcminutes to account for the survey PSF of
RASS. After finding the red member galaxies, we derive the
$z^\prime$-band luminosity weighted center for each cluster candidate,
which then defines the distance between the optical counterpart and
the X-ray source position (hereafter Opt--X-ray distance). Figure
\ref{361_4paper.eps} shows the red sequence finder results for a
cluster at a redshift of 0.19.

In order to distinguish between X-ray sources associated by
  clusters and other X-ray sources, we used Opt--X-ray distance and
  $\alpha$ parameters. With a comparison between properties of RASS
  sources and random sources, we try to find X-ray clusters among RASS
  sources.  We similarly apply the red sequence finder on 300 random
  sources in CFHTLS fields. Figure \ref{rass_sigma_z_dist.eps} shows
  $\alpha$ parameter versus Opt--X-ray and redshift for RASS and
  random sources.  The red circles represent the 245 RASS sources and
  black circles are random points in CFHTLS fields. While only a
  handful of random sources can have high $\alpha$ values (15 or more)
  and low Opt--X-ray distance, tens of RASS sources achieve such
  values. This suggests that a combination of $\alpha$ and Opt--X-ray
  distance can discriminate between clusters and other sources of
  X-ray emission.  We select the X-ray clusters by cuts of $13<\alpha$
  and Opt--X-ray distance less than 2.5 arcminutes. The left panel in
  Figure \ref{rass_sigma_z_dist.eps} illustrates the cuts in blue
  dashed line. Nine random sources and 32 RASS X-ray sources located
  in selection region that shows the purity ($\sim$ 80 \%) in selected
  sample of X-ray clusters with this method. We will show in
  $\S$\ref{catalogs} that by adopted criteria, we can detected all XMM
  clusters with X-ray flux above the RASS X-ray detection
  threshold. By increasing $\alpha$ value one can achieve a purer
  sample. For instance, 20 RASS sources have $20<\alpha$ but only one
  random source has such high $20<\alpha$ value.

\section{RASS-CFHTLS and XMM-CFHTLS catalogs of X-ray selected clusters}\label{catalogs}

In this section, we present the RASS and XMM (X-ray) selected cluster
catalogs. The first catalog, Table \ref{xmm_cat}, belongs to the 196 XMM clusters. The first 133
lines in Table \ref{xmm_cat} belong to the sample with X-ray detection
threshold above 4.6 sigma and the last 63 lines are those with lower
detection threshold. Column 1 in Table \ref{xmm_cat} shows the
cluster ID for the XMM-CFHTLS sample with the first digit referring to
the CFHTLS wide field (1,2 or 4).  Columns 2 and 3 are respectively
R.A. and Dec. of the X-ray source centers. Columns 4 and 5 are the red
sequence redshift and red sequence significance, $\alpha$, of the
clusters.  Column 6 lists cluster flux and one sigma error in flux
corresponding to the 0.5--2 keV band in units of 10$^{-14}$ ergs
cm$^{-2}$ s$^{-1}$. Column 7 reports the rest-frame X-ray luminosity,
$L_\mathrm{X}$, in the 0.1--2.4 keV band. The total mass
$M_\mathrm{200c}$, estimated from the X-ray luminosity using the Lx--M
scaling relation and its evolution from \cite{Leauthaud10}, is given
in column 8. Column 9 lists corresponding radius, $r_\mathrm{200c}$,
in arcminutes. Spectroscopic redshifts of the clusters 
are provided in column 10. For clusters with a
spectroscopic redshift in this column, $M_\mathrm{200c}$, and,
$r_\mathrm{200c}$ are computed using spectroscopic redshift. Column 11
reports the visual flag described in Sect. \ref{red_seq_xmm}.
Velocity dispersion and number of spectroscopic members (both
described in Sect. \ref{Velocity_dispersion}) for clusters having more
than five spectroscopic members are given in columns 12 and 13,
respectively.

The RASS-CFHTLS cluster catalog is listed in Table \ref{rass_cat}. This catalog
includes 32 clusters with selection shown in the left-hand panel of 
Figure \ref{rass_sigma_z_dist.eps}. Column 1 is the cluster ID. The coordinates (RA
and DEC, Equinox J2000) of the clusters are given in columns 2 and 3.
The red sequence redshift and significance ($\alpha$) are listed in
column 4 and 5. The position of the optical center is reported in columns 6 and 7.
 Columns 8 and 9 report ROSAT X-ray flux and luminosity in units of
$10^{-13}$ erg s$^{-1}$ cm$^{-2}$ and $10^{42}$ erg s$^{-1}$ respectively. The spectroscopic redshifts which
were also verified visually are given in column 10. The columns 11 and
12 present the velocity dispersion and the number of spectroscopic
member galaxies from Sect. \ref{Velocity_dispersion}.  Based on the
derived relation between $L_\mathrm{X}$ and $L_\mathrm{S}$, we
estimate the $L_\mathrm{X}(L_\mathrm{S})$ for 32 RASS clusters. We
measured $L_\mathrm{S}$ within 1~Mpc from the optical center of
clusters (column 6 and 7 in table \ref{rass_cat}). The estimated
cluster $L_\mathrm{X}$ using orthogonal fitting result (in Table \ref{fitting_result_Lx_Ls_z} with 0.23 and 0.26 dex
scatter in X-ray luminosity for low and high redshifts respectively) are listed in column 13 of table
\ref{rass_cat}.

The inferred mass and X-ray luminosity of the XMM clusters as a
function of redshift are illustrated in Figure \ref{xmass_xlum_z}. We
mark 16 XMM clusters in common to RASS clusters as squares. This
sub-sample of RASS clusters stems from our targeted follow-up
observations of RASS clusters found inside the part of CFHTLS survey
publicly released in T0005 and presents an effective search for
massive clusters in the area of 90 square degrees.  The two curves in
Figure \ref{xmass_xlum_z} show the detection boundary related to
$2\times10^{-13} erg$ $cm^{-2} s^{-1}$ in X-ray flux. This flux is
associated with a detection limit over 85\% of the survey area. 
All XMM clusters more luminous or massive than these two curves
are also identified as RASS clusters using adopted criteria for selection 
of clusters among RASS sources (see $\S$ \ref{red_seq_rass}). The
mass (and luminosity) detection limit, shown with a curve in Figure
\ref{xmass_xlum_z}, also implies that only extreme clusters ($\sim 10^{15}
M_{\odot}$) at redshift $\sim1$ are detectable in RASS data.

We added COSMOS X-ray selected galaxy clusters
(\citealt{Finoguenov07}, \citealt{George11}) to the plots, to show the
difference in the cluster sample. At a fixed redshift, the typical
mass (and luminosity) of XMM-CFHTLS clusters are an order of magnitude
more massive (and more luminous) in comparison with a typical group in
deep surveys such as COSMOS. For example, at the redshift range of
$0.2\le z \le 0.3$, the median of the $M_\mathrm{200c}$ of XMM-CFHTLS
and COSMOS clusters are respectively 1.1$\times 10^{14}$ M$_{\odot}$
and 2.6$\times 10^{13}$ M$_{\odot}$. This difference between the mean
total mass (and luminosity) is even larger between COSMOS and
RASS-CFHTLS clusters.

A comparison of the X-ray fluxes from RASS and XMM is presented in
Figure \ref{xmmf_rassf}. At low flux levels, the RASS flux estimates are
subject to the Malmquist bias, as shown by a model curve. We also 
report that the mean of distances between the centre of RASS and 
XMM X-ray emissions is 0.6 arcminutes for 16 clusters in overlap 
between RASS and XMM samples.

\section{Summary}\label{summary}

We have presented the results of an X-ray search for bright clusters
in the CFHTLS fields. In this work we presented the cluster
identification in RASS and XMM data. We developed a method for
identifying clusters at the limits of RASS data, reaching flux
levels of $2\times10^{-13}$ erg cm$^{-2}$ s$^{-1}$, with the help of
deep photometric data, such as that of CFHTLS.

We have described a multi-color red sequence finder and calibrated it
for CFHTLS $u^\ast g^\prime r^\prime i^\prime z^\prime$ filters and
the redshift below 1.1. The spectroscopic follow-up was done using the
Hectospec spectrograph on MMT, with higher priority for clusters with
high X-ray flux. To increase the efficiency of spectroscopic
follow-up, target galaxies were selected within a range of photometric
redshift around the red sequence redshift of clusters. In this work we
also used SDSS spectroscopic data in the CFHTLS wide fields. We
applied our red sequence finder on RASS and XMM-Newton X-ray sources
in the W1, W2, and W4 CFHTLS fields. In total, we identified 32
clusters associated with RASS sources and 196 clusters among XMM X-ray
sources, with a 100\% identification rate achieved for the
high-significance XMM sample. We computed the X-ray luminosity and
mass from the X-ray flux and the scaling relations from the
literature. In comparison to other XMM samples, the clusters in our
sample are typically of $\sim$ $10^{14}$ M$_{\odot}$ masses, while
e.g., COSMOS X-ray selected groups are of an order of magnitude lower
mass. We calculated the velocity dispersions with an iterative gapper
method and derive the scaling relation between velocity dispersion and
X-ray luminosity of clusters.

We also explored a correlation of integrated optical luminosity and
X-ray luminosity.  We showed that multi-color red sequence reduces the
scatter in relation with X-ray luminosity. This set of optical methods
for cluster finding are particularly useful for providing large
samples of X-ray luminous (or massive) clusters (especially for
cosmological studies) using shallow X-ray data and wide optical
surveys. First, by applying the red sequence finder and maximising
$\alpha$, we can extract a pure sample of clusters out of a list of
X-ray sources. Second, by measuring the optical luminosity of clusters
within an appropriate fixed radius we can estimate the cluster total
mass, allowing an efficient separation of high X-ray luminous
(high-mass) clusters for further studies.

\begin{table*}[t]
\begin{center}
\renewcommand{\arraystretch}{1.1}\renewcommand{\tabcolsep}{0.05cm}
\caption{\footnotesize
{Catalog of XMM-CFHTLS X-ray Selected Galaxy Clusters }
\label{xmm_cat}}
\tiny
\centering
\begin{tabular}{ccccccccccccc}
\hline
\hline
ID         & RA        & DEC          & R.S. z & $\alpha$ & X-ray  Flux                   & $L_{X}$               & $M_{200c}$          & $r_{200c}$& specz&visual flag&$\sigma(v)$ & $N_{\sigma}$\\
XMM-CFHTLS & (degrees)  & (degrees)   &        &          &$10^{-14} erg$ $cm^{-2} s^{-1}$& $10^{42} erg$ $s^{-1}$& $10^{13}M_{\odot}$ & arcmin    &       &          &($km s^{-1})$ &  \\
(1)         & (2)       & (3)         & (4)    & (5)      & (6)                           & (7)                   & (8)                & (9)       &   (10) & (11)    &(12)&(13)\\

\hline
XCC J0210.4-0343 & 32.6184 & -3.7202 & 0.45 & 0.86$\pm$0.92 & 3.39$\pm$0.54 & 37.08$\pm$5.96 & 14.42$\pm$1.44 & 2.73 & 0.4417 & 1 & -  & 1 \\
XCC J0211.0-0905 & 32.7665 & -9.0977 & 0.45 & 0.86$\pm$1.45 & 4.28$\pm$1.70 & 48.05$\pm$19.15 & 16.90$\pm$4.04 & 2.838 &   -   & 2 & -  & - \\
XCC J0211.2-0343 & 32.8079 & -3.7226 & 0.78 & 20.48$\pm$2.30 & 7.52$\pm$1.04 & 271.79$\pm$37.86 & 37.77$\pm$3.28 & 2.526 &   -   & 1 & -  & - \\
XCC J0211.3-0927 & 32.8338 & -9.4567 & 0.71 & 1.38$\pm$1.17 & 2.73$\pm$1.05 & 88.06$\pm$33.76 & 19.60$\pm$4.52 & 2.16 &   -   & 2 & -  & - \\
XCC J0211.4-0920 & 32.8551 & -9.3366 & 0.84 & 1.20$\pm$0.59 & 3.83$\pm$1.95 & 173.23$\pm$88.38 & 26.77$\pm$8.08 & 2.148 &   -   & 2 & -  & - \\
XCC J0211.5-0939 & 32.8894 & -9.6606 & 0.51 & 4.82$\pm$1.30 & 3.03$\pm$1.41 & 40.33$\pm$18.78 & 14.70$\pm$4.07 & 2.586 & 0.4801 & 2 & -  & 1 \\
XCC J0212.2-0852 & 33.0716 & -8.8752 & 0.09 & 1.93$\pm$1.72 & 7.93$\pm$3.20 & 2.62$\pm$1.05 & 3.52$\pm$0.85 & 6.318 & 0.094 & 2 & -  & 3 \\
XCC J0213.4-0813 & 33.37 & -8.2204 & 0.26 & 2.76$\pm$1.02 & 6.96$\pm$1.82 & 18.04$\pm$4.72 & 10.86$\pm$1.74 & 4.08 & 0.2358 & 1 & 446$\pm$84  & 12 \\
XCC J0213.6-0552 & 33.4141 & -5.8707 & 1.1 & 4.40$\pm$0.75 & 2.14$\pm$0.91 & 186.85$\pm$80.19 & 22.11$\pm$5.67 & 1.704 &   -   & 2 & -  & - \\
XCC J0214.1-0630 & 33.545 & -6.516 & 0.88 & 1.03$\pm$0.93 & 1.24$\pm$0.47 & 70.62$\pm$26.81 & 14.52$\pm$3.32 & 1.698 &   -   & 2 & -  & - \\
XCC J0214.4-0627 & 33.6064 & -6.4605 & 0.25 & 13.80$\pm$1.26 & 27.67$\pm$49.35 & 69.48$\pm$123.92 & 25.73$\pm$23.81 & 5.424 & 0.2366 & 1 & 329$\pm$108  & 12 \\
XCC J0215.6-0702 & 33.9113 & -7.0478 & 1.02 & -1.78$\pm$0.33 & 1.33$\pm$0.43 & 104.43$\pm$33.88 & 16.39$\pm$3.22 & 1.614 &   -   & 2 & -  & - \\
XCC J0215.7-0654 & 33.9274 & -6.9051 & 0.23 & 2.76$\pm$1.02 & 5.79$\pm$1.48 & 17.83$\pm$4.56 & 10.62$\pm$1.66 & 3.804 & 0.2544 & 1 & 377$\pm$65  & 10 \\
XCC J0216.1-0935 & 34.0291 & -9.5885 & 0.62 & 21.55$\pm$1.39 & 5.95$\pm$0.86 & 121.63$\pm$17.66 & 26.81$\pm$2.43 & 2.706 & 0.5955 & 1 & -  & 4 \\
XCC J0216.5-0658 & 34.1416 & -6.9691 & 1.1 & -2.59$\pm$0.25 & 1.26$\pm$0.54 & 118.05$\pm$50.52 & 16.48$\pm$4.22 & 1.542 &   -   & 2 & -  & - \\
XCC J0216.7-0934 & 34.1938 & -9.5702 & 0.94 & 7.82$\pm$0.69 & 1.58$\pm$0.51 & 101.42$\pm$33.12 & 17.31$\pm$3.43 & 1.728 &   -   & 2 & -  & - \\
XCC J0217.5-0655 & 34.3779 & -6.9236 & 1.01 & 0.30$\pm$0.34 & 1.38$\pm$0.78 & 106.01$\pm$60.14 & 16.70$\pm$5.56 & 1.632 &   -   & 2 & -  & - \\
XCC J0217.5-0936 & 34.3788 & -9.6136 & 0.39 & 3.87$\pm$1.31 & 1.70$\pm$0.60 & 14.43$\pm$5.15 & 8.25$\pm$1.78 & 2.49 &   -   & 2 & -  & - \\
XCC J0217.5-0927 & 34.3874 & -9.462 & 0.46 & 5.85$\pm$3.15 & 6.80$\pm$2.38 & 120.11$\pm$42.06 & 27.48$\pm$5.82 & 2.85 & 0.56 & 1 & -  & 2 \\
XCC J0217.8-0641 & 34.4574 & -6.689 & 1.04 & -0.03$\pm$0.73 & 2.13$\pm$0.99 & 165.26$\pm$76.94 & 21.59$\pm$5.98 & 1.746 &   -   & 2 & -  & - \\
XCC J0217.9-0648 & 34.4871 & -6.8068 & 0.84 & -0.79$\pm$0.51 & 1.28$\pm$0.51 & 64.99$\pm$26.26 & 14.29$\pm$3.46 & 1.74 &   -   & 2 & -  & - \\
XCC J0218.0-0937 & 34.5029 & -9.6256 & 0.15 & 1.79$\pm$1.07 & 4.46$\pm$0.89 & 4.75$\pm$0.95 & 4.91$\pm$0.61 & 4.374 & 0.1598 & 1 & 232$\pm$73  & 9 \\
XCC J0218.3-0942 & 34.5814 & -9.7028 & 0.45 & 3.86$\pm$3.18 & 2.62$\pm$0.73 & 21.93$\pm$6.11 & 10.78$\pm$1.83 & 2.718 & 0.3908 & 1 & -  & 1 \\
XCC J0219.6-0759 & 34.9039 & -7.9882 & 0.86 & 1.13$\pm$0.94 & 1.63$\pm$0.52 & 85.26$\pm$27.47 & 16.69$\pm$3.26 & 1.806 &   -   & 1 & -  & - \\
XCC J0220.1-0836 & 35.031 & -8.6072 & 0.07 & 0.97$\pm$3.36 & 16.87$\pm$8.29 & 2.98$\pm$1.46 & 3.90$\pm$1.13 & 8.604 &   -   & 2 & -  & - \\
XCC J0220.3-0730 & 35.0849 & -7.5027 & 0.99 & -0.47$\pm$0.68 & 3.79$\pm$0.89 & 247.01$\pm$58.32 & 29.23$\pm$4.24 & 1.992 &   -   & 2 & -  & - \\
XCC J0220.6-0839 & 35.167 & -8.6639 & 0.07 & 0.97$\pm$2.91 & 5.16$\pm$2.38 & 2.48$\pm$1.14 & 3.36$\pm$0.92 & 5.292 & 0.1121 & 1 & 410$\pm$174  & 5 \\
XCC J0220.9-0838 & 35.2279 & -8.6402 & 0.51 & 6.82$\pm$4.29 & 1.21$\pm$0.41 & 21.01$\pm$7.28 & 9.30$\pm$1.95 & 2.076 & 0.5251 & 1 & -  & 1 \\
XCC J0221.2-0846 & 35.3163 & -8.7702 & 0.09 & 1.93$\pm$1.72 & 7.00$\pm$1.88 & 2.10$\pm$0.56 & 3.07$\pm$0.50 & 6.282 &   -   & 2 & -  & - \\
XCC J0221.5-0630 & 35.3822 & -6.515 & 1.03 & 3.10$\pm$0.31 & 2.18$\pm$1.03 & 165.18$\pm$78.29 & 21.78$\pm$6.13 & 1.764 &   -   & 2 & -  & - \\
XCC J0221.5-0830 & 35.3904 & -8.5108 & 1.07 & -0.68$\pm$0.41 & 1.89$\pm$0.68 & 157.79$\pm$57.54 & 20.39$\pm$4.49 & 1.686 &   -   & 2 & -  & - \\
XCC J0221.6-0618 & 35.4101 & -6.316 & 0.61 & 7.58$\pm$1.42 & 2.32$\pm$0.74 & 54.07$\pm$17.39 & 15.74$\pm$3.07 & 2.226 &   -   & 2 & -  & - \\
XCC J0221.6-0825 & 35.4113 & -8.4271 & 1.04 & -0.03$\pm$0.89 & 1.55$\pm$0.55 & 124.93$\pm$44.92 & 18.05$\pm$3.92 & 1.644 &   -   & 2 & -  & - \\
XCC J0221.9-0857 & 35.4915 & -8.9622 & 0.28 & 2.71$\pm$1.88 & 3.34$\pm$0.78 & 14.36$\pm$3.35 & 8.95$\pm$1.28 & 3.198 & 0.2933 & 1 & -  & 4 \\
XCC J0222.8-0623 & 35.7098 & -6.3935 & 0.39 & 0.87$\pm$1.31 & 3.16$\pm$0.96 & 26.20$\pm$8.01 & 12.09$\pm$2.25 & 2.826 &   -   & 2 & -  & - \\
XCC J0223.2-0830 & 35.8101 & -8.514 & 0.14 & 1.80$\pm$2.08 & 5.97$\pm$1.80 & 4.75$\pm$1.43 & 4.99$\pm$0.91 & 4.944 &   -   & 2 & -  & - \\
XCC J0223.8-0826 & 35.9663 & -8.4449 & 0.71 & 3.38$\pm$1.20 & 1.08$\pm$0.37 & 38.15$\pm$13.11 & 11.47$\pm$2.38 & 1.806 &   -   & 1 & -  & - \\
XCC J0223.8-0821 & 35.967 & -8.3552 & 0.22 & 8.8$\pm$3.06 & 13.08$\pm$2.30 & 31.25$\pm$5.50 & 15.53$\pm$1.69 & 4.716 & 0.2287 & 1 & 435$\pm$109  & 7 \\
XCC J0223.9-0830 & 35.9826 & -8.5069 & 0.16 & 2.78$\pm$1.20 & 3.96$\pm$0.96 & 4.43$\pm$1.08 & 4.69$\pm$0.70 & 4.218 & 0.1635 & 1 & 407$\pm$214  & 9 \\
XCC J0224.0-0835 & 35.998 & -8.5956 & 0.27 & 15.75$\pm$3.01 & 39.57$\pm$1.97 & 130.36$\pm$6.49 & 37.44$\pm$1.18 & 5.514 & 0.2701 & 1 & 675$\pm$142  & 10 \\
XCC J0224.1-0816 & 36.0234 & -8.2682 & 0.26 & 5.76$\pm$2.36 & 6.63$\pm$2.74 & 21.38$\pm$8.85 & 11.87$\pm$2.94 & 3.876 &   -   & 2 & -  & - \\
XCC J0224.3-0917 & 36.0903 & -9.289 & 0.67 & 0.44$\pm$0.70 & 2.05$\pm$0.66 & 59.62$\pm$19.29 & 15.85$\pm$3.11 & 2.094 &   -   & 2 & -  & - \\
XCC J0224.4-0924 & 36.0983 & -9.4054 & 0.49 & 2.85$\pm$1.22 & 1.92$\pm$0.57 & 27.29$\pm$8.18 & 11.37$\pm$2.08 & 2.346 & 0.4874 & 2 & -  & 1 \\
XCC J0224.4-0827 & 36.1046 & -8.4578 & 0.09 & 1.93$\pm$3.97 & 9.66$\pm$6.04 & 2.91$\pm$1.82 & 3.79$\pm$1.38 & 6.738 &   -   & 2 & -  & - \\
XCC J0224.6-0931 & 36.1586 & -9.5279 & 0.71 & 1.38$\pm$1.17 & 2.23$\pm$0.66 & 73.34$\pm$21.69 & 17.43$\pm$3.14 & 2.076 &   -   & 2 & -  & - \\
XCC J0224.6-0919 & 36.1606 & -9.3302 & 1.08 & 2.01$\pm$0.94 & 2.02$\pm$1.13 & 170.78$\pm$96.03 & 21.25$\pm$7.02 & 1.698 &   -   & 2 & -  & - \\
XCC J0224.7-0924 & 36.1888 & -9.4073 & 0.93 & 0.88$\pm$0.43 & 1.63$\pm$0.65 & 101.87$\pm$40.78 & 17.52$\pm$4.21 & 1.746 &   -   & 2 & -  & - \\
XCC J0224.8-0620 & 36.2207 & -6.3371 & 1.05 & 7.75$\pm$0.47 & 1.35$\pm$0.37 & 113.31$\pm$31.63 & 16.80$\pm$2.86 & 1.596 &   -   & 2 & -  & - \\
XCC J0225.0-0950 & 36.2713 & -9.8381 & 0.15 & 9.79$\pm$2.70 & 34.41$\pm$1.79 & 36.81$\pm$1.91 & 18.22$\pm$0.60 & 6.786 & 0.1594 & 1 & 528$\pm$69  & 18 \\
XCC J0225.2-0623 & 36.3021 & -6.3837 & 0.2 & 7.78$\pm$1.73 & 18.65$\pm$1.55 & 34.42$\pm$2.86 & 16.85$\pm$0.88 & 5.334 & 0.2041 & 1 & 414$\pm$78  & 10 \\
XCC J0225.5-0619 & 36.3929 & -6.3228 & 0.95 & 2.76$\pm$0.66 & 1.44$\pm$0.66 & 95.43$\pm$44.13 & 16.50$\pm$4.54 & 1.686 &   -   & 2 & -  & - \\
XCC J0225.5-0612 & 36.3953 & -6.2134 & 0.31 & 2.77$\pm$1.66 & 3.87$\pm$1.67 & 16.58$\pm$7.17 & 9.81$\pm$2.53 & 3.3 & 0.2932 & 2 & -  & 1 \\
XCC J0225.6-0946 & 36.4034 & -9.7797 & 0.34 & 5.84$\pm$1.95 & 2.24$\pm$0.81 & 13.93$\pm$5.05 & 8.41$\pm$1.84 & 2.766 & 0.3429 & 1 & 452$\pm$152  & 5 \\
XCC J0225.9-0830 & 36.479 & -8.5086 & 1.05 & 0.75$\pm$0.28 & 1.51$\pm$0.61 & 124.59$\pm$50.99 & 17.85$\pm$4.38 & 1.632 &   -   & 2 & -  & - \\
XCC J0226.4-0845 & 36.6144 & -8.766 & 0.33 & 2.84$\pm$1.18 & 2.70$\pm$1.88 & 15.29$\pm$10.64 & 9.02$\pm$3.63 & 2.922 &   -   & 2 & -  & - \\
XCC J0229.0-0549 & 37.2606 & -5.8297 & 1.02 & 0.21$\pm$0.53 & 0.76$\pm$0.42 & 64.61$\pm$36.26 & 12.05$\pm$3.97 & 1.458 &   -   & 2 & -  & - \\
XCC J0229.2-0553 & 37.3203 & -5.8983 & 0.3 & 1.75$\pm$1.16 & 5.21$\pm$0.57 & 21.88$\pm$2.42 & 11.73$\pm$0.81 & 3.522 & 0.2915 & 1 & 505$\pm$94  & 7 \\
XCC J0229.5-0553 & 37.3826 & -5.8998 & 0.3 & 4.75$\pm$1.93 & 3.31$\pm$0.41 & 14.46$\pm$1.79 & 8.97$\pm$0.69 & 3.186 & 0.295 & 1 & 322$\pm$64  & 13 \\
XCC J0230.1-0540 & 37.5371 & -5.6803 & 0.47 & 7.84$\pm$4.63 & 2.84$\pm$0.87 & 41.40$\pm$12.69 & 14.7$\pm$2.74 & 2.514 & 0.4991 & 1 & -  & 3 \\
XCC J0230.8-0421 & 37.7203 & -4.3507 & 0.16 & 2.78$\pm$2.09 & 11.74$\pm$1.96 & 9.63$\pm$1.61 & 7.83$\pm$0.81 & 5.712 & 0.1408 & 1 & 427$\pm$132  & 9 \\
XCC J0230.9-0431 & 37.7413 & -4.5285 & 0.39 & 0.87$\pm$0.93 & 3.41$\pm$2.14 & 28.08$\pm$17.65 & 12.64$\pm$4.63 & 2.868 &   -   & 2 & -  & - \\
XCC J0231.7-0452 & 37.927 & -4.8814 & 0.2 & 24.78$\pm$1.22 & 128.74$\pm$3.52 & 183.53$\pm$5.02 & 49.94$\pm$0.87 & 8.328 & 0.1852 & 1 & 426$\pm$194  & 9 \\
XCC J0232.6-0449 & 38.1654 & -4.8331 & 0.17 & 2.76$\pm$1.67 & 5.70$\pm$2.33 & 7.04$\pm$2.88 & 6.27$\pm$1.53 & 4.494 &   -   & 2 & -  & - \\
XCC J0233.4-0540 & 38.3625 & -5.6749 & 0.51 & 3.82$\pm$1.30 & 2.52$\pm$0.63 & 42.21$\pm$10.66 & 14.48$\pm$2.24 & 2.4 & 0.5287 & 1 & -  & 4 \\
XCC J0233.6-0542 & 38.4048 & -5.701 & 0.3 & 2.75$\pm$1.65 & 2.42$\pm$0.87 & 11.02$\pm$3.99 & 7.51$\pm$1.64 & 2.964 &   -   & 2 & -  & - \\
XCC J0233.6-0941 & 38.4183 & -9.6995 & 0.26 & 5.76$\pm$1.18 & 11.33$\pm$1.48 & 37.40$\pm$4.89 & 16.91$\pm$1.38 & 4.302 & 0.2646 & 1 & 395$\pm$56  & 20 \\
XCC J0233.8-0939 & 38.4607 & -9.6656 & 0.25 & 1.80$\pm$1.34 & 4.41$\pm$1.50 & 15.54$\pm$5.30 & 9.60$\pm$1.98 & 3.51 & 0.2695 & 1 & 475$\pm$117  & 7 \\
XCC J0234.3-0940 & 38.5794 & -9.6711 & 0.79 & 20.41$\pm$3.94 & 12.28$\pm$1.78 & 439.67$\pm$63.87 & 50.91$\pm$4.61 & 2.766 &   -   & 1 & -  & - \\
XCC J0234.3-0936 & 38.5851 & -9.6158 & 1.09 & 2.70$\pm$1.00 & 3.00$\pm$0.98 & 247.16$\pm$81.09 & 26.68$\pm$5.31 & 1.824 &   -   & 2 & -  & - \\
XCC J0234.3-0951 & 38.586 & -9.8583 & 0.65 & 13.51$\pm$1.13 & 14.29$\pm$2.40 & 291.32$\pm$48.93 & 46.18$\pm$4.82 & 3.18 & 0.6119 & 1 & -  & 4 \\
XCC J0234.7-0548 & 38.6768 & -5.8095 & 1.0 & 0.4$\pm$0.58 & 2.59$\pm$1.18 & 179.6$\pm$82.2 & 23.62$\pm$6.44 & 1.842 &   -   & 2 & -  & - \\
XCC J0234.9-0400 & 38.7257 & -4.013 & 0.61 & 1.58$\pm$1.42 & 0.90$\pm$0.35 & 22.80$\pm$8.84 & 9.06$\pm$2.11 & 1.854 &   -   & 2 & -  & - \\
XCC J0849.2-0252 & 132.307 & -2.8775 & 0.25 & 7.80$\pm$2.10 & 11.30$\pm$1.56 & 26.36$\pm$3.65 & 13.96$\pm$1.20 & 4.602 & 0.2259 & 1 & 491$\pm$124  & 11 \\
XCC J0849.9-0312 & 132.473 & -3.2009 & 0.6 & 4.57$\pm$1.35 & 2.18$\pm$0.84 & 49.22$\pm$18.94 & 14.96$\pm$3.46 & 2.214 &   -   & 1 & -  & - \\
XCC J0849.9-0159 & 132.491 & -1.9925 & 0.05 & 0.97$\pm$0.98 & 28.05$\pm$10.69 & 2.44$\pm$0.93 & 3.48$\pm$0.80 & 11.394 &   -   & 2 & -  & - \\
XCC J0850.0-0149 & 132.5 & -1.8261 & 0.38 & 3.86$\pm$1.27 & 1.09$\pm$0.32 & 8.80$\pm$2.61 & 6.07$\pm$1.09 & 2.292 &   -   & 1 & -  & - \\
XCC J0850.0-0235 & 132.504 & -2.5955 & 0.22 & 5.8$\pm$2.07 & 6.01$\pm$2.96 & 14.19$\pm$6.98 & 9.39$\pm$2.74 & 4.032 & 0.226 & 1 & -  & 3 \\
XCC J0850.1-0149 & 132.53 & -1.8279 & 1.02 & 8.21$\pm$0.99 & 0.68$\pm$0.18 & 58.78$\pm$16.23 & 11.34$\pm$1.91 & 1.428 &   -   & 2 & -  & - \\
XCC J0850.7-0140 & 132.674 & -1.682 & 0.26 & 2.76$\pm$1.02 & 2.66$\pm$1.06 & 8.65$\pm$3.46 & 6.65$\pm$1.60 & 3.198 &   -   & 1 & -  & - \\
XCC J0852.0-0134 & 133.022 & -1.5738 & 0.61 & 4.58$\pm$1.21 & 2.93$\pm$1.27 & 67.02$\pm$29.12 & 18.06$\pm$4.69 & 2.334 &   -   & 1 & -  & - \\
XCC J0852.2-0533 & 133.066 & -5.5651 & 0.21 & 21.79$\pm$3.64 & 88.92$\pm$2.82 & 133.65$\pm$4.24 & 40.64$\pm$0.82 & 7.638 & 0.1891 & 1 & 385$\pm$85  & 7 \\
XCC J0852.2-0101 & 133.067 & -1.0261 & 0.49 & 12.85$\pm$1.99 & 28.46$\pm$1.36 & 298.44$\pm$14.35 & 53.97$\pm$1.64 & 4.122 & 0.4587 & 1 & -  & 4 \\
XCC J0852.4-0345 & 133.118 & -3.752 & 0.98 & 5.56$\pm$0.38 & 3.09$\pm$1.09 & 200.94$\pm$71.07 & 25.85$\pm$5.52 & 1.926 &   -   & 2 & -  & - \\
XCC J0852.5-0112 & 133.123 & -1.2128 & 0.57 & 9.79$\pm$2.68 & 4.75$\pm$1.37 & 89.64$\pm$25.91 & 22.58$\pm$3.98 & 2.634 &   -   & 1 & -  & - \\
XCC J0852.6-0152 & 133.154 & -1.8826 & 0.81 & 3.31$\pm$1.62 & 2.49$\pm$1.25 & 108.55$\pm$54.40 & 20.41$\pm$6.06 & 2.01 &   -   & 2 & -  & - \\
XCC J0852.9-0503 & 133.236 & -5.0593 & 0.06 & 3.97$\pm$4.75 & 21.01$\pm$5.00 & 1.33$\pm$0.31 & 2.37$\pm$0.34 & 11.592 & 0.043 & 1 & 300$\pm$100  & 9 \\
\hline
Continued on next page.
\end{tabular}
\end{center}
\end{table*}

\setcounter{table}{4}
\begin{table*}[t]
\begin{center}
\renewcommand{\arraystretch}{1.1}\renewcommand{\tabcolsep}{0.05cm}
\caption{\footnotesize
 {Continued from previous page.}
 \label{}}
 \tiny
 \centering
\begin{tabular}{ccccccccccccc}
 \hline
 \hline
ID         & RA        & DEC          & R.S. z & $\alpha$ & X-ray  Flux                   & $L_{X}$               & $M_{200c}$          & $r_{200c}$& specz&visual flag&$\sigma(v)$ & $N_{\sigma}$\\
XMM-CFHTLS & (degrees)  & (degrees)   &        &          &$10^{-14} erg$ $cm^{-2} s^{-1}$& $10^{42} erg$ $s^{-1}$& $10^{13}M_{\odot}$ & arcmin    &       &          &($km s^{-1})$ &  \\
(1)         & (2)       & (3)         & (4)    & (5)      & (6)                           & (7)                   & (8)                & (9)       &   (10) & (11)    &(12)&(13)\\
 \hline
XCC J0853.0-0344 & 133.269 & -3.7363 & 0.94 & 26.82$\pm$2.07 & 3.76$\pm$0.82 & 218.92$\pm$47.64 & 28.33$\pm$3.80 & 2.034 &   -   & 1 & -  & - \\
XCC J0853.1-0459 & 133.282 & -4.9996 & 0.45 & 1.86$\pm$1.36 & 1.65$\pm$0.70 & 19.57$\pm$8.39 & 9.51$\pm$2.44 & 2.346 &   -   & 2 & -  & - \\
XCC J0853.3-0144 & 133.333 & -1.7485 & 0.62 & 10.55$\pm$1.39 & 1.59$\pm$0.54 & 39.87$\pm$13.64 & 12.84$\pm$2.66 & 2.058 &   -   & 1 & -  & - \\
XCC J0853.4-0341 & 133.361 & -3.6866 & 0.73 & 6.35$\pm$0.97 & 5.01$\pm$0.94 & 162.38$\pm$30.58 & 28.46$\pm$3.32 & 2.4 &   -   & 1 & -  & - \\
XCC J0853.6-0348 & 133.403 & -3.8094 & 0.85 & 7.16$\pm$0.57 & 1.23$\pm$0.42 & 64.58$\pm$22.21 & 14.10$\pm$2.93 & 1.722 &   -   & 1 & -  & - \\
XCC J0853.6-0532 & 133.416 & -5.543 & 0.61 & 1.58$\pm$0.74 & 0.91$\pm$0.26 & 23.07$\pm$6.65 & 9.13$\pm$1.60 & 1.86 &   -   & 2 & -  & - \\
XCC J0853.9-0503 & 133.484 & -5.0618 & 0.36 & 2.87$\pm$2.51 & 1.80$\pm$0.66 & 12.60$\pm$4.64 & 7.77$\pm$1.72 & 2.598 &   -   & 2 & -  & - \\
XCC J0854.1-0342 & 133.524 & -3.7154 & 0.73 & 8.35$\pm$1.68 & 4.26$\pm$1.28 & 139.87$\pm$42.23 & 25.87$\pm$4.75 & 2.328 &   -   & 1 & -  & - \\
XCC J0854.2-0221 & 133.555 & -2.3499 & 0.37 & 12.87$\pm$3.88 & 22.48$\pm$1.77 & 147.73$\pm$11.66 & 37.30$\pm$1.85 & 4.308 & 0.3679 & 1 & 451$\pm$133  & 9 \\
XCC J0854.8-0530 & 133.702 & -5.4999 & 0.06 & 0.97$\pm$2.91 & 4.95$\pm$1.53 & 0.64$\pm$0.19 & 1.47$\pm$0.27 & 7.188 &   -   & 2 & -  & - \\
XCC J0854.9-0147 & 133.743 & -1.7927 & 0.74 & 2.34$\pm$0.69 & 1.09$\pm$0.35 & 42.34$\pm$13.71 & 11.93$\pm$2.34 & 1.782 &   -   & 2 & -  & - \\
XCC J0856.4-0146 & 134.105 & -1.7725 & 0.15 & 2.79$\pm$1.75 & 5.34$\pm$1.86 & 4.95$\pm$1.72 & 5.08$\pm$1.07 & 4.674 &   -   & 2 & -  & - \\
XCC J0857.1-0106 & 134.293 & -1.1138 & 0.63 & 6.54$\pm$1.79 & 1.94$\pm$0.55 & 45.78$\pm$13.14 & 14.16$\pm$2.48 & 2.154 & 0.609 & 1 & -  & 1 \\
XCC J0857.4-0532 & 134.367 & -5.5371 & 0.08 & 0.95$\pm$0.97 & 11.57$\pm$4.55 & 2.71$\pm$1.06 & 3.64$\pm$0.86 & 7.416 &   -   & 2 & -  & - \\
XCC J0858.3-0438 & 134.595 & -4.6448 & 0.71 & 14.38$\pm$0.69 & 3.33$\pm$0.73 & 105.18$\pm$23.25 & 21.96$\pm$2.99 & 2.244 &   -   & 1 & -  & - \\
XCC J0858.6-0525 & 134.661 & -5.4212 & 0.09 & 8.93$\pm$2.81 & 37.10$\pm$2.22 & 11.63$\pm$0.69 & 9.19$\pm$0.34 & 9.048 &   -   & 1 & -  & - \\
XCC J0859.7-0419 & 134.923 & -4.3263 & 0.75 & 2.55$\pm$1.50 & 1.22$\pm$0.67 & 48.06$\pm$26.64 & 12.81$\pm$4.17 & 1.806 &   -   & 2 & -  & - \\
XCC J0900.3-0318 & 135.083 & -3.3071 & 0.15 & 4.79$\pm$1.07 & 2.29$\pm$0.79 & 2.09$\pm$0.72 & 2.92$\pm$0.61 & 3.888 &   -   & 2 & -  & - \\
XCC J0901.5-0139 & 135.377 & -1.6532 & 0.34 & 10.84$\pm$3.01 & 50.38$\pm$3.53 & 231.1$\pm$16.20 & 51.94$\pm$2.30 & 5.412 & 0.3163 & 1 & 456$\pm$69  & 20 \\
XCC J0901.6-0154 & 135.406 & -1.9074 & 0.29 & 6.73$\pm$1.95 & 5.18$\pm$0.32 & 25.91$\pm$1.63 & 12.81$\pm$0.51 & 3.408 & 0.3151 & 1 & 454$\pm$46  & 7 \\
XCC J0901.6-0158 & 135.415 & -1.9799 & 0.36 & 8.87$\pm$3.70 & 6.28$\pm$0.31 & 30.95$\pm$1.56 & 14.37$\pm$0.46 & 3.546 & 0.3141 & 1 & 516$\pm$80  & 11 \\
XCC J0901.7-0228 & 135.437 & -2.4809 & 0.93 & 1.88$\pm$1.11 & 3.32$\pm$1.15 & 190.98$\pm$66.45 & 26.20$\pm$5.51 & 1.998 &   -   & 2 & -  & - \\
XCC J0901.7-0208 & 135.439 & -2.1378 & 0.42 & 5.84$\pm$3.06 & 1.47$\pm$0.28 & 13.29$\pm$2.58 & 7.77$\pm$0.93 & 2.394 & 0.3994 & 1 & 334$\pm$64  & 5 \\
XCC J0901.8-0143 & 135.45 & -1.7226 & 0.25 & 5.80$\pm$1.09 & 8.33$\pm$2.34 & 24.49$\pm$6.87 & 13.06$\pm$2.23 & 4.134 &   -   & 2 & -  & - \\
XCC J0901.9-0200 & 135.494 & -2.0115 & 1.01 & 18.30$\pm$1.74 & 1.06$\pm$0.15 & 84.33$\pm$12.14 & 14.42$\pm$1.29 & 1.554 &   -   & 2 & -  & - \\
XCC J0902.0-0228 & 135.5 & -2.4734 & 0.95 & 0.76$\pm$1.03 & 2.75$\pm$1.45 & 169.29$\pm$89.67 & 23.81$\pm$7.44 & 1.908 &   -   & 2 & -  & - \\
XCC J0902.3-0230 & 135.582 & -2.5034 & 0.36 & 0.87$\pm$0.93 & 5.90$\pm$2.84 & 39.50$\pm$19.00 & 16.15$\pm$4.61 & 3.312 &   -   & 2 & -  & - \\
XCC J0902.4-0219 & 135.604 & -2.3188 & 0.3 & 1.75$\pm$1.16 & 2.58$\pm$0.90 & 11.74$\pm$4.11 & 7.82$\pm$1.66 & 3.006 &   -   & 2 & -  & - \\
XCC J0903.5-0518 & 135.873 & -5.3151 & 0.19 & 2.79$\pm$2.04 & 3.17$\pm$1.21 & 4.99$\pm$1.91 & 4.95$\pm$1.14 & 3.774 &   -   & 2 & -  & - \\
XCC J0904.0-0151 & 136.01 & -1.8636 & 0.72 & 2.34$\pm$0.67 & 6.45$\pm$2.54 & 198.57$\pm$78.17 & 32.68$\pm$7.73 & 2.538 &   -   & 2 & -  & - \\
XCC J0904.0-0142 & 136.02 & -1.7036 & 0.26 & 1.76$\pm$1.32 & 11.66$\pm$4.01 & 36.98$\pm$12.73 & 16.86$\pm$3.51 & 4.356 &   -   & 2 & -  & - \\
XCC J0904.1-0329 & 136.026 & -3.492 & 0.71 & 1.38$\pm$0.60 & 1.76$\pm$0.67 & 59.28$\pm$22.66 & 15.21$\pm$3.50 & 1.986 &   -   & 2 & -  & - \\
XCC J0904.1-0202 & 136.043 & -2.0333 & 0.29 & 7.73$\pm$0.97 & 14.88$\pm$4.81 & 58.20$\pm$18.83 & 22.02$\pm$4.32 & 4.392 & 0.2874 & 1 & 546$\pm$70  & 17 \\
XCC J0904.6-0202 & 136.154 & -2.0496 & 0.41 & 7.85$\pm$3.07 & 9.31$\pm$3.42 & 80.48$\pm$29.58 & 24.4$\pm$5.41 & 3.45 & 0.4087 & 1 & 553$\pm$243  & 5 \\
XCC J0904.6-0200 & 136.161 & -2.0138 & 1.03 & 7.10$\pm$0.90 & 5.18$\pm$1.65 & 356.26$\pm$113.66 & 35.62$\pm$6.90 & 2.076 &   -   & 2 & -  & - \\
XCC J2202.1+0142 & 330.539 & 1.716 & 0.21 & 4.79$\pm$1.23 & 6.05$\pm$2.03 & 13.41$\pm$4.51 & 9.10$\pm$1.85 & 4.08 & 0.2199 & 2 & 522$\pm$153  & 10 \\
XCC J2206.3+0146 & 331.576 & 1.7725 & 1.04 & 5.96$\pm$0.81 & 1.76$\pm$0.78 & 139.47$\pm$61.86 & 19.36$\pm$5.12 & 1.686 &   -   & 2 & -  & - \\
XCC J2206.4+0139 & 331.603 & 1.6554 & 0.32 & 8.80$\pm$1.09 & 7.33$\pm$1.49 & 28.21$\pm$5.75 & 13.92$\pm$1.75 & 3.828 & 0.2818 & 1 & -  & 1 \\
XCC J2212.1-0010 & 333.029 & -0.168 & 0.8 & 4.35$\pm$1.19 & 2.11$\pm$1.10 & 90.77$\pm$47.42 & 18.37$\pm$5.67 & 1.956 &   -   & 2 & -  & - \\
XCC J2212.1-0008 & 333.045 & -0.1348 & 0.36 & 4.87$\pm$3.30 & 5.61$\pm$1.01 & 38.75$\pm$7.02 & 15.88$\pm$1.78 & 3.264 & 0.3647 & 1 & -  & 3 \\
XCC J2212.2+0005 & 333.072 & 0.0957 & 0.8 & 1.35$\pm$1.02 & 2.01$\pm$1.06 & 86.90$\pm$46.13 & 17.87$\pm$5.59 & 1.938 &   -   & 1 & -  & - \\
XCC J2214.3+0047 & 333.59 & 0.7857 & 0.32 & 8.80$\pm$1.09 & 3.56$\pm$0.76 & 18.72$\pm$4.00 & 10.36$\pm$1.36 & 3.132 & 0.3202 & 1 & -  & 2 \\
XCC J2214.8+0047 & 333.706 & 0.7837 & 0.34 & 5.84$\pm$2.29 & 3.92$\pm$0.57 & 19.83$\pm$2.90 & 10.79$\pm$0.98 & 3.21 & 0.3155 & 1 & -  & 3 \\
XCC J2214.9-0039 & 333.736 & -0.6541 & 0.9 & 6.96$\pm$0.73 & 2.53$\pm$1.12 & 139.33$\pm$61.74 & 22.02$\pm$5.82 & 1.926 &   -   & 2 & -  & - \\
XCC J2217.7+0017 & 334.436 & 0.2914 & 0.71 & 2.38$\pm$1.70 & 0.48$\pm$0.07 & 18.29$\pm$2.72 & 7.17$\pm$0.66 & 1.548 &   -   & 2 & -  & - \\
XCC J2217.8+0023 & 334.458 & 0.3835 & 0.91 & 5.93$\pm$1.87 & 0.27$\pm$0.06 & 20.62$\pm$4.69 & 6.42$\pm$0.90 & 1.266 &   -   & 1 & -  & - \\
XCC J2217.8+0016 & 334.471 & 0.279 & 0.83 & 1.23$\pm$0.51 & 0.17$\pm$0.04 & 10.76$\pm$2.57 & 4.56$\pm$0.67 & 1.2 &   -   & 2 & -  & - \\
XCC J0210.4-0345 & 32.6203 & -3.7545 & 0.54 & 3.77$\pm$1.04 & 0.94$\pm$0.34 & 17.85$\pm$6.56 & 8.26$\pm$1.83 & 1.956 &   -   & 2 & -  & - \\
XCC J0211.0-0853 & 32.7522 & -8.8984 & 0.45 & 8.86$\pm$2.41 & 1.81$\pm$0.87 & 21.06$\pm$10.11 & 10.00$\pm$2.85 & 2.4 & 0.4459 & 2 & -  & 1 \\
XCC J0214.1-0808 & 33.5342 & -8.1451 & 0.25 & 2.80$\pm$1.67 & 2.90$\pm$1.01 & 8.57$\pm$2.98 & 6.67$\pm$1.40 & 3.312 & 0.2495 & 1 & 180$\pm$55  & 13 \\
XCC J0214.7-0618 & 33.676 & -6.309 & 0.25 & 5.80$\pm$1.79 & 3.06$\pm$1.31 & 8.23$\pm$3.53 & 6.55$\pm$1.68 & 3.408 & 0.2395 & 1 & 528$\pm$30  & 25 \\
XCC J0214.7-0804 & 33.6919 & -8.069 & 0.7 & 3.39$\pm$1.54 & 0.86$\pm$0.62 & 29.99$\pm$21.66 & 9.93$\pm$4.13 & 1.74 &   -   & 1 & -  & - \\
XCC J0215.0-0626 & 33.764 & -6.4469 & 0.2 & 0.78$\pm$1.73 & 9.97$\pm$3.85 & 2.43$\pm$0.94 & 3.40$\pm$0.79 & 7.11 & 0.0817 & 1 & -  & - \\
XCC J0216.1-0702 & 34.0324 & -7.0483 & 0.43 & 6.85$\pm$1.22 & 0.80$\pm$0.40 & 7.86$\pm$3.91 & 5.49$\pm$1.62 & 2.088 & 0.4114 & 1 & 232$\pm$74  & 9 \\
XCC J0216.7-0648 & 34.1805 & -6.8093 & 0.19 & 2.79$\pm$2.04 & 3.19$\pm$1.48 & 5.01$\pm$2.32 & 4.97$\pm$1.37 & 3.774 &   -   & 2 & -  & - \\
XCC J0216.7-0935 & 34.1859 & -9.585 & 0.58 & 5.78$\pm$1.73 & 0.69$\pm$0.34 & 16.70$\pm$8.22 & 7.53$\pm$2.20 & 1.776 & 0.5941 & 1 & -  & 3 \\
XCC J0216.8-0918 & 34.2012 & -9.3103 & 0.73 & 11.35$\pm$0.68 & 2.95$\pm$1.46 & 78.05$\pm$38.83 & 19.15$\pm$5.64 & 2.274 & 0.652 & 2 & -  & 3 \\
XCC J0219.3-0735 & 34.8297 & -7.5998 & 0.58 & 9.78$\pm$2.73 & 2.37$\pm$0.69 & 47.02$\pm$13.73 & 14.96$\pm$2.66 & 2.304 & 0.5684 & 1 & -  & 2 \\
XCC J0221.5-0626 & 35.394 & -6.4377 & 0.3 & 7.75$\pm$1.65 & 2.10$\pm$0.86 & 10.67$\pm$4.37 & 7.27$\pm$1.78 & 2.826 & 0.314 & 1 & -  & 2 \\
XCC J0222.2-0617 & 35.5689 & -6.2946 & 0.75 & 2.55$\pm$1.59 & 1.27$\pm$0.67 & 53.36$\pm$28.24 & 13.43$\pm$4.19 & 1.8 & 0.7716 & 2 & -  & 1 \\
XCC J0223.5-0828 & 35.8887 & -8.4739 & 0.25 & 5.80$\pm$1.09 & 0.50$\pm$0.63 & 1.98$\pm$2.47 & 2.54$\pm$1.72 & 2.166 & 0.2826 & 1 & 333$\pm$72  & 13 \\
XCC J0224.4-0915 & 36.1137 & -9.2658 & 0.33 & 1.84$\pm$2.27 & 2.93$\pm$1.29 & 16.56$\pm$7.29 & 9.50$\pm$2.49 & 2.97 &   -   & 2 & -  & - \\
XCC J0225.7-0828 & 36.4279 & -8.4744 & 0.52 & 3.80$\pm$1.09 & 1.62$\pm$1.19 & 31.10$\pm$22.95 & 11.65$\pm$4.94 & 2.16 & 0.5531 & 2 & -  & 2 \\
XCC J0230.9-0418 & 37.7467 & -4.3051 & 0.14 & 1.80$\pm$1.09 & 5.02$\pm$2.49 & 4.16$\pm$2.06 & 4.57$\pm$1.34 & 4.716 & 0.1428 & 1 & 396$\pm$119  & 5 \\
XCC J0233.3-0550 & 38.3245 & -5.8364 & 0.32 & 1.80$\pm$1.34 & 1.65$\pm$1.49 & 8.02$\pm$7.24 & 6.08$\pm$3.10 & 2.7 & 0.3086 & 2 & -  & 3 \\
XCC J0233.8-0543 & 38.4688 & -5.7187 & 0.39 & 3.87$\pm$2.14 & 1.07$\pm$0.48 & 7.37$\pm$3.30 & 5.53$\pm$1.47 & 2.334 & 0.3566 & 2 & -  & 4 \\
XCC J0234.7-0542 & 38.6846 & -5.7055 & 0.12 & 0.80$\pm$1.09 & 4.09$\pm$2.70 & 3.28$\pm$2.16 & 3.93$\pm$1.50 & 4.53 & 0.1412 & 1 & -  & 1 \\
XCC J0849.2-0157 & 132.309 & -1.9545 & 0.2 & 0.78$\pm$0.88 & 1.64$\pm$0.63 & 2.87$\pm$1.10 & 3.44$\pm$0.79 & 3.198 &   -   & 2 & -  & - \\
XCC J0850.3-0324 & 132.593 & -3.4152 & 0.52 & 5.80$\pm$1.26 & 2.71$\pm$1.53 & 46.82$\pm$26.56 & 15.36$\pm$5.11 & 2.418 & 0.537 & 1 & -  & 3 \\
XCC J0850.4-0312 & 132.61 & -3.2091 & 0.45 & 5.86$\pm$2.41 & 2.21$\pm$0.89 & 29.85$\pm$12.10 & 12.14$\pm$2.95 & 2.43 & 0.4789 & 1 & -  & 2 \\
XCC J0851.2-0528 & 132.806 & -5.4789 & 0.83 & 6.23$\pm$0.51 & 2.34$\pm$1.15 & 108.65$\pm$53.76 & 20.02$\pm$5.87 & 1.962 & 0.8311 & 1 & -  & 1 \\
XCC J0851.4-0532 & 132.861 & -5.5427 & 0.22 & 1.8$\pm$1.08 & 2.43$\pm$1.09 & 9.14$\pm$4.13 & 6.79$\pm$1.83 & 3.06 & 0.2768 & 2 & -  & 3 \\
XCC J0851.4-0537 & 132.869 & -5.6233 & 0.72 & 2.34$\pm$2.00 & 2.02$\pm$1.01 & 69.06$\pm$34.56 & 16.62$\pm$4.92 & 2.028 &   -   & 2 & -  & - \\
XCC J0851.5-0104 & 132.884 & -1.0747 & 0.8 & 3.35$\pm$2.13 & 1.11$\pm$1.24 & 51.49$\pm$57.04 & 12.78$\pm$7.81 & 1.734 &   -   & 2 & -  & - \\
XCC J0851.5-0451 & 132.893 & -4.8642 & 0.06 & 0.97$\pm$3.36 & 9.00$\pm$3.79 & 2.05$\pm$0.86 & 3.05$\pm$0.77 & 7.062 & 0.0792 & 1 & -  & 2 \\
XCC J0851.6-0451 & 132.913 & -4.8567 & 0.62 & 6.55$\pm$1.39 & 1.79$\pm$1.01 & 46.06$\pm$26.01 & 13.93$\pm$4.62 & 2.088 & 0.6309 & 1 & -  & 2 \\
XCC J0851.9-0507 & 132.981 & -5.1185 & 0.39 & 4.87$\pm$1.31 & 5.39$\pm$3.18 & 45.42$\pm$26.81 & 17.08$\pm$5.90 & 3.126 & 0.3981 & 1 & -  & 4 \\
XCC J0852.8-0152 & 133.203 & -1.8718 & 0.93 & 2.88$\pm$1.22 & 1.34$\pm$0.95 & 85.73$\pm$61.03 & 15.69$\pm$6.44 & 1.686 &   -   & 1 & -  & - \\
XCC J0852.8-0137 & 133.219 & -1.6214 & 0.39 & 8.87$\pm$1.51 & 2.44$\pm$0.76 & 20.45$\pm$6.39 & 10.32$\pm$1.96 & 2.682 &   -   & 1 & -  & - \\
XCC J0852.9-0529 & 133.244 & -5.493 & 0.58 & 8.78$\pm$2.12 & 4.53$\pm$2.09 & 89.27$\pm$41.30 & 22.31$\pm$6.14 & 2.592 &   -   & 1 & -  & - \\
XCC J0853.1-0348 & 133.296 & -3.8084 & 0.8 & 4.35$\pm$1.19 & 1.56$\pm$0.49 & 69.43$\pm$21.88 & 15.47$\pm$2.96 & 1.848 &   -   & 2 & -  & - \\
XCC J0853.8-0223 & 133.448 & -2.392 & 0.36 & 2.87$\pm$1.31 & 2.87$\pm$1.83 & 24.40$\pm$15.57 & 11.52$\pm$4.27 & 2.76 & 0.3938 & 2 & -  & 3 \\
XCC J0854.5-0140 & 133.646 & -1.6745 & 0.62 & 6.55$\pm$2.49 & 3.36$\pm$1.37 & 68.67$\pm$28.14 & 18.81$\pm$4.62 & 2.442 & 0.5833 & 1 & -  & 1 \\
XCC J0855.7-0146 & 133.933 & -1.7745 & 0.52 & 0.80$\pm$2.53 & 1.09$\pm$0.33 & 20.03$\pm$6.20 & 8.92$\pm$1.68 & 2.022 &   -   & 2 & -  & - \\
XCC J0856.4-0136 & 134.098 & -1.6031 & 0.45 & 4.86$\pm$2.06 & 2.33$\pm$0.60 & 26.40$\pm$6.83 & 11.58$\pm$1.83 & 2.526 & 0.4443 & 1 & -  & 3 \\
XCC J0856.4-0107 & 134.122 & -1.1329 & 0.58 & 4.78$\pm$1.06 & 1.22$\pm$0.78 & 26.80$\pm$17.27 & 10.33$\pm$3.87 & 2.004 &   -   & 1 & -  & - \\
 \hline
 Continued on next page.
\end{tabular}
\end{center}
\end{table*}

 \setcounter{table}{4}
 \begin{table*}[t]
 \begin{center}
 \renewcommand{\arraystretch}{1.1}\renewcommand{\tabcolsep}{0.05cm}
 \caption{\footnotesize
 {Continued from previous page.}
 \label{}}
 \tiny
 \centering
 \begin{tabular}{ccccccccccccc}
 \hline
 \hline
ID         & RA        & DEC          & R.S. z & $\alpha$ & X-ray  Flux                   & $L_{X}$               & $M_{200c}$          & $r_{200c}$& specz&visual flag&$\sigma(v)$ & $N_{\sigma}$\\
XMM-CFHTLS & (degrees)  & (degrees)   &        &          &$10^{-14} erg$ $cm^{-2} s^{-1}$& $10^{42} erg$ $s^{-1}$& $10^{13}M_{\odot}$ & arcmin    &       &          &($km s^{-1})$ &  \\
(1)         & (2)       & (3)         & (4)    & (5)      & (6)                           & (7)                   & (8)                & (9)       &   (10) & (11)    &(12)&(13)\\
 \hline
XCC J0858.1-0342 & 134.532 & -3.7164 & 0.77 & 2.50$\pm$0.79 & 1.60$\pm$0.59 & 65.22$\pm$24.15 & 15.29$\pm$3.41 & 1.884 &   -   & 2 & -  & - \\
XCC J0858.9-0433 & 134.737 & -4.5637 & 0.09 & 1.93$\pm$1.72 & 4.40$\pm$3.04 & 1.32$\pm$0.91 & 2.28$\pm$0.91 & 5.688 &   -   & 1 & -  & - \\
XCC J0859.4-0432 & 134.872 & -4.5438 & 0.11 & 0.84$\pm$0.91 & 8.17$\pm$2.95 & 3.82$\pm$1.38 & 4.44$\pm$0.96 & 5.904 &   -   & 1 & -  & - \\
XCC J0859.6-0416 & 134.906 & -4.2764 & 0.14 & 0.80$\pm$1.09 & 2.54$\pm$1.06 & 1.98$\pm$0.83 & 2.85$\pm$0.71 & 4.104 &   -   & 2 & -  & - \\
XCC J0859.9-0422 & 134.993 & -4.3686 & 0.16 & 5.78$\pm$2.09 & 11.66$\pm$5.25 & 12.70$\pm$5.72 & 9.22$\pm$2.47 & 5.388 &   -   & 1 & -  & - \\
XCC J0900.7-0306 & 135.173 & -3.1143 & 0.25 & 4.80$\pm$2.19 & 3.32$\pm$0.85 & 8.08$\pm$2.08 & 6.53$\pm$1.03 & 3.528 & 0.2292 & 1 & -  & 4 \\
XCC J0901.7-0138 & 135.435 & -1.6384 & 0.3 & 9.75$\pm$1.16 & 6.77$\pm$2.06 & 30.01$\pm$9.14 & 14.26$\pm$2.64 & 3.672 &   -   & 1 & -  & - \\
XCC J0902.3-0226 & 135.583 & -2.4422 & 0.14 & 1.80$\pm$1.34 & 6.22$\pm$2.16 & 4.96$\pm$1.72 & 5.12$\pm$1.07 & 4.986 &   -   & 1 & -  & - \\
XCC J0902.8-0213 & 135.708 & -2.2305 & 0.94 & 3.82$\pm$0.69 & 1.58$\pm$1.33 & 101.26$\pm$85.60 & 17.30$\pm$8.30 & 1.728 &   -   & 1 & -  & - \\
XCC J0903.1-0537 & 135.793 & -5.6259 & 0.3 & 2.75$\pm$1.16 & 5.41$\pm$3.69 & 24.16$\pm$16.5 & 12.41$\pm$4.90 & 3.504 &   -   & 1 & -  & - \\
XCC J0904.0-0343 & 136.0 & -3.7195 & 1.0 & 11.4$\pm$1.22 & 1.44$\pm$0.51 & 107.14$\pm$38.14 & 16.97$\pm$3.65 & 1.65 &   -   & 1 & -  & - \\
XCC J0904.2-0158 & 136.065 & -1.9808 & 0.15 & 2.79$\pm$1.07 & 15.31$\pm$8.29 & 14.47$\pm$7.83 & 10.10$\pm$3.22 & 5.88 &   -   & 1 & -  & - \\
XCC J2200.4+0058 & 330.103 & 0.9804 & 0.09 & 1.93$\pm$1.72 & 6.93$\pm$3.67 & 2.64$\pm$1.39 & 3.53$\pm$1.10 & 5.94 & 0.1005 & 2 & -  & 3 \\
XCC J2200.9+0125 & 330.241 & 1.4297 & 0.09 & 0.93$\pm$1.72 & 8.21$\pm$2.99 & 3.87$\pm$1.41 & 4.48$\pm$0.98 & 5.898 & 0.1105 & 1 & 356$\pm$177  & 6 \\
XCC J2201.4+0152 & 330.37 & 1.8668 & 0.19 & 2.79$\pm$1.23 & 4.88$\pm$2.59 & 4.46$\pm$2.36 & 4.75$\pm$1.49 & 4.596 & 0.1492 & 2 & -  & 1 \\
XCC J2202.3+0148 & 330.573 & 1.8152 & 0.19 & 0.79$\pm$1.23 & 2.25$\pm$1.45 & 3.51$\pm$2.27 & 3.95$\pm$1.48 & 3.498 &   -   & 1 & -  & - \\
XCC J2204.5+0239 & 331.131 & 2.6648 & 0.58 & 5.78$\pm$1.06 & 2.63$\pm$0.89 & 67.63$\pm$22.99 & 17.66$\pm$3.63 & 2.238 & 0.6404 & 1 & -  & 1 \\
XCC J2210.4+0203 & 332.605 & 2.0554 & 0.74 & 12.34$\pm$1.14 & 1.58$\pm$0.70 & 59.09$\pm$26.29 & 14.76$\pm$3.92 & 1.914 &   -   & 2 & -  & - \\
XCC J2211.3+0200 & 332.832 & 2.0111 & 0.46 & 5.85$\pm$1.99 & 3.06$\pm$2.02 & 37.17$\pm$24.62 & 14.19$\pm$5.45 & 2.628 & 0.4617 & 2 & -  & 2 \\
XCC J2211.3+0000 & 332.834 & 0.0103 & 0.81 & 3.31$\pm$0.66 & 0.61$\pm$0.28 & 30.88$\pm$14.34 & 9.13$\pm$2.52 & 1.536 &   -   & 2 & -  & - \\
XCC J2211.9-0001 & 332.981 & -0.0311 & 0.06 & 2.97$\pm$3.36 & 7.09$\pm$6.51 & 0.90$\pm$0.83 & 1.83$\pm$0.94 & 7.734 &   -   & 2 & -  & - \\
XCC J2214.0+0057 & 333.512 & 0.9574 & 0.76 & 8.53$\pm$0.83 & 1.65$\pm$1.04 & 65.15$\pm$41.12 & 15.42$\pm$5.67 & 1.908 &   -   & 1 & -  & - \\
XCC J2214.0-0055  & 333.52 & -0.9273 & 0.25 & 4.80$\pm$2.10 & 2.16$\pm$0.90 & 4.77$\pm$1.99 & 4.69$\pm$1.17 & 3.264 & 0.2207 & 1 & -  & 4 \\
XCC J2217.0+0016 & 334.258 & 0.2693 & 0.96 & 7.71$\pm$0.76 & 0.08$\pm$0.03 & 9.12$\pm$3.90 & 3.64$\pm$0.93 & 1.014 &   -   & 2 & -  & - \\
 \hline
 \end{tabular}
 \end{center}
 \end{table*}

\begin{table*}
\begin{center}
\renewcommand{\arraystretch}{1.1}\renewcommand{\tabcolsep}{0.01cm}
\caption{\footnotesize
{Catalog of RASS-CFHTLS X-ray Selected Galaxy Clusters }
\label{rass_cat}}
\tiny
\centering
\begin{tabular}{ccccccccccccc}
\hline
\hline
ID                                 & RA            & DEC         & R.S. z      & $\alpha$    & Opt. R.A.   & Opt. Dec.      & X-ray Flux     &  $L_{X}$    & specz  & $\sigma(v)$   & $N_{\sigma}$ & log[ $L_\mathrm{X}(L_\mathrm{S})$ ] \\
RASS-CFHTLS        & (degrees) & (degrees) &                 &                     & (degrees) & (degrees)     & $10^{-13} erg$ $cm^{-2} s^{-1}$ & $10^{42} erg$ $s^{-1}$ &      & ($km s^{-1})$ &              & $erg$ $s^{-1}$ \\
(1)                               & (2)             & (3)               & (4)          & (5)                & (6)             & (7)                  & (8)                   & (9)                                                    & (10)          & (11)         & (12)   & (13) \\

\hline

  RCC  J0202.7-0700 &    30.556  &    -7.00291  &    0.06 &    26.1$\pm$2.2 &     30.5658 &     -7.01287 &     12.05$\pm$2.98 & 15.0$\pm$3.78    &     -           &     -                   &     -    &     43.0297 \\ 
  RCC  J0203.2-0949 &    30.8715 &    -9.82674  &    0.33 &    33.7$\pm$2.9 &     30.859  &     -9.82075 &     5.15$\pm$1.38 &  325.6$\pm$80.3   &   0.3216  &     -                   &     1    &     43.9727 \\ 
  RCC  J0204.6-0931 &    31.151  &    -9.53     &    0.58 &    15.8$\pm$1.5 &     31.179  &     -9.52669 &     1.11$\pm$0.59      &  253.4$\pm$141.3 &    0.6232 &     -                   &     2    &     43.7222 \\ 
  RCC  J0205.2-0544 &    31.3627 &    -5.7708   &    0.32 &    24.2$\pm$2.1 &     31.3708 &     -5.73796 &     3.15$\pm$1.12 &  156.9$\pm$53.2   &    0.2972 &     -                   &     2    &     44.0285 \\ 
  RCC  J0206.6-0943 &    31.5231 &    -9.74276  &    0.1  &    13.3$\pm$1.9 &     31.5168 &     -9.72768 &     2.36$\pm$1.09 &  7.8$\pm$3.4          &   0.0857  &     -                   &     4    &     43.0459 \\ 
  RCC  J0208.6-0554 &    32.0915 &    -5.90348  &    0.07 &    25.3$\pm$4.8 &     32.0871 &     -5.91055 &     2.91$\pm$1.35 & 4.2$\pm$2.2          &     -           &     -                    &     -    &     43.3055  \\ 
  RCC  J0210.9-0633 &    32.6952 &    -6.58427  &    0.06 &    31.4$\pm$10.5 &     32.6897 &     -6.56617 &     7.82$\pm$1.82 & 4.6$\pm$1.7        &     0.0416 &     -                  &     1     &     43.4136 \\ 
  RCC  J0211.0-0454 &    32.7836 &    -4.9044   &    0.14 &    15.1$\pm$2.1 &     32.7694 &     -4.90051 &     1.52$\pm$0.78 &  11.8$\pm$6.4        &    0.1379 &     -                   &     3     &     43.3059 \\ 
  RCC  J0214.9-0627 &    33.6162 &    -6.47615  &    0.24 &    28.8$\pm$0.6 &     33.6165 &     -6.46603 &     4.00$\pm$1.40 &  65.0$\pm$27.7    &    0.2366 &    344$\pm$93 &     13    &     43.857 \\ 
  RCC  J0214.0-0433 &    33.661  &    -4.56333  &    0.16 &    31.3$\pm$2.1 &     33.6726 &     -4.55127 &     8.18$\pm$1.76 &   93.9$\pm$20.7     &   0.1456 &     -                    &     4     &     44.023 \\ 
  RCC  J0214.6-0355 &    33.6875 &    -3.94368  &    0.16 &    16.7$\pm$1.3 &     33.7144 &     -3.92789 &     16.60$\pm$3.19 & 96.0$\pm$20.5    &     0.1402 &     -                 &     1    &     43.601 \\ 
  RCC  J0214.3-0349 &    33.7193 &    -3.83385  &    0.71 &    17.1$\pm$1.8 &     33.7035 &     -3.82267 &     2.68$\pm$1.19 &  583.4$\pm$270.8 &    -          &     -                      &     -    &     44.0768 \\ 
  RCC  J0221.5-0545 &    35.5108 &    -5.76391  &    0.26 &    17.0$\pm$1.0 &     35.4915 &     -5.75868 &     2.03$\pm$0.94 &  92.0$\pm$37.1     &    0.2591 &     -                    &     2     &     43.4434 \\ 
  RCC  J0223.3-0851 &    35.8678 &    -8.86884  &    0.19 &    24.4$\pm$1.4 &     35.8689 &     -8.85491 &     2.13$\pm$0.93 &  46.2$\pm$16.5     &    0.1632 &  381$\pm$154 &     5     &     43.6504 \\ 
  RCC  J0223.8-0857 &    35.8981 &    -8.98174  &    0.43 &    24.4$\pm$2.4 &     35.8944 &     -8.96471 &     1.76$\pm$0.87 & 92.7$\pm$57.0      &    0.4145 &     -                      &     3     &     43.7293  \\ 
  RCC  J0223.6-0821 &    35.9671 &    -8.36164  &    0.24 &    17.2$\pm$1.4 &     35.9666 &     -8.36037 &     2.04$\pm$0.90 & 44.4$\pm$21.0      &    0.2287 &   435$\pm$109 &     7     &     43.6225 \\ 
  RCC  J0223.7-0835 &    35.977  &    -8.60424  &    0.27 &    32.8$\pm$2.3 &     35.9944 &     -8.59622 &     3.51$\pm$1.38 &  102.3$\pm$39.3    &  0.2701 &     675$\pm$142 &     10    &     43.7919 \\ 
  RCC  J0225.4-0949 &    36.285  &    -9.83422  &    0.18 &    22.1$\pm$1.9 &     36.2683 &     -9.82379 &     3.45$\pm$1.16 &  62.3$\pm$18.6      &  0.1594 &       516$\pm$74 &     18    &     43.6489 \\ 
  RCC  J0225.6-0623 &    36.3029 &    -6.39542  &    0.2  &    21.0$\pm$1.1 &     36.3018 &     -6.3939  &     5.65$\pm$1.77 &   97.7$\pm$32.3      & 0.2041 &        452$\pm$82 &     12    &     43.6016 \\ 
  RCC  J0231.6-0452 &    37.9465 &    -4.86471  &    0.21 &    46.6$\pm$2.5 &     37.9256 &     -4.87712 &     9.69$\pm$2.44 & 204.7$\pm$49.8   &   0.1852 &    425$\pm$194 &     9     &     44.0886 \\ 
  RCC  J0233.0-0942 &    38.4439 &    -9.70813  &    0.25 &    21.4$\pm$2.1 &     38.426  &     -9.70045 &     2.80$\pm$1.29 &   145.6$\pm$56.5  &   0.2646 &       382$\pm$57 &     19    &     43.4411 \\ 
  RCC  J0849.7-0252 &    132.292 &    -2.89579  &    0.25 &    13.5$\pm$1.6 &     132.306 &     -2.87977 &     1.52$\pm$0.85 &   52.1$\pm$23.9   &   0.2259 &     516$\pm$125 &     12    &     43.2838 \\ 
  RCC  J0851.3-0416 &    132.949 &    -4.26784  &    0.26 &    17.0$\pm$1.3 &     132.915 &     -4.27156 &     1.33$\pm$0.71 &  28.1$\pm$19.0     &    -           &    -                          &     -    &     43.5926 \\ 
  RCC  J0852.7-0101 &    133.048 &    -1.02552  &    0.49 &    40.6$\pm$2.3 &     133.055 &     -1.02899 &     4.11$\pm$1.46 &  247.0$\pm$113.7 &    0.4587 &    415$\pm$143 &     5     &     44.1486 \\ 
  RCC  J0852.5-0534 &    133.063 &    -5.57346  &    0.19 &    46.9$\pm$1.9 &     133.059 &     -5.57561 &     8.99$\pm$2.06 &  130.2$\pm$33.3  &    0.1893 &     620$\pm$156 &     9     &     43.9612 \\ 
  RCC  J0854.1-0221 &    133.569 &    -2.34827  &    0.36 &    29.5$\pm$2.5 &     133.562 &     -2.35264 &     3.39$\pm$1.19 &  241.3$\pm$87.5  &    0.3679 &     451$\pm$133 &     9     &     43.7084 \\ 
  RCC  J0856.6-0108 &    134.114 &    -1.15371  &    0.59 &    14.5$\pm$1.7 &     134.113 &     -1.14425 &     3.28$\pm$1.17 &  644.7$\pm$250.2 &    0.623  &   -                        &     1    &     43.3342 \\ 
  RCC  J0857.1-0343 &    134.315 &    -3.7199   &    0.22 &    26.2$\pm$1.9 &     134.332 &     -3.71843 &     3.03$\pm$1.22 &   21.1$\pm$13.5     &   -            &       -                        &     -   &     43.7693 \\ 
  RCC  J0858.6-0525 &    134.668 &    -5.41993  &    0.07 &    30.4$\pm$1.6 &     134.674 &     -5.42784 &     4.70$\pm$1.45 &  13.0$\pm$4.2       &    -          &      -                        &     -    &     43.5323 \\ 
  RCC  J0901.8-0138 &    135.387 &    -1.64256  &    0.3  &    38.5$\pm$0.9 &     135.391 &     -1.6481  &     3.96$\pm$1.23 &     145.8$\pm$55.3 & 0.3163 &     358$\pm$45      &     17    &     43.9377 \\ 
  RCC  J0901.9-0158 &    135.424 &    -1.96503  &    0.35 &    14.6$\pm$1.1 &     135.412 &     -1.98222 &     1.87$\pm$0.87 &   106.1$\pm$47.0 &   0.3131 &     387$\pm$104 &     8     &     43.709  \\ 
  RCC  J2214.4-0055 &    333.571 &    -0.953427 &    0.26 &    18.2$\pm$1.2 &     333.575 &     -0.9233  &     4.76$\pm$1.82 &   114.9$\pm$47.0 &   -         &           -             &     -     &     43.6338 \\ 

\hline
\end{tabular}
\end{center}
\end{table*}

\acknowledgments

  We thank Mara Salvato, Daniele Pierini, Claudia Maraston, and
  Natascha Greisel for help in providing stellar population models.
We thank Charles IV Kirkpartick for the comments on the manuscript.

This work is based on observations obtained with MegaPrime/MegaCam, a
joint project of CFHT and CEA/DAPNIA, at the Canada-France-Hawaii
Telescope (CFHT) which is operated by the National Research Council
(NRC) of Canada, the Institut National des Sciences de l'Univers of
the Centre National de la Recherche Scientifique (CNRS) of France, and
the University of Hawaii. This research used the facilities of the
Canadian Astronomy Data Centre operated by the National Research
Council of Canada with the support of the Canadian Space Agency.
CFHTLenS data processing was made possible thanks to significant
computing support from the NSERC Research Tools and Instruments grant
program.
This work has been supported by a DLR project 50~OR~1013 to MPE.
KK acknowledges support from the Magnus Ehrnrooth foundation.  AF \&
KK wish to acknowledge Finnish Academy award, decision 266918.
 Funding for SDSS-III has been provided by the Alfred P. Sloan Foundation,
  the Participating Institutions, the National Science Foundation, and the
  U.S. Department of Energy Office of Science. The SDSS-III web site is
  http://www.sdss3.org/.
 SDSS-III is managed by the Astrophysical Research Consortium for the
  Participating Institutions of the SDSS-III Collaboration including
 the University of Arizona, the Brazilian Participation Group,
  Brookhaven National Laboratory, University of Cambridge, Carnegie
  Mellon University, University of Florida, the French Participation
  Group, the German Participation Group, Harvard University, the
  Instituto de Astrofisica de Canarias, the Michigan State/Notre
  Dame/JINA Participation Group, Johns Hopkins University, Lawrence
  Berkeley National Laboratory, Max Planck Institute for Astrophysics,
  Max Planck Institute for Extraterrestrial Physics, New Mexico State
  University, New York University, Ohio State University, Pennsylvania
  State University, University of Portsmouth, Princeton University,
  the Spanish Participation Group, University of Tokyo, University of Utah,
  Vanderbilt University, University of Virginia, University of Washington,
  and Yale University. 

\appendix

\section{Color variation of early-type galaxies as a function of redshift}\label{app1}

\begin{figure}[h]
\centering
\includegraphics[width=0.45\textwidth]{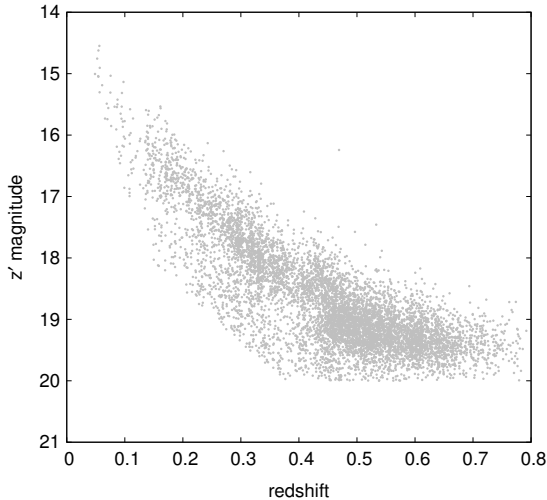}
\caption{Magnitude--redshift distribution of the ETGs
brighter than $z^\prime$=20 and $m_\mathrm{\ast}(z)$.
\label{zband_z}}
\end{figure}

In this section, we show that intrinsic colors of ETGs have a linear like
evolution through the redshift. In section \ref{red_seq_method} we derived
the color evolution of ETGs using a sample of galaxies with spectroscopic
redshift. Here, we used a sub-sample of those galaxies to show that
the linear assumption about evolution of intrinsic color dispersion for
red sequence galaxies is acceptable. To reduce the effect of error in galaxies
observed magnitude, the sample of section \ref{red_seq_method} was cut by
20 $<$ $z^\prime$ and brighter than $m_\mathrm{\ast}(z)$. Figure \ref{zband_z}
shows magnitude and the redshift distribution of this sample.

\begin{figure}
\centering
\includegraphics[width=0.45\textwidth]{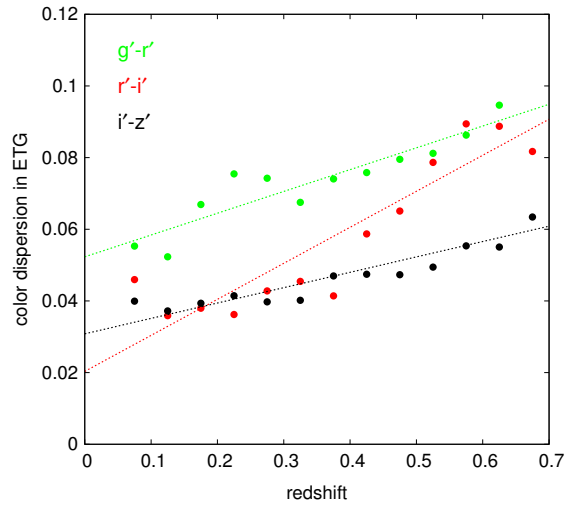}
\caption{Color evolution of ETGs as functions of redshifts.
The green, red and black dots respectively shows the evolution
of $g^\prime$-$r^\prime$, $r^\prime$-$i^\prime$ and  $i^\prime$-$z^\prime$ 
in ETGs with spectroscopic redshift. The dashed lines with the same color codes
are linear fit to each galaxy color.
\label{dcolor_etg_z}}
\end{figure} 

For the faintest galaxies in the sample ($z^\prime$=20), the typical
error in $g^\prime$, $r^\prime$, and $z^\prime$ are $\sim$0.01, 0.008,
0.005, and 0.01. Thus magnitude errors can not induce significant
effect on dispersion of colors. Dots in Figure \ref{dcolor_etg_z}
illustrates the color evolution of ETGs as a function of different
redshift bins.  Dashed lines are linear fitted lines on the color
evolution. The mean difference between linear fits and measured
color dispersions at a given redshift are 10\%, 18\% and, 7\%
respectively for $g^\prime$-$r^\prime$, $r^\prime$-$i^\prime$ and
$i^\prime$-$z^\prime$. Thus, considering linear evolution for intrinsic
color dispersion of ETGs is acceptable and we generalised this
assumption to red sequence galaxies.

\bibliographystyle{apj}


\end{document}